\documentclass[fleqn,usenatbib]{mnras}

\usepackage{newtxtext,newtxmath}

\usepackage[T1]{fontenc}

\DeclareRobustCommand{\VAN}[3]{#2}
\let\VANthebibliography\thebibliography
\def\thebibliography{\DeclareRobustCommand{\VAN}[3]{##3}\VANthebibliography}

\usepackage{graphicx}	
\usepackage{amsmath}	
\usepackage[british]{babel}
\usepackage{csquotes}
\usepackage{changepage}
\usepackage{caption}
\usepackage[export]{adjustbox}
\usepackage{mwe}
\usepackage{float}
\usepackage{multirow}
\usepackage{bold-extra}
\usepackage{bm}	 
\usepackage{color}
\usepackage{pdflscape}
\usepackage{rotating}
\usepackage[table,xcdraw]{xcolor}

\title[Energy dependent Time variability]{Time-Lag properties associated with LFQPO in X-ray variability classes of GRS 1915+105: Findings from \textit{AstroSat}}
\author[P. Majumder et al.]{
Prajjwal Majumder$^{1}$\thanks{E-mail: \href{mailto:majumderprajjwal@gmail.com}
{majumderprajjwal@gmail.com}},
Broja G. Dutta$^{1}$\thanks{E-mail: \href{mailto:brojadutta@gmail.com}
{brojadutta@gmail.com}},
Anuj Nandi$^{2}$\thanks{E-mail: \href{mailto:anuj@ursc@gov.in}
{anuj@ursc.gov.in}}
\\
$^{1}$Department of Physics, Rishi Bankim Chandra College, Naihati, West Bengal 743165, India. \\
$^{2}$Space Astronomy Group, ISITE Campus, U. R. Rao Satellite Centre, Outer Ring Road, Marathahalli, Bangalore, 560037, India.
}

\date{Accepted XXX. Received YYY; in original form ZZZ}

\pubyear{\the\year{}}

\begin{document}
\label{firstpage}
\pagerange{\pageref{firstpage}--\pageref{lastpage}}
\maketitle

\begin{abstract}
We present a comprehensive analysis of Low Frequency Quasi-periodic Oscillation (LFQPO) associated time-lags in the persistently variable black hole binary GRS 1915+105 using 441 ks of \textit{AstroSat} observations from March 2016 to March 2019. LFQPO frequency ($1.38-7.38$ Hz) are detected across the $\theta$, $\beta$, $\rho$, and $\chi$ classes, with the $\chi$ class further subdivided into $\chi_1$, $\chi_2$, $\chi_3$, and $\chi_4$ based on spectro-temporal characteristics. Class transitions occur on timescales of a few hours, appearing either as a simultaneous increase in X-ray count rate and QPO frequency, or vice versa, indicating rapid changes in the accretion flow geometry. The $\text{rms}_{\rm QPO}$ increases with QPO frequency up to $\sim 3.4$ Hz and declines at higher frequencies, a trend similar to \textit{RXTE} observations, where peak occurred at $\sim 2$ Hz.
Spectro-temporal correlations reveal that increasing $F_{\rm Comp}$ drives higher $\text{rms}_{\rm QPO}$ and decreases the soft-lag magnitude, while $\nu_{\rm QPO}$ and $\Gamma$ also decline, suggesting that the observed time lag may result from the combined effects of multiple physical mechanisms.
The consistent increase of $\text{rms}_{\rm QPO}$ with $F_{\rm Comp}$ provides clear evidence that modulated Comptonized photons enhance the rms power ($\text{rms}_{\rm QPO}$). Moreover, the soft-lag ($1.59-13.49$ ms) observed across all QPO frequencies, without the sign reversal at $\sim$ 2 Hz observed in \textit{RXTE} observations, is interpreted within the framework of a dynamical accretion disk model around the black hole.

\end{abstract}




\section{Introduction}

Compact objects accrete material from their companion stars either through Roche lobe overflow or by capturing stellar wind. Such systems are typically bright in X-rays \citep{frank2002} and are classified as X-ray binaries (XRBs; \citealt{seward_charles2010}). When the compact object is a black hole, these systems are referred to as black hole XRBs (BH-XRBs). Some BH-XRBs show variability in their X-ray flux on timescales ranging from milliseconds to a few days and are known as persistently variable sources, e.g., GRS 1915+105 \citep[][and references therein]{muno_1999,belloni2000,athulya2022}. Studying the time variability properties of these BH-XRB sources is crucial for understanding key black hole characteristics such as spin and mass, as well as the geometry of accretion flows and jet formation \citep{vadawale_rao_nandi2001,reig_kylafis2015_bh-jet,radhika_2016}.

The variability properties of BH-XRBs can be understood from time-series analysis of their lightcurves. A sharp rise in frequency-dependent power in the Power Density Spectrum (PDS) is identified as a Quasi-periodic Oscillation (QPO), with frequencies ($\nu$) ranging from $\sim0.1$ to 450 Hz. Based on their frequency, QPOs are classified into two categories: (a) low-frequency QPOs (LFQPOs), with $\nu < 40$ Hz, and (b) high-frequency QPOs (HFQPOs), with $\nu$ in the range of $40$–$450$ Hz \citep{remillard_mcClintock2006}.

HFQPOs are detected only in a few BH-XRBs \citep{belloni2012,sreehari2020}, whereas LFQPOs are ubiquitous across nearly all systems. Based on their PDS morphology and time-lag properties, LFQPOs are categorized into three types, type-A, type-B, and type-C \citep[][and references therein]{wijnands_1999_QPO-type,casella_2004_QPO-type,casella_2005,anuj_nandi_gx339_2012}. Of the three types, type-C QPOs occur most frequently, and observed in both transient and persistent BH systems.
Several models have been proposed to explain the origin of LFQPOs. \cite{stella_vietri1998,ingram2009} proposed the Lense-Thirring precession of the inner accretion disk as the origin of LFQPOs due to the relativistic frame-dragging effect around rotating black holes. Additionally, the jet precession model \citep{ferreira2022} proposes that LFQPOs may arise from instabilities at the base of the jet. Such precession changes the jet’s orientation, which in turn modulates the observed X-ray emission, giving rise to LFQPOs.
The most promising explanation comes from the two-component advective flow model \citep{chakrabarti_titarchuk1995, chakrabarti_manickam2000,rao2000,vadawale_rao_chakrabarti2001,iyer_nandi2015}, supported by numerical simulations \citep{molteni1994,garain2014,santa_das2014,aktar2018}, which attributes LFQPOs to the oscillatory behaviour of the post-shock region in the accretion flow around a black hole. This region acts as a Comptonizing cloud.

GRS 1915+105 has exhibited both LFQPOs and HFQPOs since its discovery in 1992. The LFQPOs of this source have been extensively studied using \textit{RXTE} observations \citep{chen1997,muno_1999,markwardt1999,reig2000,nandi2001}. In fact, type-C QPOs are present in the majority of \textit{RXTE} pointing with $\sim 620$ observations \citep{zhang_2020_grs}. However, type-B QPOs have also been reported by \citet{soleri_2008_grs_type-b} during the $\beta$ and $\mu$ classes in \textit{RXTE} data. The rms amplitude of type-C QPOs increases with frequency up to $\sim$2.2 Hz and then decreases as the QPO frequency rises further \citep{yan_2013, zhang_2020_grs}. 
However, the frequency of type-B QPOs does not show any correlation with their rms amplitude \citep{soleri_2008_grs_type-b}. \cite{muno_1999} studied the spectro-temporal correlations of LFQPOs in GRS 1915+105 and found that the centroid frequency increases with the temperature of the inner disc. They also noted that the power-law component dominates the spectrum when QPOs are present, whereas the disc blackbody component dominates when QPOs are absent. Additionally, \cite{rodriguez_2004} investigated the rms amplitude as a function of energy and concluded that QPOs could originate from the modulation of the Comptonized X-ray flux. 
Owing to its rich variability and frequent detection of QPOs, GRS 1915+105 provides an excellent opportunity for studying the connection between the evolution of spectral and timing properties such as time-lag behaviour at QPO frequency. 

BH-XRBs exhibit complex time-lag behaviour associated with LFQPOs, which remains challenging to interpret as it depends on the relative contributions of various physical processes such as Comptonization, reflection and on the geometry of the system \citep{dutta2016}. Using Monte Carlo simulations considering the two-component advective flow geometry, and incorporating Comptonization, gravitational light bending and reflection, \cite{arka-chatterjee2017b} demonstrated that the time-lag decreases with the shrinking size of the Compton cloud during the rising phase, and increases again during the declining phase. 
The phase-lag of type-C QPOs strongly depends on inclination and QPO frequency \citep[see][]{van_eijnden_2017, choudhury_etal2025}. In case of high inclination sources, phase-lag decreases with QPO frequency while for low inclination sources, phase-lag increases with QPO frequency \citep{reig2000,pahari_2013,van_eijnden_2017,zhang_2020_grs}.
This trend, along with the role of multiple physical mechanisms, has been observationally confirmed in both transient and persistent sources using \textit{RXTE} data \citep{dutta2016, dutta2018,patra2019,arka-chatterjee2020,nandi_prantik2021,debnath2024,prajjwal_2024,prajjwal2025,pragati_sahu2024, choudhury_etal2025}.

In GRS 1915+105, the time-lag switches sign from positive to negative at a QPO frequency of $\sim2.2$ Hz during the \textit{RXTE} era \citep{reig2000,pahari_2013,dutta2018,zhang_2020_grs}. Despite these temporal studies, the correlation between spectral properties and phase/time-lags at LFQPOs have not been extensively studied. However, \cite{prajjwal_2024} showed through a broadband (0.7$-$50 keV) spectral study that the Comptonizing medium has a high optical depth ($\tau \sim 6.90-12.55$), and the magnitude of the soft-lag associated with HFQPOs increases linearly with increasing $\tau$. This naturally motivates a systematic investigation of correlation between spectral properties and time-lag at LFQPOs, which could provide crucial insights into the accretion geometry and radiative processes in black hole systems.

GRS 1915+105 has also been extensively observed with \textit{AstroSat}, where both LFQPOs and HFQPOs have been detected \citep{yadav2016, rawat_2019_grs, belloni2019, anubhab2021_grs1915, athulya2022, athulya2023,sreehari2020, seshadri2022,sreehari_2025,seshadri2025}. The first detection of LFQPO was observed during the $\chi$ class, with frequencies varying between $2$–$8$ Hz, and detailed time-lag analysis revealed a soft-lag at the corresponding QPO frequency \citep{yadav2016}. \cite{rawat_2019_grs} further studied the source during $\chi$, IMS, and $\rho$ class observations, reporting strong LFQPOs in the $3$–$5$ Hz range, with soft-lags at the fundamental QPO and hard-lags at the harmonic. Additionally, \cite{anubhab2021_grs1915} reported LFQPOs in the $4$–$5$ Hz range during $\theta$ class observations.

\cite{athulya2022} performed the first comprehensive analysis of all \textit{AstroSat} observations (MJD 57705–58648) of GRS 1915+105, identifying six variability classes ($\chi$, $\rho$, $\omega$, $\delta$, $\kappa$, and $\gamma$), with LFQPOs detected mainly in the $\chi$ and $\rho$ classes. The $\beta$ class was later reported by \cite{seshadri2022}, though without detailed LFQPO analysis. Spectral studies further revealed accretion dynamics during LFQPOs. \cite{misra2020_grs1915} found an anti-correlation between LFQPO frequency and the inner disc radius during $\chi$, IMS, and $\rho$ classes. More recently, \cite{belloni2024} explored the time-lag properties of GRS 1915+105 during $\chi$ class observations and concluded that the size of the corona is correlated with the QPO frequency. Despite these important contributions, time-lag studies at LFQPOs with \textit{AstroSat} remain incomplete. In particular, a systematic spectro-temporal investigation of time-lag properties associated with type-C QPOs is still absent in literature. Moreover, no sub-classification of $\chi$ class observations has yet been attempted using the rich \textit{AstroSat} datasets.

In this work, we present a comprehensive spectro-temporal study of 441 ks observations of all LFQPOs of GRS 1915+105 using \textit{AstroSat}. Light curves and colour–colour diagrams (CCDs) were generated to identify the variability classes. The $\chi$ class observations were further sub-classified into $\chi_1$, $\chi_2$, $\chi_3$, and $\chi_4$, following the scheme of \cite{belloni2000}. Fourier analysis was then carried out to characterize the LFQPO features and to investigate the associated time-lag properties at the corresponding QPO frequencies. In addition, detailed wide-band spectral analysis was performed to correlate the time-lag properties and spectral parameters.

We organize this paper as follows. In \S \ref{sec2:obs_reduction}, we describe the data reduction procedures for \textit{MAXI}, \textit{BAT}, and \textit{AstroSat}. The methods of timing and spectral analysis are presented in \S \ref{sec3:analysis_modelling}. The results and the spectro-temporal correlations are discussed in \S \ref{sec4:results}. In \S \ref{sec5:discussion}, we interpret our findings in the context of existing theoretical models. Finally, we summarize our conclusions in \S \ref{sec6:conclusion}.

\section{Observation and Data Reduction}
\label{sec2:obs_reduction}
We analyzed the \textit{AstroSat} observations obtained between 4 March 2016 and 22 March 2019, when LFQPOs were detected. 
However, we excluded the analysis of the observations during the re-brightening period of obscured phase \citep{athulya2023}.
The considered observations were compared against the long-term simultaneous flux variations and `colour' diagram from \textit{MAXI/GSC} ($2-10$ keV) and \textit{Swift/BAT} ($15-50$ keV), as shown in the lightcurve in Fig.\ref{fig1:maxi_lightcurve}. We defined the `colour' as the ratio of \textit{BAT} ($15-50$ keV) to \textit{MAXI} ($2-10$ keV) counts/flux, expressed in Crab units. The selected \textit{AstroSat} observations are indicated by coloured vertical lines in the same figure. In this work, we considered a total of 42 observations, with a cumulative exposure time of 441 ks. The details of these observations are provided in Table \ref{tab:log_table}.

\subsection{\textit{MAXI} Data Reduction}
The Gas Slit Camera (\textit{GSC}) \citep{mihara_maxi_2011}, onboard the Monitor of All-sky X-ray Image (\textit{MAXI}) mission, has been continuously monitoring GRS 1915+105 since 11 August 2009. In this work, we utilize data from the \textit{MAXI/GSC} website\footnote{\label{web:maxi}\url{http://maxi.riken.jp/top/lc.html}} to analyze the source's lightcurve in the 2–10 keV energy range. The \textit{GSC} count rates are converted to Crab units\footnote{\label{web:maxi_crab}\url{http://maxi.riken.jp/star_data/J0534+220/J0534+220.html}} by dividing count rate by $\sim$ 3.74 photons/s/cm$^{2}$ \citep[see also][]{motta2021}.

\subsection{\textit{BAT} Data Reduction}
The Burst Alert Telescope (\textit{BAT}), onboard the Neil Gehrels Swift Observatory \citep{krim2013_BAT}, has been monitoring the source since 15 February 2005. The long-term \textit{BAT} lightcurve of GRS 1915+105 in the 15–50 keV energy band was downloaded from the \textit{Swift/BAT} website\footnote{\label{web:bat}\url{https://swift.gsfc.nasa.gov/results/transients/}}. The lightcurve was then converted to Crab units, assuming an average Crab count rate of 0.22 counts/s/cm$^2$.

\subsection{\textit{AstroSat} Data Reduction}

\textit{AstroSat} \citep{agrawal2006,kp_singh2014_astrosat} is the first multi-wavelength astronomy mission launched by India, in 2015. We utilized observations from the Soft X-ray Telescope (\textit{SXT}) \citep{singh2017} and the Large Area X-ray Proportional Counter (\textit{LAXPC}) \citep{yadav2016,antia2017} for spectro-temporal analysis of the source.

The \textit{SXT} \citep{singh2017} observes X-ray sources in the 0.3–8 keV energy range with a timing resolution of 2.3775 s. Due to this limited time resolution, we performed only spectral analysis using the Level-2 \textit{SXT} data available from the \textit{AstroSat} public archive\footnote{\label{fn3}\url{https://webapps.issdc.gov.in/astro_archive/archive/Home.jsp}}.

For the \textit{LAXPC} \citep{yadav2017}, we used Level-1 data from the same archive$^{\ref{fn3}}$ for timing and spectral analysis. Timing analysis was performed by processing the Level-1 data into Level-2 using the \textsc{laxpcsoftware} pipeline. The resulting Level-2 data were then used to study the temporal properties of all observations.

Wide-band (0.7–60 keV) spectral analysis was performed for the \textit{AstroSat} observations by combining data from \textit{SXT} (0.7–7 keV) and \textit{LAXPC} (3–60 keV). Due to high count rate in observations AS1 to AS25 (see Table \ref{tab:log_table}), we extracted top-layer single-event \textit{LAXPC20} spectra using \textsc{laxpcsoftv3.4.3} as suggested by \cite{antia2021}. The corresponding \textit{SXT} spectra were generated considering an annulus region with inner and outer radii of 3$'$ and 7$'$, respectively, to minimize pile-up effects caused by high count rates during these observations (see Table \ref{tab:log_table}).
For the remaining observations, we used all layers of \textit{LAXPC20}, and the \textit{SXT} spectra were extracted using a circular region with a radius of 12$'$, appropriate for the relatively lower count rates.

All \textit{SXT} spectra were grouped to have a minimum of 30 counts per bin, while no grouping was applied to the \textit{LAXPC20} spectra. A systematic error of 3\% was added to both sets of spectra \citep{antia2021}. To account for instrumental feature near the Si and Au edges at 1.8 keV and 2.2 keV \citep{singh2017}, respectively, we applied the \texttt{gain fit} \citep[see][and references therein]{prajjwal_2024} command in \texttt{XSPEC} version 12.12.0 \citep{k.arnaud_1996_xspec}.

\section{Analysis and Modelling}
\label{sec3:analysis_modelling}
We carried out a detail spectro-temporal analysis of selected archival \textit{AstroSat} observations when LFQPOs were detected. Our aim is to explore the generic time-lag properties associated with LFQPOs across different phases and variability classes and then to compare the results with those obtained from \textit{RXTE} observations.

\begin{figure*}
 \includegraphics[width=0.95\textwidth]{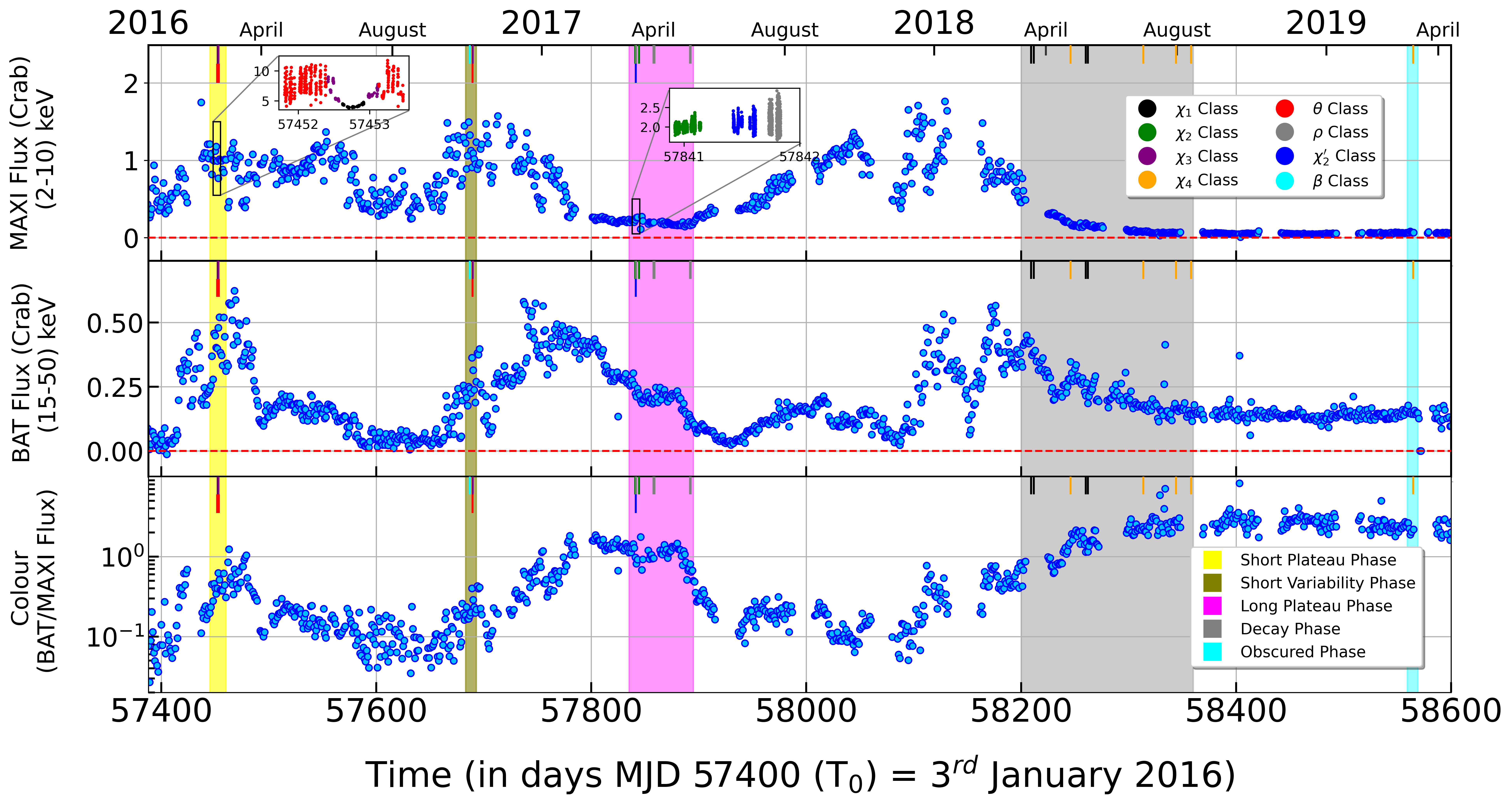}
 \caption{The \textit{MAXI/GSC} (2$-$10 keV) and \textit{Swift/BAT} ($15-50$ keV) lightcurves of GRS 1915+105 from January 2016 to April 2019 are plotted in the unit of Crab, shown in the upper and middle panels respectively. The `Colour' in the lower panel is defined as the ratio of \textit{BAT} flux to the \textit{MAXI} flux. The coloured vertical lines represent the considered \textit{AstroSat} observations of different variability classes mentioned in the legend in the upper panel. All observations can be classified as five different phases, shown in the legend in lower panel. See text for details. }
 \label{fig1:maxi_lightcurve}
\end{figure*}

\subsection{Timing Analysis}

\subsubsection{Lightcurve and CCD}
\begin{figure*}
 \includegraphics[width=0.95\textwidth]{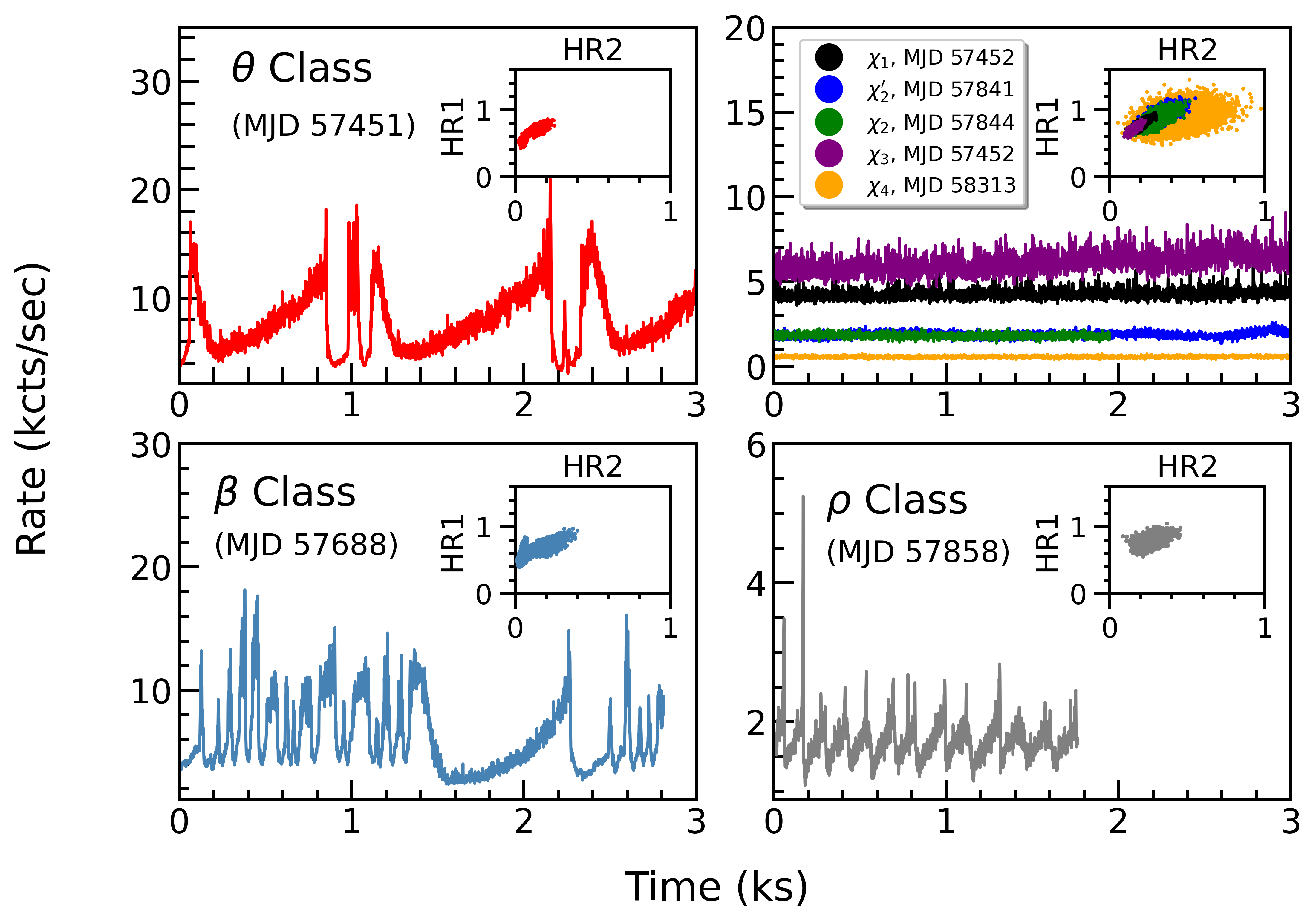}
 \caption{Lightcurve and CCD of variability classes ($\theta$, $\chi$, $\beta$, $\rho$) of the source GRS 1915+105 during LFQPO observations using \textit{AstroSat}. The background subtracted and dead-time corrected 1s binned \textit{LAXPC} lightcurves are plotted in the 3–60 keV energy range with CCD (top-right inset). The sub-classifications of $\chi$ class is shown in the top-right panel. See text for detail.
}
 \label{fig2:lc_ccd}
\end{figure*}

We generate 1s binned background subtracted lightcurves for the \textit{AstroSat} observations, in the 3$-$6 keV (band A), 6$-$15 keV (band B), and 15$-$60 keV (band C) energy bands by combining \textit{LAXPC10} and \textit{LAXPC20} data \citep{sreehari2020}. To plot the CCD, we define soft and hard colours as HR1=B/A and HR2=C/A, respectively. The background subtracted and dead-time corrected lightcurve in 3$-$60 keV and CCD for \textit{AstroSat} observations are shown in Fig. \ref{fig2:lc_ccd}. The average count rate, HR1 and HR2 for all observations are tabulated in Table \ref{tab:log_table}.

\subsubsection{Power Spectrum and Cross Spectrum Analysis }

We performed Fourier analysis using 1 ms binned lightcurves from \textit{AstroSat} observations to generate power density spectra (PDS). The lightcurves were produced by combining data from \textit{LAXPC10} and \textit{LAXPC20} in the 3–60 keV energy band and PDS is generated using the \textsc{powspec} tool in \textsc{heasoft} v6.29c. Each segment consisted of 65,536 bins (i.e., 65.536 s in length), and the resulting PDS were further averaged to obtain the final spectrum. A geometric rebinning factor of 1.02 was applied for the power spectral analysis. The dead-time corrected Poisson noise was then subtracted following \citet{agrawal2018}, and the final PDS was normalized in squared rms units \citep{belloni_hasinger1990}. 
The PDS are modelled in \texttt{XSPEC V12.12.0} using multiple \texttt{Lorentzian}. The \texttt{Lorentzian} feature with Q-factor $\geq$ 3 and significance ($\sigma$ = $\frac{Norm}{Norm_{err}}$) $\geq$ 3 \citep{belloni2012,sreehari2019} are classified as QPOs. We further calculated the percentage rms amplitude of the QPO as $100\times\sqrt{Norm}$ \citep{van_der_Klis1989_Fourier_technic,wang_2024_rms_formula}. 

\begin{figure*}
 \includegraphics[width=0.95\textwidth]{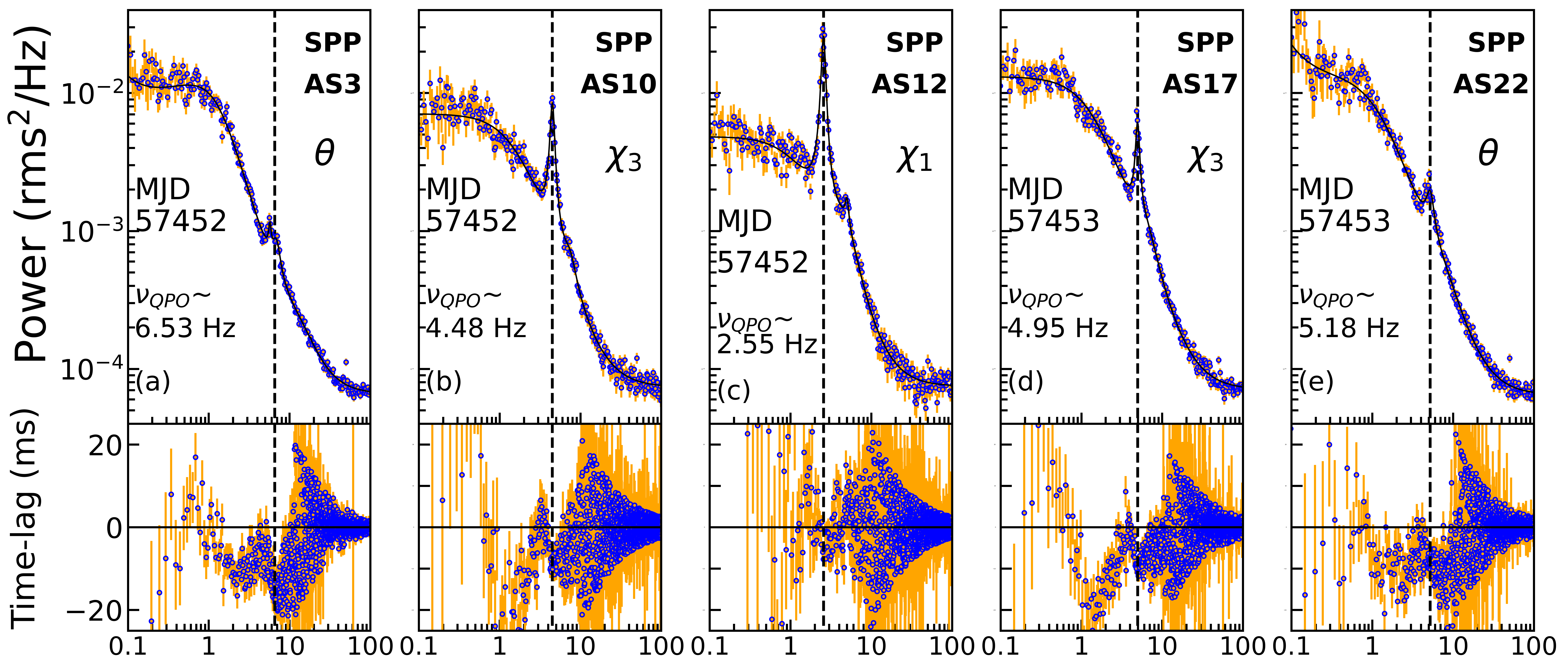}
 \includegraphics[width=0.95\textwidth]{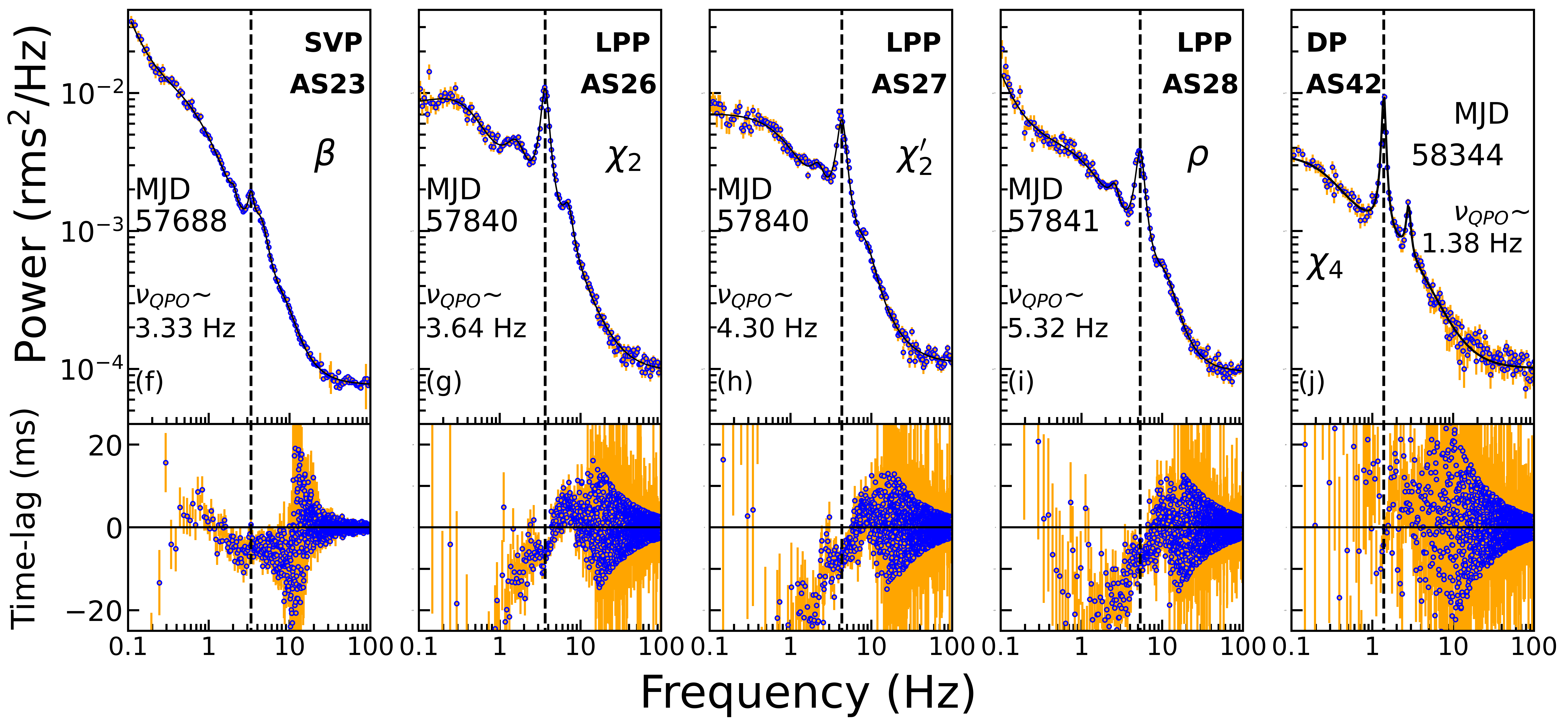}
 \caption{The broadband PDS (3$-$60 keV) and time-lag spectra ($6-20$ keV w.r.t $3-6$ keV) corresponding to different variability classes of \textit{AstroSat} observations. The MJD of each observation and the QPO frequency are mentioned in each panel. The dashed vertical black line corresponds to the QPO frequency. See text for details.}
 \label{fig3:pds}
\end{figure*}

The time lag is calculated from the argument of the complex cross-spectrum \citep[see][for details]{vaughan_nowak1997}, which is defined as $C(j) = X^{*}_{1}(j) X_{2}(j)$, where $X_{1}(j)$ and $X_{2}(j)$ are the Fourier coefficients corresponding to two different energy bands at frequency $\nu(j)$ \citep{van_der_klis1987}. We computed the cross-spectrum between the $6-20$ keV and $3-6$ keV energy bands using the \texttt{LAXPCsoftware}, based on data from \textit{LAXPC10} and \textit{LAXPC20} \citep{prajjwal_2024}. The frequency resolution and Nyquist frequency were set to 0.1 Hz and 250 Hz, respectively. The time lag was then averaged over the FWHM of the QPO feature, following the method described by \citet{reig2000}. 
The resulting power density spectrum (PDS) and time-lag spectrum are shown in Fig. \ref{fig3:pds}, while the centroid frequency, FWHM, fractional rms amplitude, and average time-lag are listed in Table \ref{tab:log_table}.

To ensure a consistent comparison of time lags between \textit{RXTE} and \textit{AstroSat}, we adopted the closest possible reference energy bands. Since the $3-6$ keV band could not be used for most \textit{RXTE} observations due to data structure limitations, we selected the lowest available band ($2-5$ keV) as the reference and computed the time lag for the $5-20$ keV band. For the energy-dependent analysis, we divided the \textit{AstroSat} data into seven bands ($3-6$ keV, $6-9$ keV, $9-12$ keV, $12-15$ keV, $15-18$ keV, $18-21$ keV and $21-25$ keV), calculated the QPO fractional rms (${\rm rms}_{\rm QPO}$) as a function of energy, and derived the average time lag from the cross-spectrum relative to the $3-6$ keV band.

\begin{figure}
 \includegraphics[width=\columnwidth]{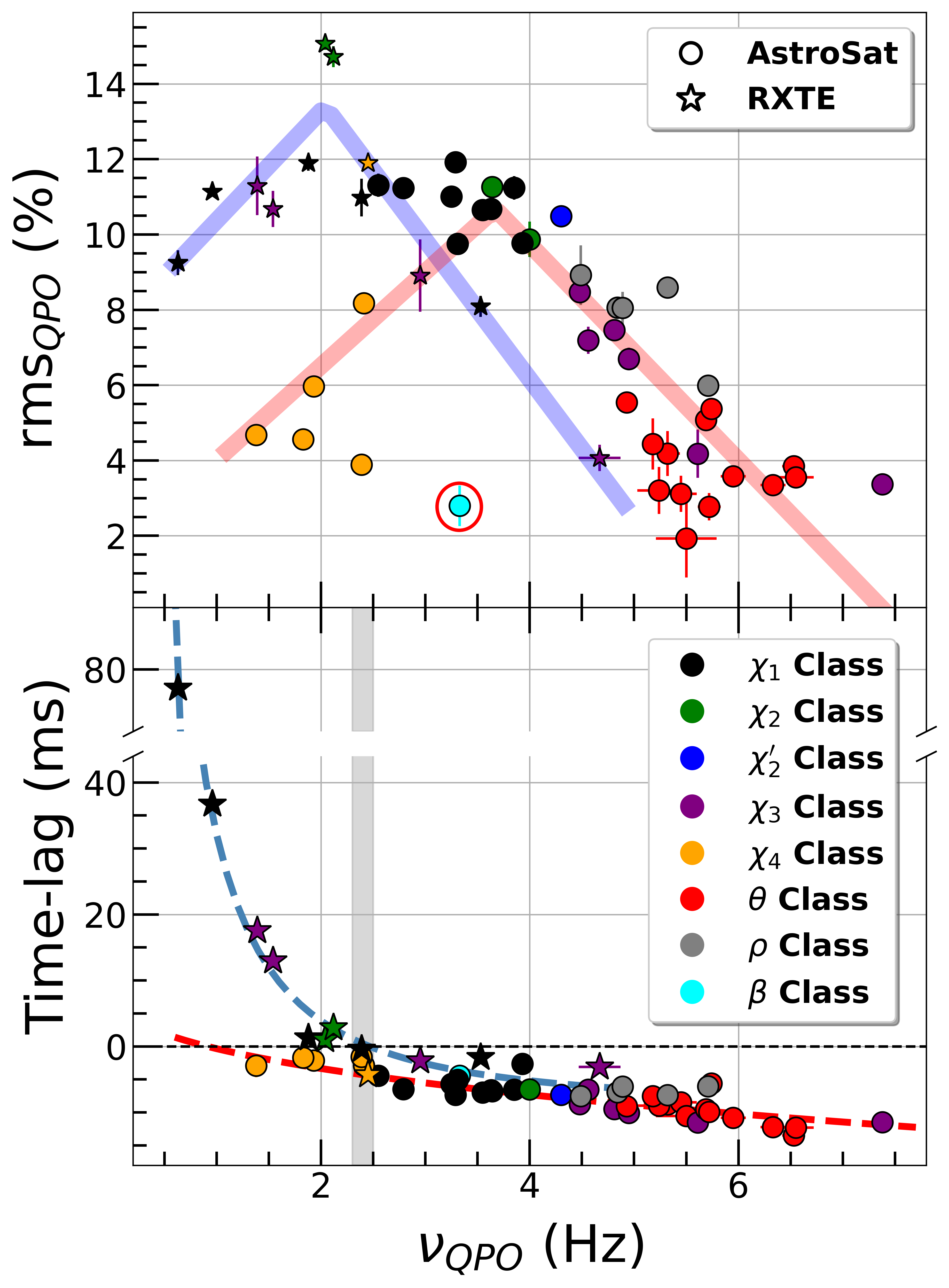}
 \caption{Variation of percentage rms amplitude ($\text{rms}_\text{QPO}$) and time-lag as a function of the centroid frequency ($\nu_\text{QPO}$) of the LFQPO. The star and circle represent \textit{RXTE} and \textit{AstroSat} observations respectively. The filled colours correspond to different variability classes shown in the legend. See text for details. }
 \label{fig4:rms_lag_v_qpo}
\end{figure}

\begin{figure}
 \includegraphics[width=\columnwidth]{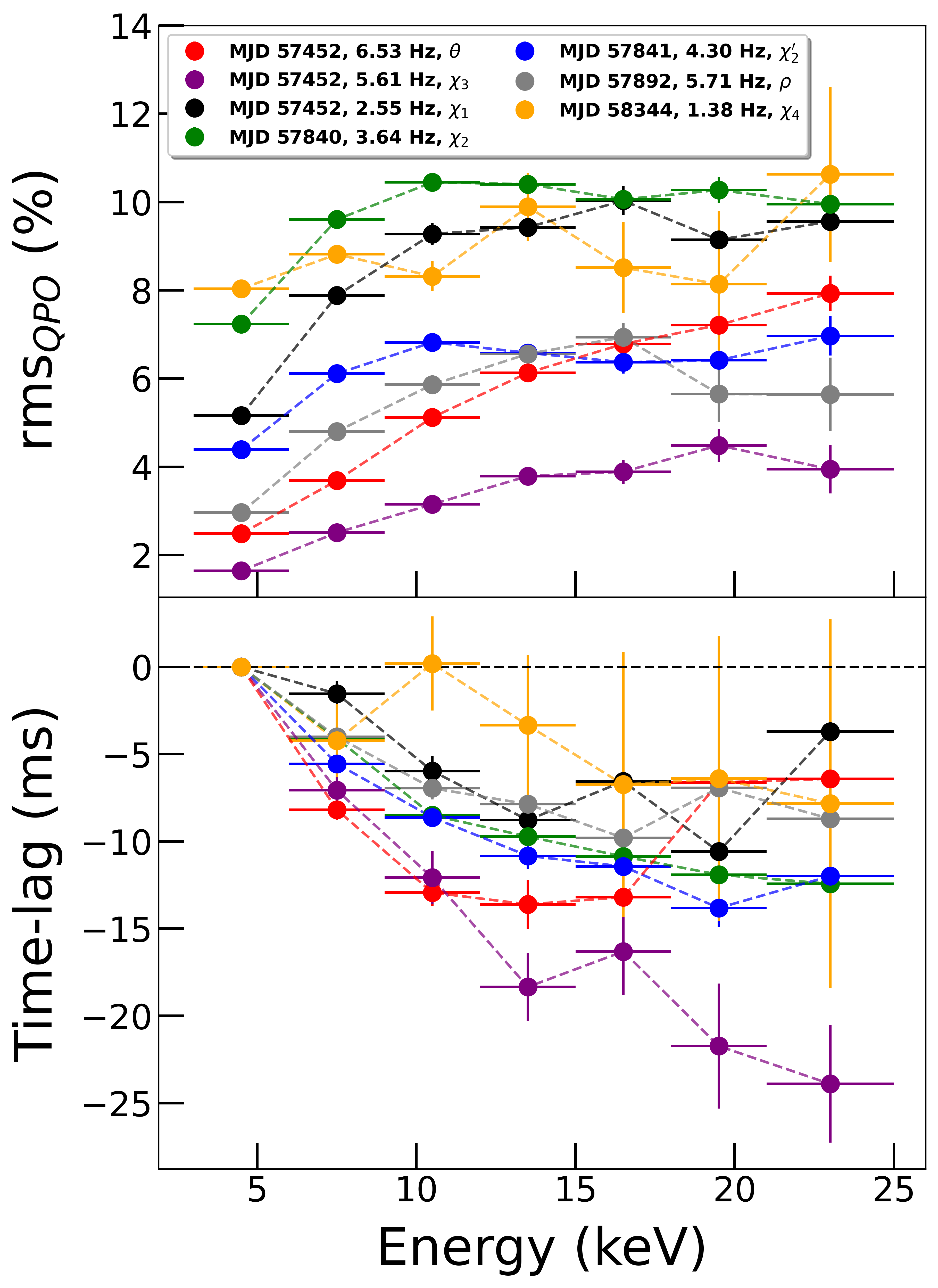}
 \caption{Energy dependent percentage rms amplitude ($\text{rms}_\text{QPO}$) and time-lag at LFQPO using \textit{AstroSat} shown in the upper and lower panel respectively. Different colours correspond to different observations. The MJD, QPO frequency and variability class are shown in the legend. See text for details. }
 \label{fig5:rms_lag_v_energy}
\end{figure}


\begin{figure}
 \includegraphics[width=\columnwidth]{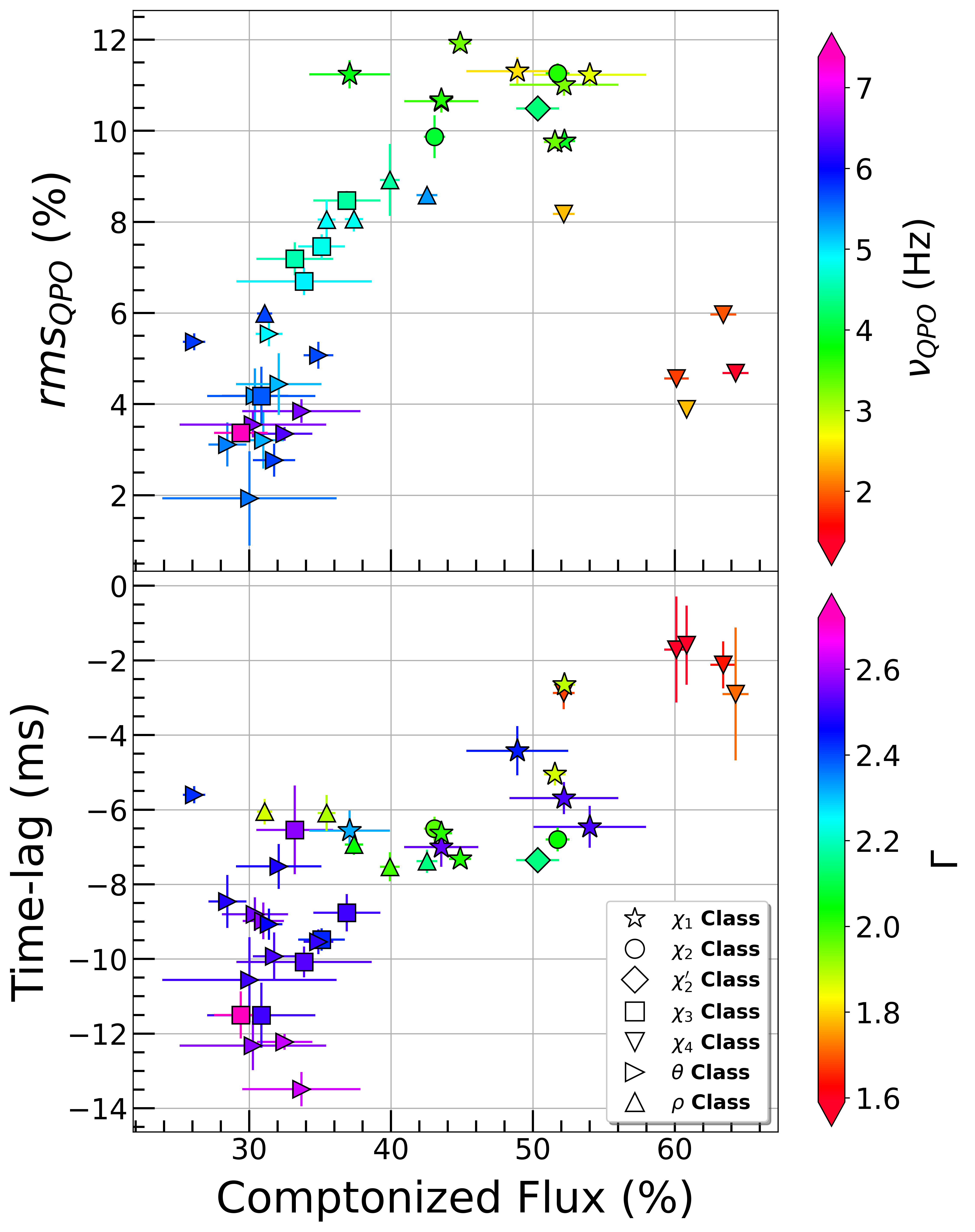}
 \caption{The correlation between $\text{rms}_\text{QPO}$ and Comptonized flux percentage is plotted in the upper panel along with the $\nu_\text{QPO}$ as a colour-bar. In lower panel, the variation of time-lag as a function of Comptonized flux is shown along with spectral index ($\Gamma$) as colour-bar. Different markers represents different variability class which is shown in the legend. See text for details.}
 \label{fig8:rms_lag_v_comp-flux}
\end{figure}



\subsection{Energy Spectral Analysis}
We performed spectral analysis of the \textit{SXT} and \textit{LAXPC20} data ($0.7-60$ keV) using \textsc{xspec} v12.12.0 \citep{k.arnaud_1996_xspec}, available as part of \textsc{heasoft} v6.29c. All spectra were fitted using the model combination \texttt{constant$\times$TBabs$\times$(thcomp$\otimes$diskbb)}. The \texttt{constant} component accounts for cross-calibration differences between the two instruments. The \texttt{TBabs} model \citep{wilms2000} represents Galactic absorption along the line of sight.
The \texttt{diskbb} component \citep{mitsuda_diskbb1984,makishima_diskbb1986} models the thermal emission from the accretion disk, while the convolution model \texttt{thcomp} \citep{zdziarski_2020_thcomp} describes the thermal Comptonization of seed photons from the disk that are reprocessed in the corona. In addition, we included an instrumental Xenon \texttt{edge} near $\sim$32 keV as required \citep{antia2017,sreehari2020,seshadri2022}.

The hydrogen column density $(N_{\mathrm{H}}$) was fixed at $6 \times 10^{22}$ atoms cm\(^{-2}\) for most of the observations \citep{muno1999,sreehari2020}. However, for observations from MJD 58246 onward, $N_{\mathrm{H}}$ was left free and was found to lie in the range $2.15 - 5.17 \times 10^{22}$ atoms cm$^{-2}$ \citep{athulya2023}. The best-fit spectral parameters, along with the error in 90\% confidence intervals, are presented in Table~\ref{tab:spectral_table}. We also calculated the bolometric flux in the $1-100$ keV energy range using the \texttt{cflux} model component and expressed it as a percentage of the Eddington luminosity, assuming a black hole mass of $12.4 M_{\odot}$ and a distance of 8.6 kpc \citep{reid2014}.
Furthermore, we calculated the individual contributions of the Comptonized flux and the disk flux to the total flux, which are also listed in Table~\ref{tab:spectral_table}. Finally, the optical depth ($\tau$) of the Comptonizing medium was calculated using the relation given by \citet{zdziarski1996}.


\begin{table*}

    \centering
    \caption{\label{tab:log_table}Observation details of the source GRS 1915+105 observed by \textit{AstroSat} during March 2016 to March 2019. In the table, Mission (ID), ObsID and MJD start are mentioned. The detected $r_{det}$ count rates for all observations with hardness ratios and variability classes are also tabulated. Furthermore, we tabulated the temporal parameters of the Fourier analysis, that is, centroid frequency ($\nu_\text{QPO}$) and FWHM of the QPO feature, along with the percentage rms amplitude in $3-60$ keV band. The time lag of $6-20$ keV w.r.t $3-6$ keV at $\nu_\text{QPO}$ is tabulated in milliseconds. See text for details.}
    
    \resizebox{\textwidth}{!}{
    \begin{tabular}{ccccccccccc}
   
    \hline
    \hline
     Mission (ID) & ObsID & MJD  & $r_{det}$ & HR1   & HR2  & Class & $\nu_{QPO}$ & FWHM & $\text{rms}_\text{QPO}$(\%) & Time-lag \\
        &    &   & (cts/s) & (B/A)* & (C/A)* &  & (Hz) & (Hz) &  & (ms)  \\

    \hline
    &&& \multicolumn{2}{c}{Short Plateau Phase (SPP)}  \\
    \hline

AS1 & T01\_030T01\_9000000358 & 57451.81 & 8336 & 0.68 & 0.12 & $\theta$ & 5.32$_{-0.09}^{+0.12}$ & 1.82$_{-0.38}^{+0.58}$ & 4.18 $\pm$ 0.60 & $-8.80\pm 0.45$   \\ 

AS2 & T01\_030T01\_9000000358 & 57451.89 & 7741 & 0.67 & 0.12 & $\theta$ & 5.24$_{-0.21}^{+0.21}$ & 1.03$_{-0.27}^{+0.32}$ & 3.21 $\pm$ 0.62 & $-8.98\pm 0.49$   \\ 

AS3 & T01\_030T01\_9000000358 & 57452.04 & 9856 & 0.69 & 0.12 & $\theta$ & 6.53$_{-0.11}^{+0.11}$ & 2.34$_{-0.32}^{+0.37}$ & 3.85 $\pm$ 0.26 & $-13.49 \pm 0.46$   \\ 

AS4 & T01\_030T01\_9000000358 & 57451.98 & 8937 & 0.67 & 0.12 & $\theta$ & 5.95$_{-0.10}^{+0.12}$ & 2.74$_{-0.35}^{+0.44}$ & 3.58 $\pm$ 0.28 & $-10.81\pm0.28$   \\ 

AS5 & T01\_030T01\_9000000358 & 57451.98 & 9487 & 0.68 & 0.12 & $\theta$ & 6.33$_{-0.12}^{+0.12}$ & 2.82$_{-0.31}^{+0.36}$ & 3.35 $\pm$ 0.15 & $-12.22\pm0.21$   \\ 

AS6 & T01\_030T01\_9000000358 & 57452.26 & 9862 & 0.68 & 0.12 & $\theta$ & 5.45$_{-0.13}^{+0.15}$ & 1.15$_{-0.35}^{+0.36}$ & 3.11 $\pm$ 0.48 & $-8.46\pm0.71$   \\ 

AS7 & T01\_030T01\_9000000358 & 57452.33 & 9666 & 0.67 & 0.12 & $\theta$ & 5.50$_{-0.10}^{+0.29}$ & 0.55$_{-0.25}^{+0.66}$ & 1.93 $\pm$ 1.04 & $-10.56\pm1.14$   \\ 

AS8 & T01\_030T01\_9000000358 & 57452.41 & 8445 & 0.68 & 0.13 & $\chi_3$ & 5.61$_{-0.07}^{+0.07}$ & 0.45$_{-0.14}^{+0.16}$ & 4.18 $\pm$ 0.64 & $-11.51\pm0.87$   \\ 

AS9 & T01\_030T01\_9000000358 & 57452.48 & 8049 & 0.68 & 0.13 & $\chi_3$ & 4.56$_{-0.04}^{+0.04}$ & 0.25$_{-0.09}^{+0.10}$ & 2.68 $\pm$ 0.36 & $-6.54\pm1.19$   \\ 

AS10 & T01\_030T01\_9000000358 & 57452.55 & 5487 & 0.68 & 0.14 & $\chi_3$ & 4.48$_{-0.02}^{+0.02}$ & 0.61$_{-0.05}^{+0.06}$ & 8.47 $\pm$ 0.21 & $-8.76\pm0.50$    \\ 

AS11 & T01\_030T01\_9000000358 & 57452.62 & 4549 & 0.71 & 0.16 & $\chi_1$ & 3.55$_{-0.02}^{+0.02}$ & 0.48$_{-0.04}^{+0.04}$ & 10.65 $\pm$ 0.25 & $-7.00\pm0.53$   \\ 

AS12 & T01\_030T01\_9000000358 & 57452.68 & 3998 & 0.78 & 0.19 & $\chi_1$ & 2.55$_{-0.01}^{+0.01}$ & 0.33$_{-0.02}^{+0.02}$ & 11.31 $\pm$ 0.31 & $-4.42\pm0.66$   \\ 

AS13 & T01\_030T01\_9000000358 & 57452.75 & 4027 & 0.77 & 0.18 & $\chi_1$ & 2.79$_{-0.01}^{+0.01}$ & 0.36$_{-0.02}^{+0.03}$ & 11.23 $\pm$ 0.26 & $-6.46\pm0.56$   \\

AS14 & T01\_030T01\_9000000358 & 57452.82 & 4286 & 0.74 & 0.17 & $\chi_1$ & 3.25$_{-0.02}^{+0.02}$ & 0.49$_{-0.03}^{+0.04}$ & 11.01 $\pm$ 0.24 & $-5.69\pm0.43$   \\ 

AS15 & T01\_030T01\_9000000358 & 57452.90 & 4731 & 0.71 & 0.15 & $\chi_1$ & 3.85$_{-0.02}^{+0.02}$ & 0.59$_{-0.04}^{+0.05}$ & 11.24 $\pm$ 0.31 & $-6.56\pm0.54$   \\ 

AS16 & T01\_030T01\_9000000358 & 57452.98 & 6181 & 0.69 & 0.14 & $\chi_3$ & 4.81$_{-0.02}^{+0.02}$ & 0.65$_{-0.06}^{+0.06}$ & 7.46 $\pm$ 0.27 & $-9.48\pm0.30$   \\ 

AS17 & T01\_030T01\_9000000358 & 57453.06 & 6606 & 0.69 & 0.14 & $\chi_3$ & 4.95$_{-0.02}^{+0.02}$ & 0.59$_{-0.06}^{+0.06}$ & 6.69 $\pm$ 0.30 &  $-10.08\pm0.41$    \\ 

AS18 & T01\_030T01\_9000000358 & 57453.13 & 7589 & 0.69 & 0.13 & $\theta$ & 5.69$_{-0.08}^{+0.08}$ & 1.65$_{-0.21}^{+0.22}$ & 5.07 $\pm$ 0.30 & $-9.54\pm0.33$   \\ 

AS19 & T01\_030T01\_9000000358 & 57453.21 & 8351 & 0.69 & 0.13 & $\theta$ & 4.93$_{-0.05}^{+0.05}$ & 1.02$_{-0.12}^{+0.12}$ & 5.54 $\pm$ 0.27 & $-9.07\pm0.42$   \\ 

AS20 & T01\_030T01\_9000000358 & 57453.28 & 9632 & 0.69 & 0.13 & $\theta$ & 5.72$_{-0.13}^{+0.11}$ & 0.77$_{-0.19}^{+0.23}$ & 2.77 $\pm$ 0.36 &  $-9.93\pm0.64$   \\ 

AS21 & T01\_030T01\_9000000358 & 57453.35 & 9996 & 0.68 & 0.12 & $\theta$ & 6.55$_{-0.14}^{+0.17}$ & 2.48$_{-0.41}^{+0.56}$ & 3.55 $\pm$ 0.28 &  $-12.32\pm0.66$   \\ 

AS22 & T01\_030T01\_9000000358 & 57453.42 & 8059 & 0.69 & 0.13 & $\theta$ & 5.18$_{-0.07}^{+0.08}$ & 1.31$_{-0.22}^{+0.43}$ & 4.44 $\pm$ 0.68 & $-7.52\pm0.60$    \\ 

\hline
    &&& \multicolumn{2}{c}{Short Variability Phase (SVP)}  \\
    \hline 

AS23 & G06\_033T01\_9000000760 & 57688.73 & 6158 & 0.62 & 0.09 & $\beta$ & $3.33_{-0.05}^{+0.06}$ & $0.61_{-0.32}^{+0.30}$ & $2.80\pm0.54$ & $-4.42\pm0.68$  \\

AS24 & G06\_033T01\_9000000760 & 57689.54 & 8142 & 0.65 & 0.11 & $\theta$ & $5.74_{-0.10}^{+0.10}$ & $3.15_{-0.24}^{+0.25}$ & $5.37\pm0.19$ & $-5.60\pm0.23$  \\

AS25 & G06\_033T01\_9000000760 & 57689.83 & 10664 & 0.67 & 0.11 & $\chi_3$ & $7.38_{-0.06}^{+0.06}$ & $1.32_{-0.15}^{+0.18}$ & $3.37\pm0.15$ & $-11.50\pm0.63$  \\

    \hline
    &&& \multicolumn{2}{c}{Long Plateau Phase (LPP)}  \\
    \hline 

AS26 & G06\_033T01\_9000001116   & 57840.92 & 1779 & 1.00 & 0.37 & $\chi_2$ & 3.64$_{-0.01}^{+0.01}$ & 0.87$_{-0.04}^{+0.03}$ & 11.26 $\pm$ 0.22 &  $-6.80\pm0.32$   \\ 

AS27 & G06\_033T01\_9000001116   & 57841.38 & 1856 & 0.91 & 0.31 & $\chi_2'$ & 4.30$_{-0.02}^{+0.02}$ & 1.30$_{-0.08}^{+0.07}$ & 10.49 $\pm$ 0.21 & $-7.35\pm0.34$   \\ 

AS28 & G06\_033T01\_9000001116   & 57841.72 & 1929 & 0.80 & 0.24 & $\rho$ & 5.32$_{-0.02}^{+0.02}$ & 1.50$_{-0.06}^{+0.07}$ & 8.59 $\pm$ 0.15 &  $-7.38\pm0.31$    \\ 

AS29 & G07\_046T01\_9000001124   & 57844.49 & 1853 & 0.85 & 0.28 & $\chi_2$ & 4.00$_{-0.02}^{+0.02}$ & 0.78$_{-0.08}^{+0.08}$ & 9.87 $\pm$ 0.47 & $-6.51\pm0.32$   \\ 

AS30 & G07\_046T01\_9000001162   & 57857.92 & 1737 & 0.74 & 0.23 & $\rho$ & 4.84$_{-0.01}^{+0.02}$ & 0.86$_{-0.04}^{+0.06}$ & 8.06 $\pm$ 0.27 &  $-6.93\pm0.28$   \\ 

AS31 & G07\_028T01\_9000001166   & 57858.82 & 1719 & 0.77 & 0.26 & $\rho$ & 4.49$_{-0.03}^{+0.03}$ & 0.55$_{-0.08}^{+0.07}$ & 8.92 $\pm$ 0.79 &  $-7.53\pm0.39$   \\ 


AS32 & G07\_028T01\_9000001232   & 57891.82 & 1451 & 0.70 & 0.21 & $\rho$ & 4.89$_{-0.02}^{+0.02}$ & 0.79$_{-0.06}^{+0.06}$ & 8.05 $\pm$ 0.43 &  $-6.09\pm0.48$   \\ 

AS33 & G07\_046T01\_9000001236   & 57892.50 & 1483 & 0.64 & 0.20 & $\rho$ & 5.71$_{-0.01}^{+0.01}$ & 1.67$_{-0.05}^{+0.05}$ & 5.99 $\pm$ 0.07 &  $-6.05\pm0.34$    \\ 

    \hline
    &&& \multicolumn{2}{c}{Decay Phase (DP)}  \\
    \hline 
AS34 & A04\_180T01\_9000002000   & 58209.13 & 2929 & 0.73 & 0.21 & $\chi_1$ & 3.29$_{-0.01}^{+0.01}$ & 0.88$_{-0.03}^{+0.03}$ & 11.92 $\pm$ 0.17 &  $-7.32\pm0.18$    \\ 

AS35 & G08\_028T01\_9000002006   & 58211.76 & 2429 & 0.69 & 0.21 & $\chi_1$ & 3.63$_{-0.01}^{+0.01}$ & 0.91$_{-0.04}^{+0.04}$ & 10.68 $\pm$ 0.14 &  $-6.63\pm0.18$    \\ 



AS36 & G08\_028T01\_9000002080   & 58246.00 & 1282 & 0.87 & 0.45 & $\chi_4$ & 2.41$_{-0.01}^{+0.01}$ & 0.38$_{-0.02}^{+0.02}$ & 8.17 $\pm$ 0.12 &  $-2.87\pm0.44$    \\ 

AS37 & G08\_078T01\_9000002110   & 58260.00 & 1112 & 0.75 & 0.25 & $\chi_1$ & 3.93$_{-0.01}^{+0.01}$ & 0.77$_{-0.03}^{+0.03}$ & 9.77 $\pm$ 0.10 &  $-2.65\pm0.20$    \\ 

AS38 & G08\_028T01\_9000002112   & 58262.00 & 1065 & 0.75 & 0.26 & $\chi_1$ & 3.31$_{-0.01}^{+0.01}$ & 0.65$_{-0.03}^{+0.03}$ & 9.75 $\pm$ 0.15 &  $-5.05\pm0.30$    \\ 


AS39 & G08\_028T01\_9000002220   & 58313.57 & 559 & 0.86 & 0.41 & $\chi_4$ & 1.93$_{-0.01}^{+0.00}$ & 0.21$_{-0.01}^{+0.01}$ & 5.97 $\pm$ 0.08 &  $-2.12\pm0.63$   \\ 

AS40 & G08\_028T01\_9000002306   & 58344.43 & 414 & 0.89 & 0.47 & $\chi_4$ & 1.38$_{-0.00}^{+0.00}$ & 0.15$_{-0.01}^{+0.01}$ & 4.68 $\pm$ 0.11 &  $-2.90\pm1.78$    \\ 

AS41 & G08\_028T01\_9000002334   & 58358.00 & 397 & 0.85 & 0.44 & $\chi_4$ & 1.83$_{-0.01}^{+0.01}$ & 0.17$_{-0.01}^{+0.01}$ & 4.56 $\pm$ 0.11 &  $-1.71\pm1.42$    \\ 

 \hline
    &&& \multicolumn{2}{c}{Obscured Phase (OP)}  \\
    \hline 
AS42 & A05\_173T01\_9000002812  & 58564.67 & 380 & 0.83 & 0.48 & $\chi_4$ & 2.39$_{-0.01}^{+0.01}$ & 0.17$_{-0.01}^{+0.01}$ & 3.89 $\pm$ 0.10 &  $-1.59\pm1.06$    \\

\hline 
        
    \end{tabular}
    }
    \begin{list}{}{}
            \item[*] For \textit{AstroSat/LAXPC}  observations, A, B and C bands are defined as $3-6$ keV, $6-15$ keV and $15-60$ keV respectively. See \cite{sreehari2020} for details.
	\end{list}
    
\end{table*}



\begin{table*}

    \centering
    \caption{\label{tab:spectral_table}Model parameters of the observations fitted with \texttt{constant$\times$TBabs$\times$(thcomp*diskbb)}. The Comptonized flux (F$_\text{Comp}$), Disc flux (F$_\text{Disc}$) and bolometric luminosity in the unit of Eddington luminosity (L$_\text{Edd}$) are calculated in the energy range $1-100$ keV. All errors are calculated with 90\% confidence range. See text for details.}
    
    \begin{tabular}{ccccccccccc}
   
    \hline
    \hline
     Mission (ID)$\dagger$ & Class & $\Gamma$  & $kT_e$ & $f_{cov}$   & $kT_{in}$  & $\chi^2$/dof & F$_\text{Comp}$(\%) & F$_\text{Disc}$(\%)  & L$_\text{Edd}$(\%) & $\tau$ \\
        &    &   & (keV) &  & (keV) &  &  &  &  &   \\

    \hline
    &&&&& \multicolumn{2}{c}{Short Plateau Phase (SPP)}  \\
    \hline

AS1 & $\theta$ & 2.55$_{-0.03}^{+0.03}$ & 20.00* & 0.49$_{-0.03}^{+0.03}$ & 1.40$_{-0.04}^{+0.04}$ & 510.64/528 & 30.41 $\pm$ 2.34 & 69.59 $\pm$ 4.06 & 34.73  & 2.12 $\pm$ 0.04   \\ 

AS2 & $\theta$ & 2.57$_{-0.07}^{+0.09}$ & 19.02$_{-3.73}^{+7.94}$ & 0.52$_{-0.07}^{+0.11}$ & 1.40$_{-0.05}^{+0.04}$ & 605.08/551 & 30.99 $\pm$ 1.45 & 69.01 $\pm$ 2.49 & 29.00  & 2.17 $\pm$ 0.32   \\ 

AS3 & $\theta$ & 2.63$_{-0.08}^{+0.08}$ & 18.81$_{-3.79}^{+7.54}$ & 0.61$_{-0.09}^{+0.14}$ & 1.33$_{-0.08}^{+0.07}$ & 333.27/347 & 33.68 $\pm$ 4.17 & 66.32 $\pm$ 6.26 & 27.19  & 2.11 $\pm$ 0.32   \\ 


AS5 & $\theta$ & 2.62$_{-0.06}^{+0.07}$ & 17.91$_{-2.50}^{+3.84}$ & 0.58$_{-0.07}^{+0.09}$ & 1.35$_{-0.04}^{+0.04}$ & 571.01/536 & 32.49 $\pm$ 1.95 & 67.51 $\pm$ 3.02 & 26.95  & 2.20 $\pm$ 0.23   \\ 

AS6 & $\theta$ & 2.49$_{-0.09}^{+0.09}$ & 13.02$_{-1.85}^{+2.99}$ & 0.43$_{-0.06}^{+0.07}$ & 1.48$_{-0.06}^{+0.05}$ & 595.36/557 & 28.47 $\pm$ 1.34 & 71.53 $\pm$ 2.58 & 22.83  & 2.96 $\pm$ 0.32   \\ 

AS7 & $\theta$ & 2.51$_{-0.09}^{+0.09}$ & 13.52$_{-2.04}^{+3.26}$ & 0.46$_{-0.07}^{+0.08}$ & 1.43$_{-0.06}^{+0.06}$ & 206.19/210 & 30.02 $\pm$ 6.14 & 69.98 $\pm$ 11.07 & 31.80  & 2.85 $\pm$ 0.33   \\

AS8 & $\chi_3$ & 2.51$_{-0.09}^{+0.09}$ & 15.05$_{-2.68}^{+4.90}$ & 0.48$_{-0.07}^{+0.09}$ & 1.46$_{-0.05}^{+0.05}$ & 570.87/479 & 30.85 $\pm$ 3.82 & 69.15 $\pm$ 6.52 & 41.70  & 2.65 $\pm$ 0.35   \\ 

AS9 & $\chi_3$ & 2.57$_{-0.11}^{+0.12}$ & 19.88$_{-5.96}^{+22.59}$ & 0.57$_{-0.11}^{+0.12}$ & 1.38$_{-0.06}^{+0.06}$ & 516.23/472 & 33.21 $\pm$ 2.70 & 66.79 $\pm$ 4.28 & 30.97  & 2.11 $\pm$ 0.47   \\ 

AS10 & $\chi_3$ & 2.51$_{-0.02}^{+0.02}$ & 19.76* & 0.58$_{-0.04}^{+0.05}$ & 1.23$_{-0.04}^{+0.04}$ & 456.87/461 & 36.88 $\pm$ 2.36 & 63.12 $\pm$ 3.16 & 23.91  & 2.19 $\pm$ 0.03   \\ 

AS11 & $\chi_1$ & 2.54$_{-0.07}^{+0.07}$ & 27.80$_{-9.09}^{+34.77}$ & 0.76$_{-0.14}^{+0.21}$ & 1.06$_{-0.06}^{+0.05}$ & 463.35/437 & 43.54 $\pm$ 2.61 & 56.46 $\pm$ 2.52 & 20.11  & 1.68 $\pm$ 0.41   \\ 

AS12 & $\chi_1$ & 2.44$_{-0.08}^{+0.07}$ & 23.55$_{-7.77}^{+23.11}$ & 0.79$_{-0.16}^{p}$ & 1.07$_{-0.09}^{+0.07}$ & 544.82/479 & 48.90 $\pm$ 3.59 & 51.10 $\pm$ 2.75 & 30.69  & 2.03 $\pm$ 0.49   \\ 

AS13 & $\chi_1$ & 2.52$_{-0.07}^{+0.02}$ & 25.98$_{-7.92}^{+8.80}$ & 0.94$_{-0.23}^{p}$ & 0.84$_{-0.07}^{+0.11}$ & 563.31/485 & 54.01 $\pm$ 3.97 & 45.99 $\pm$ 2.48 & 32.41  & 1.79 $\pm$ 0.40   \\ 

AS14 & $\chi_1$ & 2.52$_{-0.02}^{+0.02}$ & 20.92$_{-2.74}^{+3.75}$ & 0.87* & 0.84$_{-0.03}^{+0.03}$ & 613.97/543 & 52.18 $\pm$ 3.83 & 47.82 $\pm$ 2.58 & 34.96  & 2.09 $\pm$ 0.20   \\ 

AS15 & $\chi_1$ & 2.32$_{-0.07}^{+0.07}$ & 10.96$_{-1.36}^{+1.98}$ & 0.47$_{-0.06}^{+0.07}$ & 1.27$_{-0.04}^{+0.04}$ & 539.31/511 & 37.07 $\pm$ 2.85 & 62.93 $\pm$ 3.67 & 33.20  & 3.67 $\pm$ 0.34   \\ 

AS16 & $\chi_3$ & 2.43$_{-0.06}^{+0.06}$ & 13.06$_{-1.59}^{+2.29}$ & 0.50$_{-0.05}^{+0.07}$ & 1.32$_{-0.04}^{+0.03}$ & 635.18/532 & 35.10 $\pm$ 1.65 & 64.90 $\pm$ 2.34 & 25.58  & 3.06 $\pm$ 0.27   \\ 

AS17 & $\chi_3$ & 2.53$_{-0.08}^{+0.08}$ & 18.45$_{-3.65}^{+7.15}$ & 0.56$_{-0.08}^{+0.11}$ & 1.41$_{-0.05}^{+0.05}$ & 409.34/399 & 33.87 $\pm$ 4.77 & 66.13 $\pm$ 7.15 & 35.34  & 2.27 $\pm$ 0.33   \\ 

AS18 & $\theta$ & 2.50$_{-0.06}^{+0.07}$ & 14.57$_{-2.04}^{+3.11}$ & 0.54$_{-0.06}^{+0.08}$ & 1.30$_{-0.05}^{+0.05}$ & 538.45/537 & 34.88 $\pm$ 1.05 & 65.12 $\pm$ 1.46 & 18.33  & 2.72 $\pm$ 0.28   \\ 

AS19 & $\theta$ & 2.46$_{-0.08}^{+0.08}$ & 12.87$_{-1.66}^{+2.50}$ & 0.46$_{-0.06}^{+0.07}$ & 1.43$_{-0.06}^{+0.05}$ & 529.33/515 & 31.40 $\pm$ 0.94 & 68.60 $\pm$ 1.53 & 14.58  & 3.03 $\pm$ 0.30   \\ 

AS20 & $\theta$ & 2.52$_{-0.08}^{+0.08}$ & 13.70$_{-1.96}^{+3.26}$ & 0.49$_{-0.06}^{+0.09}$ & 1.39$_{-0.07}^{+0.06}$ & 528.17/548 & 31.76 $\pm$ 1.49 & 68.24 $\pm$ 2.46 & 21.51  & 2.81 $\pm$ 0.30   \\ 

AS21 & $\theta$ & 2.57$_{-0.09}^{+0.09}$ & 14.73$_{-2.54}^{+4.43}$ & 0.50$_{-0.07}^{+0.10}$ & 1.43$_{-0.07}^{+0.07}$ & 296.87/297 & 30.27 $\pm$ 5.16 & 69.73 $\pm$ 9.09 & 35.63  & 2.59 $\pm$ 0.33   \\ 

AS22 & $\theta$ & 2.48$_{-0.07}^{+0.08}$ & 14.54$_{-2.28}^{+3.81}$ & 0.48$_{-0.06}^{+0.08}$ & 1.39$_{-0.04}^{+0.04}$ & 505.02/470 & 32.08 $\pm$ 3.01 & 67.92 $\pm$ 4.90 & 31.50  & 2.76 $\pm$ 0.32   \\ 

\hline
    &&&&& \multicolumn{2}{c}{Short Variability Phase (SVP)}  \\
    \hline

AS23 & $\beta$ & $2.39_{-0.17}^{+0.17}$ & $18.14_{-4.46}^{+12.07}$ & $0.18_{-0.04}^{+0.06}$ & $2.10_{-0.05}^{+0.05}$ & 705.00/651 & $14.50\pm0.25$ & $85.50\pm1.21$ & $13.74$  & $2.51\pm0.50$   \\ 

AS24 & $\theta$ & $2.42_{-0.08}^{+0.08}$ & $12.20_{-1.38}^{+1.85}$ & $0.35_{-0.04}^{+0.05}$ & $1.55_{-0.05}^{+0.05}$ & 739.80/668 & $26.12\pm0.78$ & $73.88\pm1.65$ & $31.55$  & $3.21\pm0.28$   \\ 

AS25 & $\chi_3$ & $2.72_{-0.09}^{+0.10}$ & $23.82_{-6.58}^{+18.01}$ & $0.58_{-0.10}^{+0.16}$ & $1.39_{-0.06}^{+0.06}$ & 618.38/551 & $29.41\pm1.88$ & $70.59\pm3.53$ & $33.59$  & $1.69\pm0.35$   \\ 

\hline
    &&&&& \multicolumn{2}{c}{Long Plateau Phase (LPP)}  \\
    \hline

AS26 & $\chi_2$ & 2.04$_{-0.06}^{+0.06}$ & 29.19$_{-7.81}^{+25.66}$ & 0.49$_{-0.08}^{+0.16}$ & 1.19$_{-0.06}^{+0.04}$ & 653.75/550 & 51.75 $\pm$ 0.84 & 48.25 $\pm$ 0.62 & 16.19  & 2.34 $\pm$ 0.45   \\ 

AS27 & $\chi_2'$ & 2.15$_{-0.03}^{+0.03}$ & 29.00* & 0.56$_{-0.07}^{+0.08}$ & 1.18$_{-0.06}^{+0.05}$ & 591.47/537 & 50.34 $\pm$ 1.51 & 49.66 $\pm$ 1.11 & 16.28  & 2.15 $\pm$ 0.05   \\ 

AS28 & $\rho$ & 2.15$_{-0.04}^{+0.05}$ & 29.26* & 0.44$_{-0.06}^{+0.09}$ & 1.32$_{-0.06}^{+0.05}$ & 627.90/576 & 42.53 $\pm$ 0.74 & 57.47 $\pm$ 0.81 & 17.01  & 2.14 $\pm$ 0.08   \\ 

AS29 & $\chi_2$ & 1.97$_{-0.05}^{+0.05}$ & 15.44$_{-1.94}^{+2.68}$ & 0.32$_{-0.03}^{+0.04}$ & 1.31$_{-0.03}^{+0.04}$ & 621.91/554 & 43.06 $\pm$ 0.75 & 56.94 $\pm$ 0.81 & 15.90  & 3.79 $\pm$ 0.34   \\ 

AS30 & $\rho$ & 2.04$_{-0.05}^{+0.06}$ & 25.23$_{-6.06}^{+16.55}$ & 0.28$_{-0.03}^{+0.06}$ & 1.29$_{-0.02}^{+0.03}$ & 596.06/633 & 37.38 $\pm$ 0.65 & 62.62 $\pm$ 0.89 & 12.08  & 2.59 $\pm$ 0.44   \\ 

AS31 & $\rho$ & 1.99$_{-0.04}^{+0.04}$ & 21.76$_{-3.87}^{+6.26}$ & 0.29$_{-0.03}^{+0.03}$ & 1.31$_{-0.02}^{+0.02}$ & 621.51/558 & 39.91 $\pm$ 0.69 & 60.09 $\pm$ 0.85 & 12.89  & 2.98 $\pm$ 0.37   \\ 

AS32 & $\rho$ & 1.90$_{-0.05}^{+0.06}$ & 15.94$_{-2.62}^{+5.12}$ & 0.21$_{-0.02}^{+0.03}$ & 1.36$_{-0.02}^{+0.03}$ & 572.53/561 & 35.45 $\pm$ 0.61 & 64.55 $\pm$ 0.91 & 10.70  & 3.95 $\pm$ 0.46   \\ 

AS33 & $\rho$ & 1.87$_{-0.05}^{+0.05}$ & 19.30$_{-3.66}^{+7.24}$ & 0.16$_{-0.02}^{+0.02}$ & 1.39$_{-0.02}^{+0.02}$ & 654.02/633 & 31.09 $\pm$ 0.54 & 68.91 $\pm$ 0.97 & 10.90  & 3.58 $\pm$ 0.47   \\ 

\hline
    &&&&& \multicolumn{2}{c}{Decay Phase (DP)}  \\
    \hline

AS34 & $\chi_1$ & 2.02$_{-0.05}^{+0.05}$ & 16.91$_{-2.13}^{+3.13}$ & 0.39$_{-0.04}^{+0.05}$ & 1.38$_{-0.02}^{+0.02}$ & 716.49/650 & 44.87 $\pm$ 0.78 & 55.13 $\pm$ 0.78 & 28.05  & 3.44 $\pm$ 0.31   \\ 

AS35 & $\chi_1$ & 2.01$_{-0.05}^{+0.05}$ & 15.86$_{-1.78}^{+2.67}$ & 0.36$_{-0.04}^{+0.04}$ & 1.36$_{-0.03}^{+0.03}$ & 722.41/642 & 43.54 $\pm$ 0.75 & 56.46 $\pm$ 0.80 & 23.61  & 3.61 $\pm$ 0.30   \\ 

AS36 & $\chi_4$ & 1.68$_{-0.05}^{+0.04}$ & 11.70$_{-0.81}^{+0.94}$ & 0.32$_{-0.03}^{+0.03}$ & 1.65$_{-0.06}^{+0.06}$ & 703.30/630 & 52.17 $\pm$ 0.77 & 47.83 $\pm$ 0.52 & 11.18  & 5.89 $\pm$ 0.35   \\ 

AS37 & $\chi_1$ & 1.89$_{-0.05}^{+0.05}$ & 26.80$_{-6.52}^{+19.10}$ & 0.40$_{-0.05}^{+0.08}$ & 1.43$_{-0.04}^{+0.04}$ & 686.07/652 & 52.22 $\pm$ 0.75 & 47.78 $\pm$ 0.50 & 11.19  & 2.83 $\pm$ 0.48   \\ 

AS38 & $\chi_1$ & 1.87$_{-0.05}^{+0.05}$ & 22.47$_{-4.35}^{+9.64}$ & 0.37$_{-0.04}^{+0.06}$ & 1.41$_{-0.05}^{+0.05}$ & 670.46/619 & 51.54 $\pm$ 0.76 & 48.46 $\pm$ 0.52 & 10.62  & 3.24 $\pm$ 0.44   \\ 

AS39 & $\chi_4$ & 1.64$_{-0.04}^{+0.04}$ & 21.25$_{-3.93}^{+9.33}$ & 0.44$_{-0.05}^{+0.07}$ & 1.86$_{-0.08}^{+0.08}$ & 764.32/618 & 63.42 $\pm$ 0.91 & 36.58 $\pm$ 0.37 & 6.63  & 4.26 $\pm$ 0.53   \\ 

AS40 & $\chi_4$ & 1.71$_{-0.07}^{+0.07}$ & 16.39$_{-2.25}^{+3.67}$ & 0.64$_{-0.11}^{+0.15}$ & 2.26$_{-0.26}^{+0.25}$ & 635.21/574 & 64.29 $\pm$ 0.92 & 35.71 $\pm$ 0.36 & 5.20  & 4.64 $\pm$ 0.52   \\ 

AS41 & $\chi_4$ & 1.59$_{-0.05}^{+0.05}$ & 13.69$_{-1.36}^{+1.89}$ & 0.35$_{-0.04}^{+0.04}$ & 1.73$_{-0.09}^{+0.09}$ & 590.05/533 & 60.12 $\pm$ 0.86 & 39.88 $\pm$ 0.41 & 4.77  & 5.92 $\pm$ 0.50   \\

\hline
    &&&&& \multicolumn{2}{c}{Obscured Phase (OP)}  \\
    \hline

AS42 & $\chi_4$ & 1.59$_{-0.02}^{+0.02}$ & 14.16$_{-1.44}^{+1.61}$ & 0.37$_{-0.04}^{+0.05}$ & 1.77$_{-0.08}^{+0.08}$ & 704.44/573 & 60.82 $\pm$ 0.25 & 39.18 $\pm$ 0.16 & 4.38  & 5.81 $\pm$ 0.38   \\

\hline
\hline
     \end{tabular}
     \begin{itemize}
         \item[*] Frozen parameter 
         \item[p] Parameter pegged at higher limit. 
         \item[$\dagger$] The spectral analysis of AS4 is not performed due to the unavailability of Level 2 \textit{SXT} products.
     \end{itemize}

\end{table*}


\section{Results}
\label{sec4:results}
In this work, we present a detailed analysis of the energy-dependent fast variability properties of GRS 1915+105 during intervals exhibiting LFQPOs, using \textit{AstroSat} observations. These were correlated with long-term flux measurements from \textit{MAXI} ($2-10$ keV) and \textit{BAT} ($15-50$ keV) between January 2016 and April 2019 (see Fig.~\ref{fig1:maxi_lightcurve}) to study flux variations across energy bands. All LFQPO detections were grouped into five distinct phases, corresponding to different variability classes observed by \textit{AstroSat} as mentioned below.

\begin{enumerate}
    \item Short Plateau Phase (MJD 57451 $-$ 57453)
    \item Short Variability Phase (MJD 57688 $-$ 57689)
    \item Long Plateau Phase (MJD 57840 $-$ 57892)
    \item Decay Phase (MJD 58209 $-$ 58358)
    \item Obscured Phase (MJD 58564)
\end{enumerate}
These five phases (i) to (v) are highlighted by yellow, olive, magenta, grey, and cyan shaded regions, respectively, in Fig. \ref{fig1:maxi_lightcurve}. Interestingly, the `colour' remains nearly constant at $\sim$1 throughout all phases except the Short Variability Phase, indicating that the source persistently resides in a harder spectral state. Variability class transitions were observed only during the Short Plateau Phase (MJD 57451 $-$ 57453), Short Variability Phase (MJD 57688 $-$ 57689) and in the initial period of Long Plateau Phase (MJD 57840 $-$ 57892) due to the availability of continuous observations during these intervals. Despite the similar magnitude of `colour', each phase exhibits distinct variability classes and fast variability characteristics of each classes.

\subsection{Variability Classes during Different Phases}
\label{subsec:lc_ccd}
The source was found to be in the $\theta$, $\beta$, $\rho$, and $\chi$ variability classes of GRS 1915+105 during the considered observations, consistent with classifications reported by \citet{yadav2016}, \citet{rawat_2019_grs}, \citet{athulya2022}, and \citet{seshadri2022}.
Lightcurves in the $3-60$ keV energy band for each variability class are shown in Fig. \ref{fig2:lc_ccd}, with the corresponding CCD displayed in the top-right corner of each panel. The variability class and the respective MJD are mentioned in the top-left corner of each panel. 
The lightcurve during $\theta$ class, exhibits a characteristic `M' shaped intervals (see top-left panel in Fig. \ref{fig2:lc_ccd}) with typical durations of a few hundred seconds. Furthermore, the $\beta$ class displays highly irregular and complex variability signatures (see bottom-left panel in Fig. \ref{fig2:lc_ccd}) followed by a prolonged low-count intervals. In contrast, the $\rho$ class shows heartbeat-like variability \citep[see also][]{athulya2022}, which is shown in the bottom-right panel in Fig. \ref{fig2:lc_ccd}. 
The ‘harder’ variability class $\chi$ typically shows relatively low variability in count rates and a CCD dominated by high-energy photons, with HR2 values extending to the higher value (see top-right panel of Fig. \ref{fig2:lc_ccd}). Following \cite{belloni2000} and \cite{sreehari2020}, we further sub-classify the $\chi$ class into $\chi_1$, $\chi_2$, $\chi_3$, and $\chi_4$ based on average count rate and hardness ratios (HR1 and HR2) as shown in the top-right panel of Fig. \ref{fig2:lc_ccd}. Among these, $\chi_3$ shows the highest count rate ($8445-5487$ cts/s) with the lowest hardness, while $\chi_1$ exhibits lower count rates ($1065-4731$ cts/s) with relatively higher hardness. $\chi_2$ shows hardness greater than both $\chi_3$ and $\chi_1$, with count rates of $1779-1853$ cts/s. The $\chi_4$ sub-class exhibits the maximum hardness, with HR2 extending further to the higher value, and the lowest count rates ($380-1282$ cts/s).
The variability classes and the class transitions observed during different phases are presented as follows.

\begin{itemize}

    \item \textbf{Short Plateau Phase (SPP):} The source exhibits rapid variability over a $\sim$1.5 day period (MJD $57451-57453$), characterized by a short dip-like feature in the lightcurve persisting for $\sim 15$ hours. The count rate drops from 8336 to 3998 cts/s within a day, followed by a rise to 8059 cts/s in the next day. Interestingly, the hardness ratios exhibit an opposite trend where it increases initially and decreases further (see Table \ref{tab:log_table}). During this phase, the source undergoes a transition through the $\theta \rightarrow \chi \rightarrow \theta$ classes \citep{yadav2016,seshadri2022, belloni2024}. Moreover, the study of hardness ratio during $\chi$ class indicates that the source goes through a transition $\chi_3 \rightarrow \chi_1 \rightarrow \chi_3$. The lightcurve and CCD for $\chi_1$ and $\chi_3$ in MJD 57452 is shown in the top-right panel of Fig. \ref{fig2:lc_ccd}. 
    Hence, the class transition during SPP is observed to be: $\theta \rightarrow \chi_3 \rightarrow \chi_1 \rightarrow \chi_3 \rightarrow \theta$ (see Table \ref{tab:log_table}).

    \item \textbf{Short Variability Phase (SVP):} During this phase, the \textit{MAXI} light curve shows a variable feature, while the \textit{BAT} flux rises sharply from 0.25 to 0.37 Crab in the $15-50$ keV band (see middle panel of Fig.\ref{fig1:maxi_lightcurve}). Simultaneously, the source `colour' increases from $\sim$0.1 to $\sim$0.5 (see lower panel), indicating a hardening for short interval. During this phase, $\beta$ class is observed, as first reported by \cite{seshadri2022} (see bottom-left panel of Fig.\ref{fig2:lc_ccd}). However, our detailed analysis reveals that the source
    exhibits $\theta$ class afterwards, following by a short interval of $\chi$ class which can be sub-classified as $\chi_3$ based on hardness ratio (see Table \ref{tab:log_table}).
    Hence, the source undergoes a transition from $\beta \rightarrow \theta \rightarrow \chi_3$ during this phase. The average count rate rises from 6158 to 10,664 cts/s, accompanied by an increase in the average hardness ratio (see Table \ref{tab:log_table}). Notably, the count rate of the $\chi_3$ class in this observation is significantly higher than that observed during the Short Plateau Phase.

     \item \textbf{Long Plateau Phase (LPP):} During this phase (MJD $57840-57892$), the \textit{MAXI} flux remained nearly constant at approximately 0.2 Crab for 52 days, while the \textit{BAT} flux gradually declined from about 0.2 to 0.09 Crab (see Fig. \ref{fig1:maxi_lightcurve}). The source count rate varied between 1451 and 1929 counts/s, with an overall decrease in both HR1 and HR2.
     Between MJD 57840 and 57841 (AS26$-$AS28), the source was reported to exhibit a transition from $\chi \rightarrow IMS \rightarrow \rho$ class, where IMS represents an intermediate transitional state between $\chi$ and $\rho$ \citep{rawat_2019_grs}. 
     The sub-classification indicates the aforementioned $\chi$ to be $\chi_2$. Additionally, the IMS class identified in earlier studies, exhibits long-term, large-amplitude, irregular variability only on timescales of a few thousand seconds. This behaviour is observed only in the soft X-ray band ($3-5$ keV) and exhibits strong similarities in CCD with the $\chi_2$ class (see  the inset in top-right panel in Fig. \ref{fig2:lc_ccd}). However, it exhibits a clear modulation in the lightcurve \citep[see][for details]{rawat_2019_grs}.
     Therefore, despite the lack of significant spectral differences, the subtle timing variations provide sufficient support to conclude that the IMS is a subset of $\chi_2$, which we designate as the $\chi_2'$ variability class. Hence, within one day (MJD $57840-57841$) the source exhibits a class transition of $\chi_2 \rightarrow \chi_2' \rightarrow \theta$. Beyond MJD 57841, the source exhibited $\chi_2$ and $\rho$ variability class during this entire phase.

    \item \textbf{Decay Phase (DP):} During this phase, beginning around MJD 58209, both \textit{MAXI} flux ($2-10$ keV) and \textit{BAT} flux ($15-50$ keV) start to decline, and this decay continues for $\sim$ 150 days (see Fig. \ref{fig1:maxi_lightcurve}). The average source count rate drops from 2929 cts/s to 397 cts/s, while the hardness ratios gradually increase (see Table \ref{tab:log_table} for details). This trend is also observed in the colour variation of the source shown in the bottom panel of Fig. \ref{fig1:maxi_lightcurve}. The source remains in the $\chi$ class throughout this phase, which we further sub-classify into $\chi_1$, $\chi_2$, and $\chi_4$ variability classes. Interestingly, the $\chi_1$ class observations during this period exhibit lower count rates compared to those in the Short Plateau Phase, while maintaining similar hardness ratios.

    \item \textbf{Obscured Phase (OP):} In this phase, the \textit{MAXI} flux ($2-10$ keV) reaches its minimum, while the \textit{BAT} flux ($15-50$ keV) remains nearly constant at $\sim$0.15 Crab, leading to the maximum observed `colour' of the source. Analysis of the light curve and CCD confirms that the source corresponds to the $\chi_4$ variability class (see Fig. \ref{fig2:lc_ccd} and Table \ref{tab:log_table}).

\end{itemize}

\subsection{Timing Features in Different Phases}
\label{subsec:pds_lag}
The Fast Fourier analysis of all observations reveal the presence of QPOs with centroid frequencies ($\nu_{\text{QPO}}$) ranging from 1.38 to 7.38 Hz and corresponding percentage rms amplitudes ($\text{rms}_\text{QPO}$) between 1.93\% and 11.92\% (see Table \ref{tab:log_table}). Distinct variability characteristics associated with $\nu_{\text{QPO}}$ are observed across the five identified phases. During the SPP, $\nu_\text{QPO}$ exhibits rapid fluctuations, initially decreasing from 5.32 Hz to 2.55 Hz as the count rate declines, and then rising to 6.55 Hz as the count rate increases again (see panels (a) to (e) in Fig. \ref{fig3:pds}). In contrast, the $\text{rms}_{\text{QPO}}$ follows an opposite trend, increasing from 4.18\% to 11.31\% as the count rate decreases, and then dropping to 3.55\% with the subsequent increase in the count rate.

Interestingly, the QPOs observed during the SVP are quite broad (see panel (f) in Fig. \ref{fig3:pds} and Table \ref{tab:log_table}), with a Q-factor as low as 1.83. The $\nu_\text{QPO}$ increases steadily from 3.33 Hz to 7.38 Hz, while the $\text{rms}_\text{QPO}$ varies between 2.80\% and 5.37\% (see Table \ref{tab:log_table}). The associated time-lag is soft, ranging between $\sim$1–2 ms. During the LPP, the $\nu_\text{QPO}$ remains nearly constant at $\sim 4-5$ Hz (see panels (g)–(i) in Fig.\ref{fig3:pds}), with $\text{rms}_\text{QPO}$ in the range $\sim5.99-11.26$\% and a soft-lag of $\sim$6.5 ms. In the DP, the QPO frequency decreases to as low as 1.38 Hz (see panel (j) in Fig.\ref{fig3:pds}), with an rms amplitude of 4.68\% and associated with a soft-lag of 2.90 ms. Finally, in the OP, the QPO is detected at 2.39 Hz with an rms amplitude of 3.89\% and a soft-lag of 1.59 ms. Details of the timing properties are tabulated in Table \ref{tab:log_table}.

Apart from this, we find that different variability classes exhibit distinct ranges of QPO centroid frequencies ($\nu_{\text{QPO}}$). The QPO frequency of $\chi_4$ class observations lie in the range $1.38-2.41$ Hz. Whereas, $\chi_1$, $\chi_2$ and $\beta$ class observations show QPO frequency in the range $2.55-4.30$ Hz. However, the $\chi_2'$, $\chi_3$, $\rho$ and $\theta$ class observations lie in an overlapping range of $4.30-6.55$ Hz.

The broadband ($3-60$ keV) PDS and time-lag spectrum in the frequency domain $0.1-100$ Hz are shown in Fig. \ref{fig3:pds} for different phases. We found soft-lag for all observations for all values of $\nu_\text{QPO}$ and the magnitude of soft-lag lies in the range 1.59$-$13.49 ms. 
The soft-lag during $\chi_4$ class is observed to be in range of $1.59-2.90$ ms, which is the minimum. During $\chi_1$, $\chi_2$ and $\beta$ classes, the soft-lag lies in the range $2.65-7.32$ ms. However, the soft-lag during $\chi_2'$, $\chi_3$, $\rho$ and $\theta$ class observations lies in an overlapping range of $6.05-13.49$ ms.
Surprisingly, our analysis shows that the time-lag associated with $\nu_\text{QPO} < 2.2$ Hz exhibits a soft-lag (see panel (j) in Fig. \ref{fig3:pds}), in contrast to the hard-lag previously reported at lower $\nu_\text{QPO}$ using \textit{RXTE} observations \citep{zhang_2020_grs}. To verify this result, we rechecked the time-lag analysis of the \textit{AstroSat} data using \textsc{ghats} v3.2.0\footnote{\url{http://astrosat-ssc.iucaa.in/uploads/ghats_home.html}}, and obtained consistent findings.

To investigate the long-term evolution of the QPO properties, we analyzed \textit{RXTE} observations from MJD 50278 to 53214. A clear correlation between $\nu_\text{QPO}$ and $\text{rms}_\text{QPO}$ is observed, as shown in the upper panel of Fig. \ref{fig4:rms_lag_v_qpo}. The star and circle markers represent \textit{RXTE} and \textit{AstroSat} data, respectively, with different variability classes indicated by the colours in the legend. For \textit{AstroSat} observations, $\text{rms}_\text{QPO}$ initially increases with $\nu_\text{QPO}$ and then decreases at higher frequencies. A similar trend is also seen in the \textit{RXTE} data, consistent with the results reported by \cite{zhang_2020_grs}.
However, in case of \textit{AstroSat} observations we find an exception for the $\beta$ class (AS23), where it do not follow the similar trend as other observations (marked in red circle; see upper panel in Fig. \ref{fig4:rms_lag_v_qpo}).
We fitted the data points of $\text{rms}_\text{QPO}$ as a function of $\nu_\text{QPO}$, excluding the $\beta$ class observation for \textit{AstroSat} with broken linear function as follows.
\setlength{\abovedisplayskip}{5pt}
\setlength{\belowdisplayskip}{5pt}
\begin{equation}
\text{rms}_\text{QPO} [RXTE] = 
\begin{cases}
 +2.91\times\nu_{QPO} + 7.45 & ;\nu_{QPO} \leq 2.04 \\
-3.69\times \nu_{QPO} + 20.92 & ;\nu_{QPO} \geq 2.04
\end{cases}
\end{equation}
\vspace{-1.0em}
\begin{equation}
\text{rms}_\text{QPO} [AstroSat] = 
\begin{cases}
 +4.21\times\nu_{QPO} -2.81 &  ;\nu_{QPO} \leq 3.42 \\
-3.22\times \nu_{QPO} + 22.61 & ;\nu_{QPO} \geq 3.42
\end{cases}
\end{equation}

The model fitted broken linear function for both \textit{RXTE} (blue) and \textit{AstroSat} (red) is shown in the upper panel of Fig. \ref{fig4:rms_lag_v_qpo}. Interestingly, the source does not follow the same trend during \textit{AstroSat} observations as during \textit{RXTE} observations. The break frequency ($\nu_\text{QPO}$, where $\text{rms}_\text{QPO}$ is maximum) during \textit{RXTE} era was found to be at 2.04 Hz, which agrees with the results obtained by \cite{zhang_2020_grs}. However, during \textit{AstroSat} observations the break frequency is found to be 3.42 Hz. 
We find maximum $\text{rms}_\text{QPO}$ for $\chi_1$ and $\chi_2$ variability classes for both \textit{RXTE} and \textit{AstroSat} observations.

A correlation is also obtained between the average time-lag ($\delta t$) and $\nu_\text{QPO}$, which is shown in the lower panel of Fig. \ref{fig4:rms_lag_v_qpo}. The circle and star represent the \textit{AstroSat} and \textit{RXTE} observations respectively and the filled colours represent different variability classes.
For \textit{RXTE} observations, the time-lag increases as the $\nu_\text{QPO}$ decreases and switches it's sign from soft-lag to hard-lag at $\sim$ 2.2 Hz \citep{reig2000,dutta2018,zhang_2020_grs}. 
However, in our analysis of \textit{AstroSat} observations, the time-lag increases as $\nu_\text{QPO}$ decreases, but it does not switch sign and remains negative even for $\nu_\text{QPO} \leq 2.2$ Hz. We further fitted the data points of time-lag ($\delta t$) as a function of $\nu_\text{QPO}$ with a power-law function as follows,

\begin{equation}
\delta t [RXTE]= +41.61\times(\nu_{QPO})^{-1.63}-9.47 
\end{equation}
\vspace{-1.0em}
\begin{equation}
\delta t[AstroSat]= -6.29\times(\nu_{QPO})^{+0.53}+5.71 
\end{equation}

The fitted powerlaw for \textit{RXTE} and \textit{AstroSat} is shown as blue and red dashed line respectively in the lower panel of Fig. \ref{fig4:rms_lag_v_qpo}.
The switching frequency ($\nu_\text{QPO}$, where $\delta t=0$; $\nu_{\text{QPO}\mid \delta t=0}$) for \textit{RXTE} observation is $\sim$2.2 Hz whereas, for \textit{AstroSat} observation the $\nu_{\text{QPO}\mid \delta t=0}=0.85$ Hz.  It is also important to note that for \textit{RXTE} observation $\chi_1$ and $\chi_3$ classes were exhibited hard-lag, whereas for \textit{AstroSat} observation both classes show soft-lag. In \textit{AstroSat} era, only $\chi_4$ exhibits minimum soft-lag of 1.6 ms whereas, $\theta$ class exhibits maximum soft-lag $\sim$ 13.5 ms and both these classes show distinct lag feature. Additionally both of these classes are distinctly separated by region in the time-lag and $\nu_\text{QPO}$ correlation diagram (see lower panel in Fig. \ref{fig4:rms_lag_v_qpo}) whereas, other classes remained in overlapping region. 
 
It is noteworthy that the $\beta$ class observation do not lie in the correlation between $\text{rms}_\text{QPO}$ and $\nu_\text{QPO}$ (marked in red circle; see upper panel in Fig. \ref{fig4:rms_lag_v_qpo}) and it also exhibits soft-lag at the corresponding $\nu_\text{QPO}$ (see Table \ref{tab:log_table}).
\cite{soleri_2008_grs_type-b} reported similar behaviour for type-B QPOs in the $\beta$ and $\mu$ classes, where type-B QPOs do not exhibit the $\text{rms}_\text{QPO}$ and $\nu_\text{QPO}$ correlation as seen in type-C QPOs, and moreover type-B QPOs are associated with soft-lags. This suggests that the QPO observed in the $\beta$ class during \textit{AstroSat} corresponds to a type-B QPO, while those in other variability classes are consistent with type-C QPOs.

\subsection{Energy Dependent Variability Study}
The energy-dependent variability analysis depicts how the $\text{rms}_\text{QPO}$ evolves with energy across different variability classes, as shown in the upper panel of Fig. \ref{fig5:rms_lag_v_energy}. The observation dates (MJD), corresponding variability classes, and $\nu_\text{QPO}$ are indicated in the legend. For all variability classes, the $\text{rms}_\text{QPO}$ increases with energy, indicating a strong energy dependence. Notably, the energy dependent $\text{rms}_\text{QPO}$ variations differ distinctly among classes up to $\sim$ 15 keV. Beyond this energy, the $\text{rms}_\text{QPO}$ variation exhibits a nearly flat trend. The $\text{rms}_\text{QPO}$ is generally lower for higher $\nu_{\text{QPO}}$ values and in the $\chi_1$ and $\chi_4$ classes, while it is higher for lower $\nu_\text{QPO}$ values, particularly in the $\chi_3$ class. These distinct energy dependent $\text{rms}_\text{QPO}$ trends suggest a strong correlation between QPO properties and the underlying physical mechanisms responsible for QPO generation in the hard spectral states.

The energy-dependent time-lag measurements relative to the $3-6$ keV band for all variability classes observed by \textit{AstroSat} are presented in the lower panel of Fig. \ref{fig5:rms_lag_v_energy}. For all observations, we consistently detect soft-lags, where higher-energy photons arriving earlier than the 3–6 keV photons, regardless of the QPO frequency ($\nu_{\text{QPO}}$) or variability class. In all cases, the magnitude of the soft-lag increases with energy. Notably, the slope of soft-lag variation differs across variability classes and QPO frequencies. Specifically, the soft-lag is highest in the $\chi_3$ class and at higher $\nu_{\text{QPO}}$ values, while it is comparatively lower in the $\chi_4$ class and at lower $\nu_{\text{QPO}}$ values.
This behaviour clearly indicates that photons in higher energy bands experience larger soft lags, thereby ruling out band-selection effects as the origin of the observed soft lag.

\subsection{Results of Spectral Analysis}

In this work, we carried out a detailed spectral analysis to explore the connection between the energy spectra and time-lag properties of GRS 1915+105 during observations exhibiting LFQPOs. Such LFQPOs are generally associated with the hard and hard-intermediate spectral states of BH-XRBs. For the considered observations, the photon index vary between $\Gamma = 1.59-2.63$, while the electron temperature ($kT_e$) ranges from $10.96$ to $29.26$ keV. The covering fraction is found to vary between $f_{\rm cov} = 0.16-0.94$, and the bolometric luminosity lies in the range $4.38-41.70$\% of the Eddington luminosity (see Table \ref{tab:spectral_table}). These spectral characteristics, specifically, the relatively low photon index, higher electron temperature, and lower bolometric luminosity are consistent with the source being in the low-hard or hard-intermediate state during the LFQPO intervals \citep{titarchuk_2004}. 
Different spectral characteristics are observed across different phases of evolution of LFQPO. During the SPP, the photon index ($\Gamma$) remains relatively constant, ranging between $2.32$ and $2.63$. As the source undergoes the transition $\theta \rightarrow \chi_3 \rightarrow \chi_1$, the percentage of Comptonized flux ($F_{\rm Comp}$) increases from $30.41\%$ to $54.01\%$. Interestingly, during the subsequent reverse transition $\chi_3 \rightarrow \theta$, $F_{\rm Comp}$ decreases again to $30.27\%$ (see Table \ref{tab:spectral_table}). This evolution is consistent with the hardness ratio reported in Table \ref{tab:log_table}, which exhibits a similar trend (see \S \ref{subsec:lc_ccd}).

During the SVP, the percentage disc flux ($F_{\rm Disc}$) is found to be relatively high, ranging from $70.59\%$ to $85.50\%$. However, during the rapid transition $\beta \rightarrow \theta \rightarrow \chi_3$, the Comptonized flux ($F_{\rm Comp}$) increases from $14.50\%$ to $29.41\%$, accompanied by an increase in the covering fraction ($f_{\rm cov}$) from 0.18 to 0.58. This indicates a clear spectral hardening of the source within a single day (see Table \ref{tab:spectral_table}).

In the LPP, the spectral parameters remain relatively stable. The photon index ($\Gamma$) is observed in the range $1.87$–$2.15$, which is noticeably lower than during the SVP, indicating a comparatively harder spectrum. During this phase, $F_{\rm Comp}$ gradually decreases from $51.75\%$ to $31.09\%$, a trend also reflected in the hardness ratio (see Table \ref{tab:log_table}).

Conversely, during the DP, $F_{\rm Comp}$ gradually increases from $44.87\%$ to $60.12\%$, while $\Gamma$ decreases from 2.02 to 1.59, suggesting that the source is evolving toward a harder spectral state. The lowest bolometric luminosity ($L_{\rm Edd}\sim4.4\%$) is observed during the OP, when the source reaches its minimum count rate but still exhibits LFQPOs, implying that the mechanism responsible for QPOs do not depend on the luminosity of the source.

\subsection{Spectro-temporal Correlation Study}
A correlation between the spectral ($F_{\rm Comp}$ and $\Gamma$) and temporal parameters 
($\text{rms}_{\rm QPO}$, $\nu_{\rm {QPO}}$ and time-lag) of the source is established to account for the observed time-lag behaviour. The variations in spectral parameters during phases exhibiting LFQPOs are systematically compared with the corresponding timing parameters, as presented in Tables \ref{tab:log_table} and \ref{tab:spectral_table}.
We excluded the $\beta$ class observation from this correlation study, as it exhibits type-B LFQPOs (see \S \ref{subsec:pds_lag}). 
The upper panel of Fig.~\ref{fig8:rms_lag_v_comp-flux} presents the correlation between $\text{rms}_{\rm QPO}$ and the percentage Comptonized flux ($F_{\rm Comp}$), with the LFQPO centroid frequency ($\nu_{\rm QPO}$) shown as a colour-bar. Different variability classes are represented by distinct markers as indicated in the legend. 
The $\text{rms}_{\rm QPO}$ increases sharply with $F_{\rm Comp}$ up to $\sim$12\%, except for four $\chi_4$ class observations during the decaying (DP) and obscured (OP) phases, when the total source count rate drops to $\sim$300–600 counts/s, indicating a dependence of $\text{rms}_{\rm QPO}$ on the source flux. The maximum $\text{rms}_{\rm QPO}$, reaching at $\sim45\%$ $F_{\rm Comp}$, is observed mainly during the $\chi_1$, $\chi_2$, and $\chi_2'$ observations. A clear anti-correlation is also observed between $F_{\rm Comp}$ and $\nu_{\rm QPO}$, with $\nu_{\rm QPO}$ decreasing as $F_{\rm Comp}$ increases. The highest $F_{\rm Comp}$ values are found in the $\chi_4$ class, where $\nu_{\rm QPO}$ reaches its minimum and the source flux declines to a few percent of the bolometric luminosity ($\sim4.38\% ~L_{\rm Edd}$), consistent with this class representing the extreme hard state of the source.

In the lower panel of Fig.~\ref{fig8:rms_lag_v_comp-flux}, we find clear evidence that the magnitude of the soft-lag decreases with $F_{\rm Comp}$ as the source moves toward harder spectral states, indicated by the decreasing photon index $\Gamma$ shown in the colour bar.
Thus, a strong spectro-temporal correlation is observed, as $F_{\rm Comp}$ increases, the QPO fractional rms ($\text{rms}_{\rm QPO}$) increases and the soft-lag magnitude decreases, while both $\nu_{\rm QPO}$ and the photon index $\Gamma$ decrease.
This suggests that as the source transitions toward a harder spectral state, the corona expands (leading to a decline in $\nu_{\rm QPO}$), Comptonization ($F_{\rm Comp}$) becomes more dominant, and the time-lag increases, though it remains negative.
We also find a similar correlation between time-lag and hardness ratio, as hardness primarily traces the relative contribution of hard to soft photons and thus serves as a proxy for the efficiency of the Comptonization process.
The near-linear increase in time-lag with rising $F_{\rm Comp}$ across all \textit{AstroSat} observations indicates that the lag arises from the combined effects of multiple physical processes, with Comptonization consistently contributing a positive component \citep{dutta2016}. These correlations further suggest that variations in the Comptonized flux play a key role in driving the evolution of temporal features observed in GRS 1915+105.

\section{Discussion}
\label{sec5:discussion}

In this work, we have carried out a comprehensive study of variability classes of GRS 1915+105, using \textit{AstroSat} observations spanning March 2016 to March 2019. Our primary aim was to investigate the class transitions and correlating the spectro-temporal properties in order to explain the observed time-lag behaviour, and to compare these results with those obtained from earlier \textit{RXTE} observations.

\subsection{Class Transitions: Probing Time Scale of Accretion Flow}

Based on the long-term \textit{MAXI/GSC} and \textit{BAT} light curves (see Fig.\ref{fig1:maxi_lightcurve}), all LFQPO detections are grouped into five distinct phases: a Short Plateau Phase (SPP); lasting $\sim$15 hours in the \textit{LAXPC} light curve (see inset, upper panel of Fig.\ref{fig1:maxi_lightcurve}), a Short Variability Phase (SVP); characterized by small variability features over $\sim$50 days, a Long Plateau Phase (LPP); a relatively stable period with minimal variation in the \textit{MAXI} light curve, a Decay Phase (DP); during which both \textit{MAXI} and \textit{BAT} flux decrease gradually, and finally, an Obscured Phase (OP); where the source reaches its lowest \textit{MAXI} flux. Across these phases, LFQPOs are detected in eight variability classes: $\theta$, $\chi_1$, $\chi_2$, $\chi_2'$, $\chi_3$, $\chi_4$, $\beta$, and $\rho$, with their light curves and CCDs shown in Fig.\ref{fig2:lc_ccd}.

During the SPP, the source undergoes a rapid transition, $\theta \rightarrow \chi_3 \rightarrow \chi_1 \rightarrow \chi_3 \rightarrow \theta$, within nearly two days where, the inter-class transition is observed within $\sim2$ hours (see Table \ref{tab:log_table}). At the beginning of this phase, the count rate, QPO frequency ($\nu_{\rm QPO}$), and disc flux ($F_{\rm Disc}$) decrease, while the rms amplitude ($\text{rms}_{\rm QPO}$) and Comptonized flux ($F_{\rm Comp}$) increase, indicating a recession of the inner disc edge and an expansion of the Comptonizing region within few hours (see Table \ref{tab:log_table} and \ref{tab:spectral_table}). 
The enhanced $F_{\rm Comp}$ leads to a higher $\text{rms}_{\rm QPO}$ \citep{chakrabarti_manickam2000}, consistent with the accretion dynamics predicted by the two-component advective flow (TCAF) paradigm \citep{chakrabarti_titarchuk1995, molteni1996, rao2000, nandi_2001, santa_das2014}. Subsequently, the source returns to its initial variability class ($\theta$), with all spectral and timing parameters recovering their previous values, indicating that the accretion geometry is restored to its original configuration within a few hours.
This behaviour points to a highly variable accretion flow that evolves on timescales of just a few hours, after which the system reverts to its initial variability class. Such rapid variations are consistent with previous findings, where \citet{yadav2016} reported a correlation between QPO frequency and bolometric flux, and also, \citet{belloni2024} showed that the variation of Comptonizing region as the QPO frequency increases or decreases \citep[see][]{chakrabarti_manickam2000, chakrabarti_nandi2005, dutta2018, karpouzas_2021, athulya2022,belloni2024}.

During the SVP, the source undergoes a rapid class transition $\beta \rightarrow \theta \rightarrow \chi_3$ within a day. The QPOs are broad, and \textit{AstroSat} detects a type-B QPO in the $\beta$ class for the first time. The concurrent rise in disc flux ($F_{\rm Disc}$) to its highest level further supports the association of type-B QPOs with enhanced disc emission during this variability class \citep[and references therein]{choudhury_etal2025}.
As the source progresses through $\theta \rightarrow \chi_3$, the QPO transitions from type-B to type-C, accompanied by increases in both $\nu_{\rm QPO}$ and count rate within a few hours, indicating a similar geometrical variation of the Comptonizing region as observed during the SPP.

During the LPP, the source exhibits a class transition $\chi_2 \rightarrow \chi_2' \rightarrow \rho$ again within a day where the canonical class transition occur within $\sim 7$ hours. While \citet{rawat_2019_grs} classified the interval between $\chi$ and $\rho$ as an Intermediate State (IMS), our analysis find this state closely resembles with $\chi_2$, and we therefore classify it as $\chi_2'$. During this short transition, the count rate, $\nu_{\rm QPO}$, and $F_{\rm Disc}$ increase, while the ${\rm rms}_{\rm QPO}$ and $F_{\rm Comp}$ decrease, reflecting a variation in accretion geometry similar to that observed in the SPP and SVP (see Table \ref{tab:log_table} and \ref{tab:spectral_table}).
Similarly, \citet{misra2020_grs1915} reported that the QPO frequency increases as the inner disc radius contracts inward. Beyond this short transition, the source remains in the $\chi_2$ and $\rho$ classes, showing relatively minor variations in count rate, $\nu_{\rm QPO}$, and ${\rm rms}_{\rm QPO}$, while $F_{\rm Comp}$ gradually decreases, in agreement with previous findings for the $\chi$ class \citep{dutta2018, karpouzas_2021}.

During DP, the source is observed in the $\chi_1$ and $\chi_4$ classes (see Fig. \ref{fig3:pds}). The count rate, $\nu_{\rm QPO}$, ${\rm rms}_{\rm QPO}$, and $F_{\rm Disc}$ gradually decrease, accompanied by a reduction in bolometric luminosity (see Table \ref{tab:log_table} and \ref{tab:spectral_table}). In contrast, the source hardness and $F_{\rm Comp}$ increase over time, consistent with \citet{athulya2022}. This hardening is also evident in the colour variation shown in the lower panel of Fig.~\ref{fig1:maxi_lightcurve}. The lowest count rate and bolometric flux (4.38\% $L_{\rm Edd}$) are observed during the $\chi_4$ variability class in the subsequent Obscured Phase (OP).

Therefore, we find similar variations in count rate, $\nu_{\rm QPO}$, $F_{\rm Disc}$, $\text{rms}_{\rm QPO}$, and $F_{\rm Comp}$ during rapid class transitions across the SPP, SVP, and LPP phases. Interestingly, the variability class transition occurs on timescales of just a few hours and are accompanied by an increase in count rate with rising QPO frequency or the opposite trend. The key feature is the significant change in count rate between pre- and post-transition phases of variability classes, occurring over remarkably short timescales.
Similar rapid variability class transitions have also been frequently observed in \textit{RXTE} data, where such short timescales suggest the disappearance and reformation of the inner accretion disc \citep{belloni1997a, belloni1997b}. The observed increase in count rate implies a rise in the accretion rate, driving the corresponding increase in QPO frequency, while the transition duration reflects the infall timescale \citep{nandi2001, chakrabarti_nandi2005, iyer_nandi2015, santa_das2021, athulya2022}. However, interpreting these class transitions with viscous timescales would require unrealistically low viscosity in a Keplerian disc \citep{belloni1997b}.
\citet{chakrabarti_nandi2005}, using \textit{IXAE} and \textit{RXTE} data, showed that such rapid transitions cannot occur on viscous timescales; instead, a fraction of the accretion flow must be nearly free-falling, i.e., sub-Keplerian in nature. Furthermore, \citet{athulya2022} proposed that these transitions could occur on thermal–viscous timescales, driven by variations in the sub-Keplerian flow caused by radiation-pressure–induced instabilities in the inner disc near the compact object, thereby triggering the rapid variability class transitions.

\subsection{Sub-Classification of $\chi$ class: Dynamical Hard state}
In \textit{AstroSat} observations, GRS 1915+105 is predominantly found in the $\chi$ class (23 of 42 observations), exhibiting characteristics consistent with earlier \textit{RXTE} results \citep{belloni2000,rao2000,huppenkothen_2017_GRS_ML}.
We further sub-classify these observations into $\chi_1$, $\chi_2$, $\chi_3$, and $\chi_4$ based on count rate, hardness, power spectra and time-lag spectra while \citet{belloni2000} classified them solely on the basis of hardness.
As shown in the upper-right panel of Fig. \ref{fig2:lc_ccd}, the different $\chi$ sub-classes display distinct variations in count rate and hardness (see inset of the same panel). The hardness is lowest for $\chi_1$ and $\chi_3$, and highest for $\chi_4$, while the average count rate is highest in $\chi_3$ and lowest in $\chi_4$.

Distinct features are also evident in the PDS and energy spectra across the $\chi$ sub-classes. The $\chi_3$ class shows a strong QPO frequency without harmonics (Fig.~\ref{fig3:pds}, panels b and d), while $\chi_1$ exhibits a weak harmonic without any sub-harmonic during the SPP (panel c), but shows both harmonic and sub-harmonic during the DP. Similarly, the $\chi_2$ class displays QPOs with both harmonic and sub-harmonic features (panel g). However, the PDS of $\chi_2'$ shows a subtle difference, where the sub-harmonic and harmonic humps seen in the $\chi_2$, are weaker in the PDS of $\chi_2'$ (see Fig. \ref{fig3:pds}, panels g and h). In contrast, $\chi_4$ consistently exhibits a sharp low-frequency QPO accompanied by a strong harmonic (panel j). These distinctions highlight class-dependent signatures of QPOs, harmonics, and sub-harmonics in the frequency–power domain.
Time-lag analysis clearly shows that $\chi_4$ class observations exhibit the minimum soft-lag, $\chi_3$ class displays the maximum soft-lag, while intermediate lag values are observed for the $\chi_1$ and $\chi_2$ classes. 
These differences in soft-lag correspond to variations in spectral properties, where the Comptonized flux ($F_{\rm Comp}$) is lowest in $\chi_3$ (average $\sim$33\%), moderate in $\chi_1$ and $\chi_2$ (average $\sim$47\%), and highest in $\chi_4$ (average $\sim$60\%).

The $\chi$ sub-classes also exhibit distinct radio properties, as evident from \textit{RXTE} observations. \citet{vadawale2003} modeled $\chi_{\rm RQ}$ (radio-quiet; $\chi_2$, $\chi_4$) spectra using a multi-colour disc along with Comptonized component, whereas $\chi_{\rm RL}$ (radio-loud; $\chi_1$, $\chi_3$) spectra required an additional power-law component. However in our analysis, the spectra of all $\chi$ sub-classes are modelled similarly, without incorporating the additional power-law component.
\citet{yadav2016} classified $\chi$ class observations during the Short Plateau Phase (SPP) as radio-quiet, whereas our results suggest them to be radio-loud, however, simultaneous radio data are unavailable, making definitive classification uncertain on the basis of radio emission.
Finally, a connection between time-lag and radio flux has been reported by \cite{pahari_2013} using \textit{RXTE} data, where $\chi_{\rm RL}$ (plateau states) show hard-lags and $\chi_{\rm RQ}$ exhibit soft-lags. In contrast, our \textit{AstroSat} observations show soft-lags across all $\chi$ sub-classes. The cause of this variation remains unclear, likely due to the lack of simultaneous radio coverage.

\subsection{Possible Accretion Disc Configuration}

The systematic variation between spectral and timing properties across variability classes and transition phases provides strong evidence for a coherent underlying accretion disc dynamics that governs the behaviour of this persistently variable source during the LFQPOs observed with \textit{AstroSat}. We find that class transitions consistently occur within a few hours, invariably appearing as a significant increase in the X-ray photon count rate accompanied by a rise in QPO frequency. This rapid variation suggests that changes in the accretion rate cannot occur on viscous timescales; instead, a substantial fraction of the accreting matter must be nearly less viscous in nature and follow free-fall timescales, i.e., sub-Keplerian in nature. Such fast transitions further support the presence of a two-component advective flow, a scenario already established for this source \citep{chakrabarti_titarchuk1995,chakrabarti_nandi2000,nandi_2001,chakrabarti2002,dutta2018}.
Within the TCAF paradigm \citep{chakrabarti_titarchuk1995}, each QPO frequency corresponds to a specific location of the shock front, defining the size of the oscillating Comptonizing region i.e., corona. Variations in QPO frequency during class transitions therefore indicate gradual geometrical variations in the corona, may driven by fluctuations in the sub-Keplerian accretion rate. These rapid variations likely govern the observed spectro-temporal evolution and trigger the variability class transitions \citep{chakrabarti_nandi2000, nandi2001, chakrabarti2002, chakrabarti_nandi2004,chakrabarti_nandi2005,chakrabarti2009, dutta_2010}. A similar variation in QPO frequency observed in GRS 1915+105 during the \textit{RXTE} observations has been attributed to geometric changes in the Comptonizing region or corona \citep{zhang_2020_grs, karpouzas_2021,garcia2022,mendez_nat2022,belloni2024}.

The variations in spectro-temporal parameters during variability classes exhibiting LFQPOs confirm that $\text{rms}_{\rm QPO}$ increases sharply with $F_{\rm Comp}$ up to $\sim$12\%, except for four $\chi_4$ class observations when the source flux drops dramatically to $\sim 4.38\%$ of $L_{\rm Edd}$. This behaviour provides clear evidence supporting the predictions of the TCAF paradigm, where oscillations of the corona enhance the modulation of Comptonized photons, thereby increasing the rms power ($\text{rms}_{\rm QPO}$) in the frequency domain \citep{chakrabarti_manickam2000}. Furthermore, the highest $F_{\rm Comp}$ values observed in the $\chi_4$ class, despite the lowest source flux during the hardest spectral state (lowest value of $\Gamma$), indicate an enhanced Comptonization rate, consistent with the expected geometry of the TCAF model.

The lower panel of Fig.~\ref{fig8:rms_lag_v_comp-flux} shows that as $F_{\rm Comp}$ increases, magnitude of the soft-lag decreases while the photon index $\Gamma$ also decreases, indicating a transition toward a harder spectral state.
This establishes a strong spectro-temporal interdependence, with increasing $F_{\rm Comp}$, the QPO fractional rms ($\text{rms}_{\rm QPO}$) increases and the magnitude of the soft-lag decreases, while both $\nu_{\rm QPO}$ and $\Gamma$ decrease. 
These correlations suggest that as the source hardens, the corona expands (leading to a reduction in $\nu_{\rm QPO}$), Comptonization becomes more dominant, and the magnitude of the soft-lag decreases, although net time-lag remains negative. This behaviour suggests that the observed time-lag results from the combined effects of multiple physical processes, which include Comptonization, reflection, outflows/jets, and the geometry of the accretion disc \citep{cui1999, nowak1999, poutanen_fabian1999, kara2013, dutta2016, arka-chatterjee2017b, arka-chatterjee2020, nandi_prantik2021, anuj_nandi2024}. Among these, Comptonization consistently contributes a positive (hard) component to the time-lag \citep{miyamoto1988, dutta2016}, whereas soft-lags are likely produced by the reflection of hard photons from the disc surface or by delays introduced through coronal outflows \citep{poutanen_fabian1999, arka-chatterjee2020}.

Moreover, GRS 1915+105, a persistently variable and high-inclination source, has been shown to exhibit a transition from hard to soft-lags near $\sim$2 Hz during plateau-state observations \citep{dutta2018}, consistent with extensive \textit{RXTE} studies of type-C QPOs \citep{zhang_2020_grs}. At lower QPO frequencies ($<2$ Hz), the source displays hard lags \citep{reig2000, qu2010, pahari_2013, dutta2018, zhang_2020_grs}. Interestingly, this characteristic lag sign reversal at $\sim$2 Hz, commonly observed with \textit{RXTE}, is absent in the \textit{AstroSat} observations.
\citet{dutta2016} explained that time lags can be positive across all energies in GRS 1915+105 as well as in XTE J1550–564 when the Comptonization efficiency exceeds $\sim85\%$ in both sources. The lag switching at a specific frequency, recurring nearly 450 days later at a similar coronal size, suggests that coronal geometry plays a key role, with the net time lag determined by the relative contributions of the underlying physical processes.
\citet{nobili2000} considered a two-component thermal Comptonization scenario, where a hotter, optically thick inner corona up-scatters soft disc photons to higher energies, which are then down-scattered in a cooler, optically thin outer corona, producing soft-lags at high QPO frequencies ($>2$ Hz). At lower QPO frequencies ($<2$ Hz), the inner corona is assumed optically thin, producing only hard-lags.  
In this scenario, the inner disc radius is positively correlated with the coronal size, whereas \cite{karpouzas_2021} reported an anti-correlation down to $\sim$2 Hz, followed by a positive correlation at lower QPO frequencies. 
In GRS 1915+105, \cite{mendez_nat2022} explained that at QPO frequencies of 2$-$6 Hz, a hot and extended corona covering the inner disc enhances photon feedback onto the disc, producing soft lags interpreted as delayed disc responses to returning Comptonized photons. As the QPO frequency decreases, both lag amplitude and coronal size decrease while the disc recedes, near $\sim$2 Hz, the lag approaches zero as the feedback fraction drops. At lower frequencies, the corona becomes vertically extended and may evolve into a jet-like structure, leading to positive lags despite weak feedback. However, these studies did not explain why lag switching is observed predominantly in high-inclination sources, while it is largely absent in low-inclination systems \citep{dutta2016, van_eijnden_2017, arka-chatterjee2020, choudhury_etal2025}. Moreover, the occurrence of a similar lag transition in GRS 1915+105 after $\sim$450 days, at nearly the same QPO frequency, is unexpected given its highly turbulent accretion environment and complex variability, and its physical origin remains unexplained.
Using \textit{AstroSat} observations of MAXI J1535–571, \citet{Garg_grs2022} suggested that low-frequency QPOs ($\nu_{\rm QPO}<2.2$ Hz) originate in the disc and propagate inward, whereas higher-frequency QPOs ($\nu_{\rm QPO}>2.2$ Hz) originate in the corona and propagate outward, resulting in net positive or negative lags depending on the dominant component.

Our spectro-temporal analysis (Fig. ~\ref{fig8:rms_lag_v_comp-flux}) shows that increasing fractional Comptonized flux ($F_{\rm Comp}$) leads to progressively more positive time-lags, indicating that enhanced coronal Comptonization is responsible for hard-lag production. To probe the origin of the soft lags ($1.59-13.49$ ms) observed across all QPO frequencies in the \textit{AstroSat} era without the sign reversal near $\sim2$ Hz seen in \textit{RXTE} data, we estimated the Comptonized flux ($1-100$ keV) using the model \texttt{TBabs$\times$(thcomp$\otimes$diskbb)}. For $\nu_{\rm QPO}<2.2$ Hz, the Comptonized flux is significantly lower in \textit{AstroSat} observations (ranges from $0.50-0.74 \times 10^{-8}$ erg.cm$^{-2}$.s$^{-1}$) than in \textit{RXTE} (range from $1.56-3.08 \times 10^{-8}$ erg.cm$^{-2}$.s$^{-1}$), suggesting that the hard lags seen at $\nu_{\rm QPO}$ during the \textit{RXTE} era were driven by stronger Comptonization. Although both datasets show a similar correlation between time-lag and $\nu_{\rm QPO}$ (see Fig.~\ref{fig4:rms_lag_v_qpo}), the absence of a lag sign reversal in the \textit{AstroSat} observations implies that comparable coronal geometries can yield different Comptonization strengths, resulting in either soft or hard lags. This behavior likely reflects variations in the coronal optical depth, particularly when additional processes such as enhanced reflection \citep{ross_fabian_refl_2007} or interactions with returning outflows \citep{patra2019, arka-chatterjee2020, mendez_nat2022} are weak or absent. Our previous study \citep{prajjwal_2024} demonstrated that the optical depth of the Comptonizing medium plays a key role in producing hard or soft lags. Consistent with this, the soft lags observed at LFQPO frequencies in the present analysis, particularly during phases of lower Comptonized flux, are likely attributable to a higher coronal optical depth.

\section{Conclusions}
\label{sec6:conclusion}

Understanding and interpreting the mechanisms behind the observed time-lags is challenging, as they arise from multiple non-linear and non-localized physical processes. In this work, we have correlated the spectral and temporal variability properties to investigate the time-lags associated with LFQPOs in persistently variable GRS 1915+105, using \textit{AstroSat} observations from March 4, 2016 to March 22, 2019. A summary of our key findings is as follows,

\begin{enumerate}
    \item [(i)] LFQPOs are observed across four variability classes during \textit{AstroSat} observations: $\theta$, $\beta$, $\rho$ and $\chi$ where, $\chi$ class observations are further classified into $\chi_1$, $\chi_2$, $\chi_3$, and $\chi_4$ based on their distinct spectro-temporal properties.
    
    \item[(ii)] Class transitions occur within a few hours, either as a significant increase in X-ray photon count rate accompanied by a rise in QPO frequency, or vice versa, suggesting that fluctuations in the sub-Keplerian accretion rate likely drive the observed spectro-temporal evolution and trigger the variability class transitions.

    \item[(iii)] The $\text{rms}_\text{QPO}$ initially increases with QPO frequency up to $\sim 3.4$ Hz and then decreases at higher frequencies. A similar trend was seen in \textit{RXTE} observations, though the maximum $\text{rms}_\text{QPO}$ occurred at QPO frequency $\sim 2$ Hz.

    \item[(iv)] The consistent increase of $\text{rms}_{\rm QPO}$ with $F_{\rm Comp}$ provides clear evidence that oscillations of the corona enhance the modulation of Comptonized photons, thereby increasing the rms power.

    \item[(v)] Increasing $F_{\rm Comp}$ drives to higher $\text{rms}_{\rm QPO}$ and positive time-lag, while $\nu_{\rm QPO}$ and $\Gamma$ decrease, indicating that the time-lag arises from the combined effects of multiple mechanisms, i.e., Comptonization, reflection, outflows, and disc geometry.
    
  \item[(vi)] The time-lag analysis reveals soft-lags (1.59–13.49 ms) across all QPO frequencies (1.38–7.38 Hz), with the soft-lag magnitude decreasing as $\nu_{\rm QPO}$ increases, and without the sign reversal near $\sim$2 Hz, observed in the \textit{RXTE} data.
  The soft-lags observed at low LFQPO frequencies ($\nu_{\rm QPO}<2.2$ Hz) during phases of reduced Comptonized flux, compared to the \textit{RXTE} era, are therefore likely attributable to a higher coronal optical depth.
\end{enumerate}

\section*{Acknowledgements}

We thank the anonymous reviewer for valuable suggestions and comments that helped to improve the quality of this manuscript. BGD, PM and AN acknowledge the support from ISRO sponsored project (DS-2B-1313(2)/6/2020-Sec.2). PM acknowledges the visit to Inter-University Centre for Astronomy and Astrophysics (IUCAA) where a partial amount of this work is carried out. PM also acknowledges the a short visit to Indian Institute of Astrophysics (IIA). BGD acknowledges `TARE’ scheme (Ref. No. TAR/2020/000141) under SERB, DST, Govt. of India and also acknowledges IUCAA for the Visiting Associate-ship Programme. AN thanks GH, SAG; DD, PDMSA, and Director, URSC for encouragement and continuous support to carry out this research. This work uses the data of \textit{AstroSat} mission of ISRO which is archived at the Indian Space Science Data Centre (ISSDC). We have used the data of Soft X-ray Telescope (SXT) which is calibrated and verified by \textit{AstroSat-SXT} team. We thank the SXT-POC team at TIFR for providing the necessary software tool to analyse \textit{SXT} data. This work has also used \textit{LAXPC} data which is verified by LAXPC-POC at TIFR. We thank \textit{AstroSat} Science Support Cell for providing the software \texttt{LAXPCsoftware} for the analysis of \textit{LAXPC} data.

\section*{Data Availability}
Observational data of \textit{RXTE} used for this work are available at the HEASARC website  
\url{https://heasarc.gsfc.nasa.gov/cgi-bin/W3Browse/w3browse.pl}.
\textit{AstroSat} archival data used for this work is available at the Astrobrowse (AstroSat archive) website  
\url{https://webapps.issdc.gov.in/astro_archive/archive} of the Indian Space Science Data Centre (ISSDC). 
The \textit{MAXI/GSC} data is available in the website 
\url{http://maxi.riken.jp/top/lc.html}. The \textit{BAT} data is available in the website \url{https://swift.gsfc.nasa.gov/results/transients/}.



\bibliographystyle{mnras}
\bibliography{reference} 

@ARTICLE{reid2014,
       author = {{Reid}, M.~J. and {McClintock}, J.~E. and {Steiner}, J.~F. and {Steeghs}, D. and {Remillard}, R.~A. and {Dhawan}, V. and {Narayan}, R.},
        title = "{A Parallax Distance to the Microquasar GRS 1915+105 and a Revised Estimate of its Black Hole Mass}",
      journal = {\apj},
         year = 2014,
        month = nov,
       volume = {796},
       number = {1},
          eid = {2},
        pages = {2},
          doi = {10.1088/0004-637X/796/1/2},
archivePrefix = {arXiv},
       eprint = {1409.2453},
 primaryClass = {astro-ph.GA},
       adsurl = {https://ui.adsabs.harvard.edu/abs/2014ApJ...796....2R}
}

@ARTICLE{seshadri2022,
       author = {{Majumder}, Seshadri and {Sreehari}, H. and {Aftab}, Nafisa and {Katoch}, Tilak and {Das}, Santabrata and {Nandi}, Anuj},
        title = "{Wide-band view of high-frequency quasi-periodic oscillations of GRS 1915+105 in 'softer' variability classes observed with AstroSat}",
      journal = {\mnras},
         year = 2022,
        month = may,
       volume = {512},
       number = {2},
        pages = {2508-2524},
          doi = {10.1093/mnras/stac615},
archivePrefix = {arXiv},
       eprint = {2203.02710},
 primaryClass = {astro-ph.HE},
       adsurl = {https://ui.adsabs.harvard.edu/abs/2022MNRAS.512.2508M}
}

@ARTICLE{sreehari2020,
       author = {{Sreehari}, H. and {Nandi}, Anuj and {Das}, Santabrata and {Agrawal}, V.~K. and {Mandal}, Samir and {Ramadevi}, M.~C. and {Katoch}, Tilak},
        title = "{AstroSat view of GRS 1915+105 during the soft state: detection of HFQPOs and estimation of mass and spin}",
      journal = {\mnras},
         year = 2020,
        month = dec,
       volume = {499},
       number = {4},
        pages = {5891-5901},
          doi = {10.1093/mnras/staa3135},
archivePrefix = {arXiv},
       eprint = {2010.03782},
 primaryClass = {astro-ph.HE},
       adsurl = {https://ui.adsabs.harvard.edu/abs/2020MNRAS.499.5891S}
}

@ARTICLE{belloni2000,
       author = {{Belloni}, T. and {Klein-Wolt}, M. and {M{\'e}ndez}, M. and {van der Klis}, M. and {van Paradijs}, J.},
        title = "{A model-independent analysis of the variability of GRS 1915+105}",
      journal = {\aap},
         year = 2000,
        month = mar,
       volume = {355},
        pages = {271-290},
archivePrefix = {arXiv},
       eprint = {astro-ph/0001103},
 primaryClass = {astro-ph},
       adsurl = {https://ui.adsabs.harvard.edu/abs/2000A&A...355..271B}
}

@ARTICLE{remillard_mcClintock2006,
       author = {{Remillard}, Ronald A. and {McClintock}, Jeffrey E.},
        title = "{X-Ray Properties of Black-Hole Binaries}",
      journal = {\araa},
         year = 2006,
        month = sep,
       volume = {44},
       number = {1},
        pages = {49-92},
          doi = {10.1146/annurev.astro.44.051905.092532},
archivePrefix = {arXiv},
       eprint = {astro-ph/0606352},
 primaryClass = {astro-ph},
       adsurl = {https://ui.adsabs.harvard.edu/abs/2006ARA&A..44...49R}
}

@ARTICLE{stella_vietri1998,
       author = {{Stella}, Luigi and {Vietri}, Mario},
        title = "{Lense-Thirring Precession and Quasi-periodic Oscillations in Low-Mass X-Ray Binaries}",
      journal = {\apjl},
         year = 1998,
        month = jan,
       volume = {492},
       number = {1},
        pages = {L59-L62},
          doi = {10.1086/311075},
archivePrefix = {arXiv},
       eprint = {astro-ph/9709085},
 primaryClass = {astro-ph},
       adsurl = {https://ui.adsabs.harvard.edu/abs/1998ApJ...492L..59S}
}

@ARTICLE{nandi2001,
       author = {{Nandi}, Anuj and {Manickam}, Sivakumar G. and {Rao}, A.~R. and {Chakrabarti}, Sandip K.},
        title = "{On the source of quasi-periodic oscillations of the black hole candidate GRS 1915+105: some new observations and their interpretation}",
      journal = {\mnras},
         year = 2001,
        month = jun,
       volume = {324},
       number = {1},
        pages = {267-272},
          doi = {10.1046/j.1365-8711.2001.04339.x},
archivePrefix = {arXiv},
       eprint = {astro-ph/0012527},
 primaryClass = {astro-ph},
       adsurl = {https://ui.adsabs.harvard.edu/abs/2001MNRAS.324..267N}
}

@ARTICLE{yadav2016,
       author = {{Yadav}, J.~S. and {Misra}, Ranjeev and {Verdhan Chauhan}, Jai and {Agrawal}, P.~C. and {Antia}, H.~M. and {Pahari}, Mayukh and {Dedhia}, Dhiraj and {Katoch}, Tilak and {Madhwani}, P. and {Manchanda}, R.~K. and {Paul}, B. and {Shah}, Parag and {Ishwara-Chandra}, C.~H.},
        title = "{Astrosat/LAXPC Reveals the High-energy Variability of GRS 1915+105 in the X Class}",
      journal = {\apj},
         year = 2016,
        month = dec,
       volume = {833},
       number = {1},
          eid = {27},
        pages = {27},
          doi = {10.3847/0004-637X/833/1/27},
archivePrefix = {arXiv},
       eprint = {1608.07023},
 primaryClass = {astro-ph.HE},
       adsurl = {https://ui.adsabs.harvard.edu/abs/2016ApJ...833...27Y}
}

@ARTICLE{cui1999,
       author = {{Cui}, Wei},
        title = "{Phase Lags of Quasi-periodic Oscillations in Microquasar GRS 1915+105}",
      journal = {\apjl},
         year = 1999,
        month = oct,
       volume = {524},
       number = {1},
        pages = {L59-L62},
          doi = {10.1086/312296},
       adsurl = {https://ui.adsabs.harvard.edu/abs/1999ApJ...524L..59C}
}

@ARTICLE{belloni2019,
       author = {{Belloni}, Tomaso M. and {Bhattacharya}, Dipankar and {Caccese}, Pietro and {Bhalerao}, Varun and {Vadawale}, Santosh and {Yadav}, J.~S.},
        title = "{A variable-frequency HFQPO in GRS 1915+105 as observed with AstroSat}",
      journal = {\mnras},
         year = 2019,
        month = oct,
       volume = {489},
       number = {1},
        pages = {1037-1043},
          doi = {10.1093/mnras/stz2143},
archivePrefix = {arXiv},
       eprint = {1908.00437},
 primaryClass = {astro-ph.HE},
       adsurl = {https://ui.adsabs.harvard.edu/abs/2019MNRAS.489.1037B}
}

@ARTICLE{antia2017,
       author = {{Antia}, H.~M. and {Yadav}, J.~S. and {Agrawal}, P.~C. and {Verdhan Chauhan}, Jai and {Manchanda}, R.~K. and {Chitnis}, Varsha and {Paul}, Biswajit and {Dedhia}, Dhiraj and {Shah}, Parag and {Gujar}, V.~M. and {Katoch}, Tilak and {Kurhade}, V.~N. and {Madhwani}, Pankaj and {Manojkumar}, T.~K. and {Nikam}, V.~A. and {Pandya}, A.~S. and {Parmar}, J.~V. and {Pawar}, D.~M. and {Pahari}, Mayukh and {Misra}, Ranjeev and {Navalgund}, K.~H. and {Pandiyan}, R. and {Sharma}, K.~S. and {Subbarao}, K.},
        title = "{Calibration of the Large Area X-Ray Proportional Counter (LAXPC) Instrument on board AstroSat}",
      journal = {\apjs},
         year = 2017,
        month = jul,
       volume = {231},
       number = {1},
          eid = {10},
        pages = {10},
          doi = {10.3847/1538-4365/aa7a0e},
archivePrefix = {arXiv},
       eprint = {1702.08624},
 primaryClass = {astro-ph.IM},
       adsurl = {https://ui.adsabs.harvard.edu/abs/2017ApJS..231...10A}
}

@ARTICLE{yadav2017,
       author = {{Yadav}, J.~S. and {Agrawal}, P.~C. and {Antia}, H.~M. and {Manchanda}, R.~K. and {Paul}, B. and {Misra}, Ranjeev},
        title = "{Large Area X-ray Proportional Counter instrument on AstroSat}",
      journal = {Current Science},
         year = 2017,
        month = aug,
       volume = {113},
       number = {4},
        pages = {591},
          doi = {10.18520/cs/v113/i04/591-594},
archivePrefix = {arXiv},
       eprint = {1705.06440},
 primaryClass = {astro-ph.IM},
       adsurl = {https://ui.adsabs.harvard.edu/abs/2017CSci..113..591Y}
}

@ARTICLE{agrawal2018,
       author = {{Agrawal}, V.~K. and {Nandi}, Anuj and {Girish}, V. and {Ramadevi}, M.~C.},
        title = "{Spectral and timing properties of atoll source 4U 1705-44: LAXPC/AstroSat results}",
      journal = {\mnras},
         year = 2018,
        month = jul,
       volume = {477},
       number = {4},
        pages = {5437-5446},
          doi = {10.1093/mnras/sty1005},
archivePrefix = {arXiv},
       eprint = {1804.08371},
 primaryClass = {astro-ph.HE},
       adsurl = {https://ui.adsabs.harvard.edu/abs/2018MNRAS.477.5437A}
}

@ARTICLE{sreehari2019,
       author = {{Sreehari}, H. and {Ravishankar}, B.~T. and {Iyer}, Nirmal and {Agrawal}, V.~K. and {Katoch}, Tilak B. and {Mandal}, Samir and {Nandi}, Anuj},
        title = "{AstroSat view of MAXI J1535-571: broad-band spectro-temporal features}",
      journal = {\mnras},
         year = 2019,
        month = jul,
       volume = {487},
       number = {1},
        pages = {928-941},
          doi = {10.1093/mnras/stz1327},
archivePrefix = {arXiv},
       eprint = {1905.04656},
 primaryClass = {astro-ph.HE},
       adsurl = {https://ui.adsabs.harvard.edu/abs/2019MNRAS.487..928S}
}

@ARTICLE{reig2000,
       author = {{Reig}, P. and {Belloni}, T. and {van der Klis}, M. and {M{\'e}ndez}, M. and {Kylafis}, N.~D. and {Ford}, E.~C.},
        title = "{Phase Lag Variability Associated with the 0.5-10 HZ Quasi-Periodic Oscillations in GRS 1915+105}",
      journal = {\apj},
         year = 2000,
        month = oct,
       volume = {541},
       number = {2},
        pages = {883-888},
          doi = {10.1086/309469},
       adsurl = {https://ui.adsabs.harvard.edu/abs/2000ApJ...541..883R}
}

@ARTICLE{vaughan_nowak1997,
       author = {{Vaughan}, Brian A. and {Nowak}, Michael A.},
        title = "{X-Ray Variability Coherence: How to Compute It, What It Means, and How It Constrains Models of GX 339-4 and Cygnus X-1}",
      journal = {\apjl},
         year = 1997,
        month = jan,
       volume = {474},
       number = {1},
        pages = {L43-L46},
          doi = {10.1086/310430},
archivePrefix = {arXiv},
       eprint = {astro-ph/9610257},
 primaryClass = {astro-ph},
       adsurl = {https://ui.adsabs.harvard.edu/abs/1997ApJ...474L..43V}
}

@ARTICLE{antia2021,
       author = {{Antia}, H.~M. and {Agrawal}, P.~C. and {Dedhia}, Dhiraj and {Katoch}, Tilak and {Manchanda}, R.~K. and {Misra}, Ranjeev and {Mukerjee}, Kallol and {Pahari}, Mayukh and {Roy}, Jayashree and {Shah}, P. and {Yadav}, J.~S.},
        title = "{Large Area X-ray Proportional Counter (LAXPC) in orbit performance: Calibration, background, analysis software}",
      journal = {Journal of Astrophysics and Astronomy},
         year = 2021,
        month = oct,
       volume = {42},
       number = {2},
          eid = {32},
        pages = {32},
          doi = {10.1007/s12036-021-09712-8},
archivePrefix = {arXiv},
       eprint = {2101.07514},
 primaryClass = {astro-ph.IM},
       adsurl = {https://ui.adsabs.harvard.edu/abs/2021JApA...42...32A}
}

@ARTICLE{muno1999,
       author = {{Muno}, Michael P. and {Morgan}, Edward H. and {Remillard}, Ronald A.},
        title = "{Quasi-periodic Oscillations and Spectral States in GRS 1915+105}",
      journal = {\apj},
         year = 1999,
        month = dec,
       volume = {527},
       number = {1},
        pages = {321-340},
          doi = {10.1086/308063},
       adsurl = {https://ui.adsabs.harvard.edu/abs/1999ApJ...527..321M}
}

@ARTICLE{athulya2022,
       author = {{Athulya}, M.~P. and {Radhika}, D. and {Agrawal}, V.~K. and {Ravishankar}, B.~T. and {Naik}, Sachindra and {Mandal}, Samir and {Nandi}, Anuj},
        title = "{Unravelling the foretime of GRS 1915+105 using AstroSat observations: Wide-band spectral and temporal characteristics}",
      journal = {\mnras},
         year = 2022,
        month = feb,
       volume = {510},
       number = {2},
        pages = {3019-3038},
          doi = {10.1093/mnras/stab3614},
archivePrefix = {arXiv},
       eprint = {2110.14467},
 primaryClass = {astro-ph.HE},
       adsurl = {https://ui.adsabs.harvard.edu/abs/2022MNRAS.510.3019A}
}

@ARTICLE{agrawal2006,
       author = {{Agrawal}, P.~C.},
        title = "{A broad spectral band Indian Astronomy satellite {\textquoteleft}Astrosat{\textquoteright}}",
      journal = {Advances in Space Research},
         year = 2006,
        month = jan,
       volume = {38},
       number = {12},
        pages = {2989-2994},
          doi = {10.1016/j.asr.2006.03.038},
       adsurl = {https://ui.adsabs.harvard.edu/abs/2006AdSpR..38.2989A}
}

@ARTICLE{singh2017,
       author = {{Singh}, K.~P. and {Stewart}, G.~C. and {Westergaard}, N.~J. and {Bhattacharayya}, S. and {Chandra}, S. and {Chitnis}, V.~R. and {Dewangan}, G.~C. and {Kothare}, A.~T. and {Mirza}, I.~M. and {Mukerjee}, K. and {Navalkar}, V. and {Shah}, H. and {Abbey}, A.~F. and {Beardmore}, A.~P. and {Kotak}, S. and {Kamble}, N. and {Vishwakarama}, S. and {Pathare}, D.~P. and {Risbud}, V.~M. and {Koyande}, J.~P. and {Stevenson}, T. and {Bicknell}, C. and {Crawford}, T. and {Hansford}, G. and {Peters}, G. and {Sykes}, J. and {Agarwal}, P. and {Sebastian}, M. and {Rajarajan}, A. and {Nagesh}, G. and {Narendra}, S. and {Ramesh}, M. and {Rai}, R. and {Navalgund}, K.~H. and {Sarma}, K.~S. and {Pandiyan}, R. and {Subbarao}, K. and {Gupta}, T. and {Thakkar}, N. and {Singh}, A.~K. and {Bajpai}, A.},
        title = "{Soft X-ray Focusing Telescope Aboard AstroSat: Design, Characteristics and Performance}",
      journal = {Journal of Astrophysics and Astronomy},
         year = 2017,
        month = jun,
       volume = {38},
       number = {2},
          eid = {29},
        pages = {29},
          doi = {10.1007/s12036-017-9448-7},
       adsurl = {https://ui.adsabs.harvard.edu/abs/2017JApA...38...29S}
}

@ARTICLE{van_der_klis1987,
       author = {{van der Klis}, M. and {Hasinger}, G. and {Stella}, L. and {Langmeier}, A. and {van Paradijs}, J. and {Lewin}, W.~H.~G.},
        title = "{The Complex Cross-Spectra of Cygnus X-2 and GX 5-1}",
      journal = {\apjl},
         year = 1987,
        month = aug,
       volume = {319},
        pages = {L13},
          doi = {10.1086/184946},
       adsurl = {https://ui.adsabs.harvard.edu/abs/1987ApJ...319L..13V}
}

@ARTICLE{wilms2000,
       author = {{Wilms}, J. and {Allen}, A. and {McCray}, R.},
        title = "{On the Absorption of X-Rays in the Interstellar Medium}",
      journal = {\apj},
         year = 2000,
        month = oct,
       volume = {542},
       number = {2},
        pages = {914-924},
          doi = {10.1086/317016},
archivePrefix = {arXiv},
       eprint = {astro-ph/0008425},
 primaryClass = {astro-ph},
       adsurl = {https://ui.adsabs.harvard.edu/abs/2000ApJ...542..914W}
}

@BOOK{frank2002,
       author = {{Frank}, Juhan and {King}, Andrew and {Raine}, Derek J.},
        title = "{Accretion Power in Astrophysics: Third Edition}",
         year = 2002,
       adsurl = {https://ui.adsabs.harvard.edu/abs/2002apa..book.....F}
}

@ARTICLE{zdziarski1996,
       author = {{Zdziarski}, A.~A. and {Johnson}, W.~N. and {Magdziarz}, P.},
        title = "{Broad-band {\ensuremath{\gamma}}-ray and X-ray spectra of NGC 4151 and their implications for physical processes and geometry.}",
      journal = {\mnras},
         year = 1996,
        month = nov,
       volume = {283},
       number = {1},
        pages = {193-206},
          doi = {10.1093/mnras/283.1.193},
archivePrefix = {arXiv},
       eprint = {astro-ph/9607015},
 primaryClass = {astro-ph},
       adsurl = {https://ui.adsabs.harvard.edu/abs/1996MNRAS.283..193Z}
}

@ARTICLE{miyamoto1988,
       author = {{Miyamoto}, Sigenori and {Kitamoto}, Shunji and {Mitsuda}, Kazuhisa and {Dotani}, Tadayasu},
        title = "{Delayed hard X-rays from Cygnus X-l}",
      journal = {\nat},
         year = 1988,
        month = dec,
       volume = {336},
       number = {6198},
        pages = {450-452},
          doi = {10.1038/336450a0},
       adsurl = {https://ui.adsabs.harvard.edu/abs/1988Natur.336..450M}
}

@ARTICLE{nowak1999,
       author = {{Nowak}, Michael A. and {Wilms}, J{\"o}rn and {Dove}, James B.},
        title = "{Low-Luminosity States of the Black Hole Candidate GX 339-4. II. Timing Analysis}",
      journal = {\apj},
         year = 1999,
        month = may,
       volume = {517},
       number = {1},
        pages = {355-366},
          doi = {10.1086/307189},
archivePrefix = {arXiv},
       eprint = {astro-ph/9812180},
 primaryClass = {astro-ph},
       adsurl = {https://ui.adsabs.harvard.edu/abs/1999ApJ...517..355N}
}

@ARTICLE{qu2010,
       author = {{Qu}, J.~L. and {Lu}, F.~J. and {Lu}, Y. and {Song}, L.~M. and {Zhang}, S. and {Ding}, G.~Q. and {Wang}, J.~M.},
        title = "{The Energy Dependence of the Centroid Frequency and Phase Lag of the Quasi-periodic Oscillations in GRS 1915+105}",
      journal = {\apj},
         year = 2010,
        month = feb,
       volume = {710},
       number = {1},
        pages = {836-842},
          doi = {10.1088/0004-637X/710/1/836},
archivePrefix = {arXiv},
       eprint = {0912.4769},
 primaryClass = {astro-ph.HE},
       adsurl = {https://ui.adsabs.harvard.edu/abs/2010ApJ...710..836Q}
}

@ARTICLE{dutta2016,
       author = {{Dutta}, Broja G. and {Chakrabarti}, Sandip K.},
        title = "{Temporal Variability from the Two-Component Advective Flow Solution and Its Observational Evidence}",
      journal = {\apj},
         year = 2016,
        month = sep,
       volume = {828},
       number = {2},
          eid = {101},
        pages = {101},
          doi = {10.3847/0004-637X/828/2/101},
archivePrefix = {arXiv},
       eprint = {1609.07867},
 primaryClass = {astro-ph.HE},
       adsurl = {https://ui.adsabs.harvard.edu/abs/2016ApJ...828..101D}
}

@ARTICLE{dutta2018,
       author = {{Dutta}, Broja G. and {Pal}, Partha Sarathi and {Chakrabarti}, Sandip K.},
        title = "{Evolution of accretion disc geometry of GRS 1915+105 during its {\ensuremath{\chi}} state as revealed by TCAF solution}",
      journal = {\mnras},
         year = 2018,
        month = sep,
       volume = {479},
       number = {2},
        pages = {2183-2192},
          doi = {10.1093/mnras/sty1572},
archivePrefix = {arXiv},
       eprint = {1810.03000},
 primaryClass = {astro-ph.HE},
       adsurl = {https://ui.adsabs.harvard.edu/abs/2018MNRAS.479.2183D}
}

@ARTICLE{patra2019,
       author = {{Patra}, Dusmanta and {Chatterjee}, Arka and {Dutta}, Broja G. and {Chakrabarti}, Sandip K. and {Nandi}, Prantik},
        title = "{Evidence of Outflow-induced Soft Lags of Galactic Black Holes}",
      journal = {\apj},
         year = 2019,
        month = dec,
       volume = {886},
       number = {2},
          eid = {137},
        pages = {137},
          doi = {10.3847/1538-4357/ab4c34},
archivePrefix = {arXiv},
       eprint = {1901.02245},
 primaryClass = {astro-ph.HE},
       adsurl = {https://ui.adsabs.harvard.edu/abs/2019ApJ...886..137P}
}

@ARTICLE{motta2021,
       author = {{Motta}, S.~E. and {Kajava}, J.~J.~E. and {Giustini}, M. and {Williams}, D.~R.~A. and {Del Santo}, M. and {Fender}, R. and {Green}, D.~A. and {Heywood}, I. and {Rhodes}, L. and {Segreto}, A. and {Sivakoff}, G. and {Woudt}, P.~A.},
        title = "{Observations of a radio-bright, X-ray obscured GRS 1915+105}",
      journal = {\mnras},
         year = 2021,
        month = may,
       volume = {503},
       number = {1},
        pages = {152-161},
          doi = {10.1093/mnras/stab511},
archivePrefix = {arXiv},
       eprint = {2101.01187},
 primaryClass = {astro-ph.HE},
       adsurl = {https://ui.adsabs.harvard.edu/abs/2021MNRAS.503..152M}
}

@ARTICLE{chakrabarti_manickam2000,
       author = {{Chakrabarti}, Sandip K. and {Manickam}, Sivakumar G.},
        title = "{Correlation among Quasi-Periodic Oscillation Frequencies and Quiescent-State Duration in Black Hole Candidate GRS 1915+105}",
      journal = {\apjl},
         year = 2000,
        month = mar,
       volume = {531},
       number = {1},
        pages = {L41-L44},
          doi = {10.1086/312512},
archivePrefix = {arXiv},
       eprint = {astro-ph/9910012},
 primaryClass = {astro-ph},
       adsurl = {https://ui.adsabs.harvard.edu/abs/2000ApJ...531L..41C}
}

@ARTICLE{vadawale_rao_chakrabarti2001,
       author = {{Vadawale}, S.~V. and {Rao}, A.~R. and {Chakrabarti}, S.~K.},
        title = "{Spectral differences between the radio-loud and radio-quiet low-hard states of GRS 1915+105: Possible detection of synchrotron radiation in X-rays}",
      journal = {\aap},
         year = 2001,
        month = jun,
       volume = {372},
        pages = {793-802},
          doi = {10.1051/0004-6361:20010574},
archivePrefix = {arXiv},
       eprint = {astro-ph/0104378},
 primaryClass = {astro-ph},
       adsurl = {https://ui.adsabs.harvard.edu/abs/2001A&A...372..793V}
}

@ARTICLE{rao2000,
       author = {{Rao}, A.~R. and {Naik}, S. and {Vadawale}, S.~V. and {Chakrabarti}, S.~K.},
        title = "{X-ray spectral components in the hard state of GRS 1915+105: origin of the 0.5 - 10 Hz QPO}",
      journal = {\aap},
         year = 2000,
        month = aug,
       volume = {360},
        pages = {L25-L29},
          doi = {10.48550/arXiv.astro-ph/0007405},
archivePrefix = {arXiv},
       eprint = {astro-ph/0007405},
 primaryClass = {astro-ph},
       adsurl = {https://ui.adsabs.harvard.edu/abs/2000A&A...360L..25R}
}

@ARTICLE{aktar2018,
       author = {{Aktar}, Ramiz and {Das}, Santabrata and {Nandi}, Anuj and {Sreehari}, H.},
        title = "{Advective accretion flow properties around rotating black holes - application to GRO J1655-40}",
      journal = {Journal of Astrophysics and Astronomy},
         year = 2018,
        month = feb,
       volume = {39},
       number = {1},
          eid = {17},
        pages = {17},
          doi = {10.1007/s12036-017-9507-0},
archivePrefix = {arXiv},
       eprint = {1801.04116},
 primaryClass = {astro-ph.HE},
       adsurl = {https://ui.adsabs.harvard.edu/abs/2018JApA...39...17A}
}

@ARTICLE{belloni2012,
       author = {{Belloni}, T.~M. and {Sanna}, A. and {M{\'e}ndez}, M.},
        title = "{High-frequency quasi-periodic oscillations in black hole binaries}",
      journal = {\mnras},
         year = 2012,
        month = nov,
       volume = {426},
       number = {3},
        pages = {1701-1709},
          doi = {10.1111/j.1365-2966.2012.21634.x},
archivePrefix = {arXiv},
       eprint = {1207.2311},
 primaryClass = {astro-ph.HE},
       adsurl = {https://ui.adsabs.harvard.edu/abs/2012MNRAS.426.1701B}
}

@ARTICLE{arka-chatterjee2017b,
       author = {{Chatterjee}, Arka and {Chakrabarti}, Sandip K. and {Ghosh}, Himadri},
        title = "{Temporal evolution of photon energy emitted from two-component advective flows: origin of time lag}",
      journal = {\mnras},
         year = 2017,
        month = dec,
       volume = {472},
       number = {2},
        pages = {1842-1849},
          doi = {10.1093/mnras/stx1916},
archivePrefix = {arXiv},
       eprint = {1706.07572},
 primaryClass = {astro-ph.HE},
       adsurl = {https://ui.adsabs.harvard.edu/abs/2017MNRAS.472.1842C}
}

@ARTICLE{kara2013,
       author = {{Kara}, E. and {Fabian}, A.~C. and {Cackett}, E.~M. and {Uttley}, P. and {Wilkins}, D.~R. and {Zoghbi}, A.},
        title = "{Discovery of high-frequency iron K lags in Ark 564 and Mrk 335}",
      journal = {\mnras},
         year = 2013,
        month = sep,
       volume = {434},
       number = {2},
        pages = {1129-1137},
          doi = {10.1093/mnras/stt1055},
archivePrefix = {arXiv},
       eprint = {1306.2551},
 primaryClass = {astro-ph.HE},
       adsurl = {https://ui.adsabs.harvard.edu/abs/2013MNRAS.434.1129K}
}

@ARTICLE{poutanen_fabian1999,
       author = {{Poutanen}, Juri and {Fabian}, Andrew C.},
        title = "{Spectral evolution of magnetic flares and time lags in accreting black hole sources}",
      journal = {\mnras},
         year = 1999,
        month = jul,
       volume = {306},
       number = {3},
        pages = {L31-L37},
          doi = {10.1046/j.1365-8711.1999.02735.x},
archivePrefix = {arXiv},
       eprint = {astro-ph/9811306},
 primaryClass = {astro-ph},
       adsurl = {https://ui.adsabs.harvard.edu/abs/1999MNRAS.306L..31P}
}

@ARTICLE{nandi_prantik2021,
       author = {{Nandi}, Prantik and {Chatterjee}, Arka and {Chakrabarti}, Sandip K. and {Dutta}, Broja G.},
        title = "{Long-term X-ray observations of seyfert 1 galaxy ark 120: on the origin of soft-excess}",
      journal = {\mnras},
         year = 2021,
        month = sep,
       volume = {506},
       number = {3},
        pages = {3111-3127},
          doi = {10.1093/mnras/stab1699},
archivePrefix = {arXiv},
       eprint = {2101.08043},
 primaryClass = {astro-ph.HE},
       adsurl = {https://ui.adsabs.harvard.edu/abs/2021MNRAS.506.3111N}
}

@ARTICLE{arka-chatterjee2020,
       author = {{Chatterjee}, Arka and {Dutta}, Broja G. and {Nandi}, Prantik and {Chakrabarti}, Sandip K.},
        title = "{Time-domain variability properties of XTE J1650-500 during its 2001 outburst: evidence of disc-jet connection}",
      journal = {\mnras},
         year = 2020,
        month = oct,
       volume = {497},
       number = {4},
        pages = {4222-4230},
          doi = {10.1093/mnras/staa2263},
archivePrefix = {arXiv},
       eprint = {2006.15519},
 primaryClass = {astro-ph.HE},
       adsurl = {https://ui.adsabs.harvard.edu/abs/2020MNRAS.497.4222C}
}

@ARTICLE{chakrabarti_titarchuk1995,
       author = {{Chakrabarti}, Sandip and {Titarchuk}, Lev G.},
        title = "{Spectral Properties of Accretion Disks around Galactic and Extragalactic Black Holes}",
      journal = {\apj},
         year = 1995,
        month = dec,
       volume = {455},
        pages = {623},
          doi = {10.1086/176610},
archivePrefix = {arXiv},
       eprint = {astro-ph/9510005},
 primaryClass = {astro-ph},
       adsurl = {https://ui.adsabs.harvard.edu/abs/1995ApJ...455..623C}
}

@ARTICLE{belloni1997a,
       author = {{Belloni}, T. and {M{\'e}ndez}, M. and {King}, A.~R. and {van der Klis}, M. and {van Paradijs}, J.},
        title = "{An Unstable Central Disk in the Superluminal Black Hole X-Ray Binary GRS 1915+105}",
      journal = {\apjl},
         year = 1997,
        month = apr,
       volume = {479},
       number = {2},
        pages = {L145-L148},
          doi = {10.1086/310595},
archivePrefix = {arXiv},
       eprint = {astro-ph/9702048},
 primaryClass = {astro-ph},
       adsurl = {https://ui.adsabs.harvard.edu/abs/1997ApJ...479L.145B}
}

@ARTICLE{belloni1997b,
       author = {{Belloni}, T. and {M{\'e}ndez}, M. and {King}, A.~R. and {van der Klis}, M. and {van Paradijs}, J.},
        title = "{A Unified Model for the Spectral Variability in GRS 1915+105}",
      journal = {\apjl},
         year = 1997,
        month = oct,
       volume = {488},
       number = {2},
        pages = {L109-L112},
          doi = {10.1086/310944},
archivePrefix = {arXiv},
       eprint = {astro-ph/9708113},
 primaryClass = {astro-ph},
       adsurl = {https://ui.adsabs.harvard.edu/abs/1997ApJ...488L.109B}
}

@INPROCEEDINGS{van_der_Klis1989_Fourier_technic,
       author = {{van der Klis}, M.},
        title = "{Fourier techniques in X-ray timing}",
    booktitle = {Timing Neutron Stars},
         year = 1989,
       editor = {{{\"O}gelman}, H. and {van den Heuvel}, E.~P.~J.},
       series = {NATO Advanced Study Institute (ASI) Series C},
       volume = {262},
        month = jan,
        pages = {27},
          doi = {10.1007/978-94-009-2273-0_3},
       adsurl = {https://ui.adsabs.harvard.edu/abs/1989ASIC..262...27V}
}

@ARTICLE{vadawale_rao_nandi2001,
       author = {{Vadawale}, S.~V. and {Rao}, A.~R. and {Nandi}, A. and {Chakrabarti}, S.~K.},
        title = "{Observational evidence for mass ejection during soft X-ray dips in GRS 1915+105}",
      journal = {\aap},
         year = 2001,
        month = apr,
       volume = {370},
        pages = {L17-L21},
          doi = {10.1051/0004-6361:20010318},
archivePrefix = {arXiv},
       eprint = {astro-ph/0103062},
 primaryClass = {astro-ph},
       adsurl = {https://ui.adsabs.harvard.edu/abs/2001A&A...370L..17V}
}

@ARTICLE{prajjwal_2024,
       author = {{Majumder}, Prajjwal and {Dutta}, Broja G. and {Nandi}, Anuj},
        title = "{First detection of soft-lag in GRS 1915 + 105 at HFQPO using AstroSat observations}",
      journal = {\mnras},
         year = 2024,
        month = jan,
       volume = {527},
       number = {3},
        pages = {4739-4750},
          doi = {10.1093/mnras/stad3465},
archivePrefix = {arXiv},
       eprint = {2311.04869},
 primaryClass = {astro-ph.HE},
       adsurl = {https://ui.adsabs.harvard.edu/abs/2024MNRAS.527.4739M}
}

@ARTICLE{wang_2024_rms_formula,
       author = {{Wang}, Xin-Lei and {Yan}, Zhen and {Xie}, Fu-Guo and {Wang}, Jun-Feng and {Ma}, Ren-Yi},
        title = "{An atypical low-frequency QPO detected in the hard state of MAXI J1348-630 with $Insight$-HXMT}",
      journal = {arXiv e-prints},
         year = 2024,
        month = jun,
          eid = {arXiv:2406.12477},
        pages = {arXiv:2406.12477},
          doi = {10.48550/arXiv.2406.12477},
archivePrefix = {arXiv},
       eprint = {2406.12477},
 primaryClass = {astro-ph.HE},
       adsurl = {https://ui.adsabs.harvard.edu/abs/2024arXiv240612477W}
}

@INPROCEEDINGS{k.arnaud_1996_xspec,
       author = {{Arnaud}, K.~A.},
        title = "{XSPEC: The First Ten Years}",
    booktitle = {Astronomical Data Analysis Software and Systems V},
         year = 1996,
       editor = {{Jacoby}, George H. and {Barnes}, Jeannette},
       series = {Astronomical Society of the Pacific Conference Series},
       volume = {101},
        month = jan,
        pages = {17},
       adsurl = {https://ui.adsabs.harvard.edu/abs/1996ASPC..101...17A}
}

@ARTICLE{zdziarski_2020_thcomp,
       author = {{Zdziarski}, Andrzej A. and {Szanecki}, Micha{\l} and {Poutanen}, Juri and {Gierli{\'n}ski}, Marek and {Biernacki}, Pawe{\l}},
        title = "{Spectral and temporal properties of Compton scattering by mildly relativistic thermal electrons}",
      journal = {\mnras},
         year = 2020,
        month = mar,
       volume = {492},
       number = {4},
        pages = {5234-5246},
          doi = {10.1093/mnras/staa159},
archivePrefix = {arXiv},
       eprint = {1910.04535},
 primaryClass = {astro-ph.HE},
       adsurl = {https://ui.adsabs.harvard.edu/abs/2020MNRAS.492.5234Z}
}

@ARTICLE{mihara_maxi_2011,
       author = {{Mihara}, Tatehiro and {Nakajima}, Motoki and {Sugizaki}, Mutsumi and {Serino}, Motoko and {Matsuoka}, Masaru and {Kohama}, Mitsuhiro and {Kawasaki}, Kazuyoshi and {Tomida}, Hiroshi and {Ueno}, Shiro and {Kawai}, Nobuyuki and {Kataoka}, Jun and {Morii}, Mikio and {Yoshida}, Atsumasa and {Yamaoka}, Kazutaka and {Nakahira}, Satoshi and {Negoro}, Hitoshi and {Isobe}, Naoki and {Yamauchi}, Makoto and {Sakurai}, Ikuya},
        title = "{Gas Slit Camera (GSC) onboard MAXI on ISS}",
      journal = {\pasj},
         year = 2011,
        month = nov,
       volume = {63},
        pages = {S623-S634},
          doi = {10.1093/pasj/63.sp3.S623},
archivePrefix = {arXiv},
       eprint = {1103.4224},
 primaryClass = {astro-ph.IM},
       adsurl = {https://ui.adsabs.harvard.edu/abs/2011PASJ...63S.623M}
}

@ARTICLE{ross_fabian_refl_2007,
       author = {{Ross}, R.~R. and {Fabian}, A.~C.},
        title = "{X-ray reflection in accreting stellar-mass black hole systems}",
      journal = {\mnras},
         year = 2007,
        month = nov,
       volume = {381},
       number = {4},
        pages = {1697-1701},
          doi = {10.1111/j.1365-2966.2007.12339.x},
archivePrefix = {arXiv},
       eprint = {0709.0270},
 primaryClass = {astro-ph},
       adsurl = {https://ui.adsabs.harvard.edu/abs/2007MNRAS.381.1697R}
}

@ARTICLE{anuj_nandi_gx339_2012,
       author = {{Nandi}, A. and {Debnath}, D. and {Mandal}, S. and {Chakrabarti}, S.~K.},
        title = "{Accretion flow dynamics during the evolution of timing and spectral properties of GX 339-4 during its 2010-11 outburst}",
      journal = {\aap},
         year = 2012,
        month = jun,
       volume = {542},
          eid = {A56},
        pages = {A56},
          doi = {10.1051/0004-6361/201117844},
archivePrefix = {arXiv},
       eprint = {1204.5044},
 primaryClass = {astro-ph.HE},
       adsurl = {https://ui.adsabs.harvard.edu/abs/2012A&A...542A..56N}
}

@ARTICLE{anuj_nandi2024,
       author = {{Nandi}, Anuj and {Das}, Santabrata and {Majumder}, Seshadri and {Katoch}, Tilak and {Antia}, H.~M. and {Shah}, Parag},
        title = "{Discovery of evolving low-frequency QPOs in hard X-rays ( 100 keV) observed in black hole Swift J1727.8-1613 with AstroSat}",
      journal = {\mnras},
         year = 2024,
        month = jun,
       volume = {531},
       number = {1},
        pages = {1149-1157},
          doi = {10.1093/mnras/stae1208},
archivePrefix = {arXiv},
       eprint = {2404.17160},
 primaryClass = {astro-ph.HE},
       adsurl = {https://ui.adsabs.harvard.edu/abs/2024MNRAS.531.1149N}
}

@ARTICLE{iyer_nandi2015,
       author = {{Iyer}, N. and {Nandi}, A. and {Mandal}, S.},
        title = "{Determination of the Mass of IGR J17091-3624 from ``Spectro-temporal'' Variations during the Onset Phase of the 2011 Outburst}",
      journal = {\apj},
         year = 2015,
        month = jul,
       volume = {807},
       number = {1},
          eid = {108},
        pages = {108},
          doi = {10.1088/0004-637X/807/1/108},
archivePrefix = {arXiv},
       eprint = {1505.02529},
 primaryClass = {astro-ph.HE},
       adsurl = {https://ui.adsabs.harvard.edu/abs/2015ApJ...807..108I}
}

@ARTICLE{mitsuda_diskbb1984,
       author = {{Mitsuda}, K. and {Inoue}, H. and {Koyama}, K. and {Makishima}, K. and {Matsuoka}, M. and {Ogawara}, Y. and {Shibazaki}, N. and {Suzuki}, K. and {Tanaka}, Y. and {Hirano}, T.},
        title = "{Energy spectra of low-mass binary X-ray sources observed from Tenma.}",
      journal = {\pasj},
         year = 1984,
        month = jan,
       volume = {36},
        pages = {741-759},
       adsurl = {https://ui.adsabs.harvard.edu/abs/1984PASJ...36..741M}
}

@ARTICLE{makishima_diskbb1986,
       author = {{Makishima}, K. and {Maejima}, Y. and {Mitsuda}, K. and {Bradt}, H.~V. and {Remillard}, R.~A. and {Tuohy}, I.~R. and {Hoshi}, R. and {Nakagawa}, M.},
        title = "{Simultaneous X-Ray and Optical Observations of GX 339-4 in an X-Ray High State}",
      journal = {\apj},
         year = 1986,
        month = sep,
       volume = {308},
        pages = {635},
          doi = {10.1086/164534},
       adsurl = {https://ui.adsabs.harvard.edu/abs/1986ApJ...308..635M}
}

@ARTICLE{athulya2023,
       author = {{Athulya}, M.~P. and {Nandi}, Anuj},
        title = "{Multimission view of the low-luminosity 'obscured' phase of GRS 1915+105}",
      journal = {\mnras},
         year = 2023,
        month = oct,
       volume = {525},
       number = {1},
        pages = {489-507},
          doi = {10.1093/mnras/stad2072},
archivePrefix = {arXiv},
       eprint = {2307.04206},
 primaryClass = {astro-ph.HE},
       adsurl = {https://ui.adsabs.harvard.edu/abs/2023MNRAS.525..489A}
}

@ARTICLE{rawat_2019_grs,
       author = {{Rawat}, Divya and {Pahari}, Mayukh and {Yadav}, J.~S. and {Jain}, Pankaj and {Misra}, Ranjeev and {Bagri}, Kalyani and {Katoch}, Tilak and {Agrawal}, P.~C. and {Manchanda}, R.~K.},
        title = "{Study of Timing Evolution from Nonvariable to Structured Large-amplitude Variability Transition in GRS 1915 + 105 Using AstroSat}",
      journal = {\apj},
         year = 2019,
        month = jan,
       volume = {870},
       number = {1},
          eid = {4},
        pages = {4},
          doi = {10.3847/1538-4357/aaefed},
archivePrefix = {arXiv},
       eprint = {1811.03393},
 primaryClass = {astro-ph.HE},
       adsurl = {https://ui.adsabs.harvard.edu/abs/2019ApJ...870....4R}
}

@ARTICLE{zhang_2020_grs,
       author = {{Zhang}, Liang and {M{\'e}ndez}, Mariano and {Altamirano}, Diego and {Qu}, Jinlu and {Chen}, Li and {Karpouzas}, Konstantinos and {Belloni}, Tomaso M. and {Bu}, Qingcui and {Huang}, Yue and {Ma}, Xiang and {Tao}, Lian and {Wang}, Yanan},
        title = "{A systematic analysis of the phase lags associated with the type-C quasi-periodic oscillation in GRS 1915+105}",
      journal = {\mnras},
         year = 2020,
        month = may,
       volume = {494},
       number = {1},
        pages = {1375-1386},
          doi = {10.1093/mnras/staa797},
archivePrefix = {arXiv},
       eprint = {2003.08928},
 primaryClass = {astro-ph.HE},
       adsurl = {https://ui.adsabs.harvard.edu/abs/2020MNRAS.494.1375Z}
}

@ARTICLE{belloni_hasinger1990,
       author = {{Belloni}, T. and {Hasinger}, G.},
        title = "{Variability in the noise properties of Cygnus X-1.}",
      journal = {\aap},
         year = 1990,
        month = jan,
       volume = {227},
        pages = {L33-L36},
       adsurl = {https://ui.adsabs.harvard.edu/abs/1990A&A...227L..33B}
}

@ARTICLE{belloni2024,
       author = {{Belloni}, Tomaso M. and {M{\'e}ndez}, Mariano and {Garc{\'\i}a}, Federico and {Bhattacharya}, Dipankar},
        title = "{Fast-varying time lags in the quasi-periodic oscillation in GRS 1915 + 105}",
      journal = {\mnras},
         year = 2024,
        month = jan,
       volume = {527},
       number = {3},
        pages = {7136-7143},
          doi = {10.1093/mnras/stad3639},
archivePrefix = {arXiv},
       eprint = {2311.13467},
 primaryClass = {astro-ph.HE},
       adsurl = {https://ui.adsabs.harvard.edu/abs/2024MNRAS.527.7136B}
}

@ARTICLE{reig_kylafis2015_bh-jet,
       author = {{Reig}, P. and {Kylafis}, N.~D.},
        title = "{A jet model for Galactic black-hole X-ray sources: The correlation between cutoff energy and phase lag}",
      journal = {\aap},
         year = 2015,
        month = dec,
       volume = {584},
          eid = {A109},
        pages = {A109},
          doi = {10.1051/0004-6361/201527151},
archivePrefix = {arXiv},
       eprint = {1510.05357},
 primaryClass = {astro-ph.HE},
       adsurl = {https://ui.adsabs.harvard.edu/abs/2015A&A...584A.109R}
}

@BOOK{seward_charles2010,
       author = {{Seward}, Frederick D. and {Charles}, Philip A.},
        title = "{Exploring the X-ray Universe}",
         year = 2010,
       adsurl = {https://ui.adsabs.harvard.edu/abs/2010exru.book.....S}
}

@ARTICLE{soleri_2008_grs_type-b,
       author = {{Soleri}, P. and {Belloni}, T. and {Casella}, P.},
        title = "{A transient low-frequency quasi-periodic oscillation from the black hole binary GRS 1915+105}",
      journal = {\mnras},
         year = 2008,
        month = jan,
       volume = {383},
       number = {3},
        pages = {1089-1102},
          doi = {10.1111/j.1365-2966.2007.12596.x},
archivePrefix = {arXiv},
       eprint = {0710.3030},
 primaryClass = {astro-ph},
       adsurl = {https://ui.adsabs.harvard.edu/abs/2008MNRAS.383.1089S}
}

@ARTICLE{casella_2005,
       author = {{Casella}, P. and {Belloni}, T. and {Stella}, L.},
        title = "{The ABC of Low-Frequency Quasi-periodic Oscillations in Black Hole Candidates: Analogies with Z Sources}",
      journal = {\apj},
         year = 2005,
        month = aug,
       volume = {629},
       number = {1},
        pages = {403-407},
          doi = {10.1086/431174},
archivePrefix = {arXiv},
       eprint = {astro-ph/0504318},
 primaryClass = {astro-ph},
       adsurl = {https://ui.adsabs.harvard.edu/abs/2005ApJ...629..403C}
}

@ARTICLE{muno_1999,
       author = {{Muno}, Michael P. and {Morgan}, Edward H. and {Remillard}, Ronald A.},
        title = "{Quasi-periodic Oscillations and Spectral States in GRS 1915+105}",
      journal = {\apj},
         year = 1999,
        month = dec,
       volume = {527},
       number = {1},
        pages = {321-340},
          doi = {10.1086/308063},
       adsurl = {https://ui.adsabs.harvard.edu/abs/1999ApJ...527..321M}
}

@ARTICLE{yan_2013,
       author = {{Yan}, Shu-Ping and {Ding}, Guo-Qiang and {Wang}, Na and {Qu}, Jin-Lu and {Song}, Li-Ming},
        title = "{A statistical study on the low-frequency quasi-periodic oscillation amplitude spectrum and amplitude in GRS 1915+105}",
      journal = {\mnras},
         year = 2013,
        month = sep,
       volume = {434},
       number = {1},
        pages = {59-68},
          doi = {10.1093/mnras/stt968},
archivePrefix = {arXiv},
       eprint = {1306.0640},
 primaryClass = {astro-ph.HE},
       adsurl = {https://ui.adsabs.harvard.edu/abs/2013MNRAS.434...59Y}
}

@ARTICLE{rodriguez_2004,
       author = {{Rodriguez}, J. and {Corbel}, S. and {Hannikainen}, D.~C. and {Belloni}, T. and {Paizis}, A. and {Vilhu}, O.},
        title = "{Spectral Properties of Low-Frequency Quasi-periodic Oscillations in GRS 1915+105}",
      journal = {\apj},
         year = 2004,
        month = nov,
       volume = {615},
       number = {1},
        pages = {416-421},
          doi = {10.1086/423978},
archivePrefix = {arXiv},
       eprint = {astro-ph/0407076},
 primaryClass = {astro-ph},
       adsurl = {https://ui.adsabs.harvard.edu/abs/2004ApJ...615..416R}
}

@ARTICLE{pahari_2013,
       author = {{Pahari}, Mayukh and {Neilsen}, Joseph and {Yadav}, J.~S. and {Misra}, Ranjeev and {Uttley}, Phil},
        title = "{Comparison of Time/Phase Lags in the Hard State and Plateau State of GRS 1915+105}",
      journal = {\apj},
         year = 2013,
        month = dec,
       volume = {778},
       number = {2},
          eid = {136},
        pages = {136},
          doi = {10.1088/0004-637X/778/2/136},
archivePrefix = {arXiv},
       eprint = {1310.3037},
 primaryClass = {astro-ph.HE},
       adsurl = {https://ui.adsabs.harvard.edu/abs/2013ApJ...778..136P}
}

@ARTICLE{misra2020_grs1915,
       author = {{Misra}, Ranjeev and {Rawat}, Divya and {Yadav}, J.~S. and {Jain}, Pankaj},
        title = "{Identification of QPO Frequency of GRS 1915+105 as the Relativistic Dynamic Frequency of a Truncated Accretion Disk}",
      journal = {\apjl},
         year = 2020,
        month = feb,
       volume = {889},
       number = {2},
          eid = {L36},
        pages = {L36},
          doi = {10.3847/2041-8213/ab6ddc},
archivePrefix = {arXiv},
       eprint = {2001.07452},
 primaryClass = {astro-ph.HE},
       adsurl = {https://ui.adsabs.harvard.edu/abs/2020ApJ...889L..36M}
}

@ARTICLE{anubhab2021_grs1915,
       author = {{Banerjee}, Anuvab and {Bhattacharjee}, Ayan and {Chatterjee}, Debjit and {Debnath}, Dipak and {Chakrabarti}, Sandip Kumar and {Katoch}, Tilak and {Antia}, H.~M.},
        title = "{Accretion Flow Properties of GRS 1915+105 During Its {\ensuremath{\theta}} Class Using AstroSat Data}",
      journal = {\apj},
         year = 2021,
        month = aug,
       volume = {916},
       number = {2},
          eid = {68},
        pages = {68},
          doi = {10.3847/1538-4357/ac0150},
archivePrefix = {arXiv},
       eprint = {2007.05273},
 primaryClass = {astro-ph.HE},
       adsurl = {https://ui.adsabs.harvard.edu/abs/2021ApJ...916...68B}
}

@INPROCEEDINGS{kp_singh2014_astrosat,
       author = {{Singh}, Kulinder Pal and {Tandon}, S.~N. and {Agrawal}, P.~C. and {Antia}, H.~M. and {Manchanda}, R.~K. and {Yadav}, J.~S. and {Seetha}, S. and {Ramadevi}, M.~C. and {Rao}, A.~R. and {Bhattacharya}, D. and {Paul}, B. and {Sreekumar}, P. and {Bhattacharyya}, S. and {Stewart}, G.~C. and {Hutchings}, J. and {Annapurni}, S.~A. and {Ghosh}, S.~K. and {Murthy}, J. and {Pati}, A. and {Rao}, N.~K. and {Stalin}, C.~S. and {Girish}, V. and {Sankarasubramanian}, K. and {Vadawale}, S. and {Bhalerao}, V.~B. and {Dewangan}, G.~C. and {Dedhia}, D.~K. and {Hingar}, M.~K. and {Katoch}, T.~B. and {Kothare}, A.~T. and {Mirza}, I. and {Mukerjee}, K. and {Shah}, H. and {Shah}, P. and {Mohan}, R. and {Sangal}, A.~K. and {Nagabhusana}, S. and {Sriram}, S. and {Malkar}, J.~P. and {Sreekumar}, S. and {Abbey}, A.~F. and {Hansford}, G.~M. and {Beardmore}, A.~P. and {Sharma}, M.~R. and {Murthy}, S. and {Kulkarni}, R. and {Meena}, G. and {Babu}, V.~C. and {Postma}, J.},
        title = "{ASTROSAT mission}",
    booktitle = {Space Telescopes and Instrumentation 2014: Ultraviolet to Gamma Ray},
         year = 2014,
       editor = {{Takahashi}, Tadayuki and {den Herder}, Jan-Willem A. and {Bautz}, Mark},
       series = {Society of Photo-Optical Instrumentation Engineers (SPIE) Conference Series},
       volume = {9144},
        month = jul,
          eid = {91441S},
        pages = {91441S},
          doi = {10.1117/12.2062667},
       adsurl = {https://ui.adsabs.harvard.edu/abs/2014SPIE.9144E..1SS}
}

@ARTICLE{chakrabarti_nandi2005,
       author = {{Chakrabarti}, S.~K. and {Nandi}, A. and {Chatterjee}, A.~K. and {Choudhury}, A.~K. and {Chatterjee}, U.},
        title = "{Class transitions and two component accretion flow in GRS 1915+105}",
      journal = {\aap},
         year = 2005,
        month = mar,
       volume = {431},
       number = {3},
        pages = {825-830},
          doi = {10.1051/0004-6361:20041662},
archivePrefix = {arXiv},
       eprint = {astro-ph/0501277},
 primaryClass = {astro-ph},
       adsurl = {https://ui.adsabs.harvard.edu/abs/2005A&A...431..825C}
}

@ARTICLE{choudhury_etal2025,
       author = {{Choudhury}, Sreetama Das and {Bhuvana}, G.~R. and {Das}, Santabrata and {Nandi}, Anuj},
        title = "{Revisiting disc{\textendash}jet coupling in black hole X-ray binaries: on the nature of disc dynamics and jet velocity}",
      journal = {\mnras},
         year = 2025,
        month = aug,
       volume = {541},
       number = {4},
        pages = {2934-2954},
          doi = {10.1093/mnras/staf1107},
archivePrefix = {arXiv},
       eprint = {2507.03644},
 primaryClass = {astro-ph.HE},
       adsurl = {https://ui.adsabs.harvard.edu/abs/2025MNRAS.541.2934C}
}

@ARTICLE{huppenkothen_2017_GRS_ML,
       author = {{Huppenkothen}, Daniela and {Heil}, Lucy M. and {Hogg}, David W. and {Mueller}, Andreas},
        title = "{Using machine learning to explore the long-term evolution of GRS 1915+105}",
      journal = {\mnras},
         year = 2017,
        month = apr,
       volume = {466},
       number = {2},
        pages = {2364-2377},
          doi = {10.1093/mnras/stw3190},
archivePrefix = {arXiv},
       eprint = {1611.01332},
 primaryClass = {astro-ph.HE},
       adsurl = {https://ui.adsabs.harvard.edu/abs/2017MNRAS.466.2364H}
}

@ARTICLE{vadawale2003,
       author = {{Vadawale}, S.~V. and {Rao}, A.~R. and {Naik}, S. and {Yadav}, J.~S. and {Ishwara-Chandra}, C.~H. and {Pramesh Rao}, A. and {Pooley}, G.~G.},
        title = "{On the Origin of the Various Types of Radio Emission in GRS 1915+105}",
      journal = {\apj},
         year = 2003,
        month = nov,
       volume = {597},
       number = {2},
        pages = {1023-1035},
          doi = {10.1086/378672},
archivePrefix = {arXiv},
       eprint = {astro-ph/0308096},
 primaryClass = {astro-ph},
       adsurl = {https://ui.adsabs.harvard.edu/abs/2003ApJ...597.1023V}
}

@ARTICLE{van_eijnden_2017,
       author = {{van den Eijnden}, J. and {Ingram}, A. and {Uttley}, P. and {Motta}, S.~E. and {Belloni}, T.~M. and {Gardenier}, D.~W.},
        title = "{Inclination dependence of QPO phase lags in black hole X-ray binaries}",
      journal = {\mnras},
         year = 2017,
        month = jan,
       volume = {464},
       number = {3},
        pages = {2643-2659},
          doi = {10.1093/mnras/stw2634},
archivePrefix = {arXiv},
       eprint = {1610.03469},
 primaryClass = {astro-ph.HE},
       adsurl = {https://ui.adsabs.harvard.edu/abs/2017MNRAS.464.2643V}
}

@ARTICLE{nobili2000,
       author = {{Nobili}, L. and {Turolla}, R. and {Zampieri}, L. and {Belloni}, T.},
        title = "{A Comptonization Model for Phase-Lag Variability in GRS 1915+105}",
      journal = {\apjl},
         year = 2000,
        month = aug,
       volume = {538},
       number = {2},
        pages = {L137-L140},
          doi = {10.1086/312810},
archivePrefix = {arXiv},
       eprint = {astro-ph/0006091},
 primaryClass = {astro-ph},
       adsurl = {https://ui.adsabs.harvard.edu/abs/2000ApJ...538L.137N}
}

@ARTICLE{prajjwal2025,
       author = {{Majumder}, Prajjwal and {Dutta}, Broja G. and {Nandi}, Anuj},
        title = "{Decoding the origin of HFQPOs of GRS 1915 + 105 during 'Canonical' soft states: an in-depth view using multimission observations}",
      journal = {\mnras},
         year = 2025,
        month = jun,
       volume = {540},
       number = {1},
        pages = {37-51},
          doi = {10.1093/mnras/staf698},
archivePrefix = {arXiv},
       eprint = {2504.14193},
 primaryClass = {astro-ph.HE},
       adsurl = {https://ui.adsabs.harvard.edu/abs/2025MNRAS.540...37M}
}

@ARTICLE{ferreira2022,
       author = {{Ferreira}, J. and {Marcel}, G. and {Petrucci}, P. -O. and {Rodriguez}, J. and {Malzac}, J. and {Belmont}, R. and {Clavel}, M. and {Henri}, G. and {Corbel}, S. and {Coriat}, M.},
        title = "{Are low-frequency quasi-periodic oscillations in accretion flows the disk response to jet instability?}",
      journal = {\aap},
         year = 2022,
        month = apr,
       volume = {660},
          eid = {A66},
        pages = {A66},
          doi = {10.1051/0004-6361/202040165},
archivePrefix = {arXiv},
       eprint = {2202.00438},
 primaryClass = {astro-ph.HE},
       adsurl = {https://ui.adsabs.harvard.edu/abs/2022A&A...660A..66F}
}

@ARTICLE{molteni1996,
       author = {{Molteni}, Diego and {Sponholz}, Hanno and {Chakrabarti}, Sandip K.},
        title = "{Resonance Oscillation of Radiative Shock Waves in Accretion Disks around Compact Objects}",
      journal = {\apj},
         year = 1996,
        month = feb,
       volume = {457},
        pages = {805},
          doi = {10.1086/176775},
archivePrefix = {arXiv},
       eprint = {astro-ph/9508022},
 primaryClass = {astro-ph},
       adsurl = {https://ui.adsabs.harvard.edu/abs/1996ApJ...457..805M}
}

@ARTICLE{ingram2009,
       author = {{Ingram}, Adam and {Done}, Chris and {Fragile}, P. Chris},
        title = "{Low-frequency quasi-periodic oscillations spectra and Lense-Thirring precession}",
      journal = {\mnras},
         year = 2009,
        month = jul,
       volume = {397},
       number = {1},
        pages = {L101-L105},
          doi = {10.1111/j.1745-3933.2009.00693.x},
archivePrefix = {arXiv},
       eprint = {0901.1238},
 primaryClass = {astro-ph.SR},
       adsurl = {https://ui.adsabs.harvard.edu/abs/2009MNRAS.397L.101I}
}

@ARTICLE{garain2014,
       author = {{Garain}, Sudip K. and {Ghosh}, Himadri and {Chakrabarti}, Sandip K.},
        title = "{Quasi-periodic oscillations in a radiative transonic flow: results of a coupled Monte Carlo-TVD simulation}",
      journal = {\mnras},
         year = 2014,
        month = jan,
       volume = {437},
       number = {2},
        pages = {1329-1336},
          doi = {10.1093/mnras/stt1969},
archivePrefix = {arXiv},
       eprint = {1310.6493},
 primaryClass = {astro-ph.HE},
       adsurl = {https://ui.adsabs.harvard.edu/abs/2014MNRAS.437.1329G}
}

@ARTICLE{molteni1994,
       author = {{Molteni}, Diego and {Lanzafame}, Giuseppe and {Chakrabarti}, Sandip K.},
        title = "{Simulation of Thick Accretion Disks with Standing Shocks by Smoothed Particle Hydrodynamics}",
      journal = {\apj},
         year = 1994,
        month = apr,
       volume = {425},
        pages = {161},
          doi = {10.1086/173972},
archivePrefix = {arXiv},
       eprint = {astro-ph/9310047},
 primaryClass = {astro-ph},
       adsurl = {https://ui.adsabs.harvard.edu/abs/1994ApJ...425..161M}
}

@ARTICLE{debnath2024,
       author = {{Debnath}, Dipak and {Nath}, Sujoy Kumar and {Chatterjee}, Debjit and {Chatterjee}, Kaushik and {Chang}, Hsiang-Kuang},
        title = "{Detection of QPO Soft Lag during the Outburst of Swift J1727.8-1613: Estimation of Intrinsic Parameters from Spectral Study}",
      journal = {\apj},
         year = 2024,
        month = nov,
       volume = {975},
       number = {2},
          eid = {194},
        pages = {194},
          doi = {10.3847/1538-4357/ad7a76},
archivePrefix = {arXiv},
       eprint = {2409.14286},
 primaryClass = {astro-ph.HE},
       adsurl = {https://ui.adsabs.harvard.edu/abs/2024ApJ...975..194D}
}

@ARTICLE{pragati_sahu2024,
       author = {{Sahu}, Pragati and {Chand}, Swadesh and {Thakur}, Parijat and {Dewangan}, G.~C. and {Agrawal}, V.~K. and {Tripathi}, Prakash and {Das}, Subhashish},
        title = "{2017 Outburst of H 1743{\textendash}322: AstroSat and Swift View}",
      journal = {\apj},
         year = 2024,
        month = nov,
       volume = {975},
       number = {2},
          eid = {165},
        pages = {165},
          doi = {10.3847/1538-4357/ad7a6d},
archivePrefix = {arXiv},
       eprint = {2409.10253},
 primaryClass = {astro-ph.HE},
       adsurl = {https://ui.adsabs.harvard.edu/abs/2024ApJ...975..165S}
}

@ARTICLE{santa_das2014,
       author = {{Das}, Santabrata and {Chattopadhyay}, Indranil and {Nandi}, Anuj and {Molteni}, D.},
        title = "{Periodic mass loss from viscous accretion flows around black holes}",
      journal = {\mnras},
         year = 2014,
        month = jul,
       volume = {442},
       number = {1},
        pages = {251-258},
          doi = {10.1093/mnras/stu864},
archivePrefix = {arXiv},
       eprint = {1405.4415},
 primaryClass = {astro-ph.HE},
       adsurl = {https://ui.adsabs.harvard.edu/abs/2014MNRAS.442..251D}
}

@ARTICLE{karpouzas_2021,
       author = {{Karpouzas}, Konstantinos and {M{\'e}ndez}, Mariano and {Garc{\'\i}a}, Federico and {Zhang}, Liang and {Altamirano}, Diego and {Belloni}, Tomaso and {Zhang}, Yuexin},
        title = "{A variable corona for GRS 1915+105}",
      journal = {\mnras},
         year = 2021,
        month = jun,
       volume = {503},
       number = {4},
        pages = {5522-5533},
          doi = {10.1093/mnras/stab827},
archivePrefix = {arXiv},
       eprint = {2103.09675},
 primaryClass = {astro-ph.HE},
       adsurl = {https://ui.adsabs.harvard.edu/abs/2021MNRAS.503.5522K}
}

@ARTICLE{chen1997,
       author = {{Chen}, Xingming and {Swank}, Jean H. and {Taam}, Ronald E.},
        title = "{The Pattern of Correlated X-Ray Timing and Spectral Behavior in GRS 1915+105}",
      journal = {\apjl},
         year = 1997,
        month = mar,
       volume = {477},
       number = {1},
        pages = {L41-L44},
          doi = {10.1086/310515},
archivePrefix = {arXiv},
       eprint = {astro-ph/9612128},
 primaryClass = {astro-ph},
       adsurl = {https://ui.adsabs.harvard.edu/abs/1997ApJ...477L..41C}
}

@ARTICLE{markwardt1999,
       author = {{Markwardt}, Craig B. and {Swank}, Jean H. and {Taam}, Ronald E.},
        title = "{Variable-Frequency Quasi-periodic Oscillations from the Galactic Microquasar GRS 1915+105}",
      journal = {\apjl},
         year = 1999,
        month = mar,
       volume = {513},
       number = {1},
        pages = {L37-L40},
          doi = {10.1086/311899},
archivePrefix = {arXiv},
       eprint = {astro-ph/9901050},
 primaryClass = {astro-ph},
       adsurl = {https://ui.adsabs.harvard.edu/abs/1999ApJ...513L..37M}
}

@ARTICLE{seshadri2025,
       author = {{Majumder}, Seshadri and {Das}, Santabrata and {Nandi}, Anuj},
        title = "{Possible detection of HFQPOs associated with 'unknown' variability class of GRS 1915+105}",
      journal = {\pasa},
         year = 2025,
        month = nov,
       volume = {42},
          eid = {e144},
        pages = {e144},
          doi = {10.1017/pasa.2025.10105},
archivePrefix = {arXiv},
       eprint = {2508.08594},
 primaryClass = {astro-ph.HE},
       adsurl = {https://ui.adsabs.harvard.edu/abs/2025PASA...42..144M}
}

@ARTICLE{krim2013_BAT,
       author = {{Krimm}, H.~A. and {Holland}, S.~T. and {Corbet}, R.~H.~D. and {Pearlman}, A.~B. and {Romano}, P. and {Kennea}, J.~A. and {Bloom}, J.~S. and {Barthelmy}, S.~D. and {Baumgartner}, W.~H. and {Cummings}, J.~R. and et al.},
        title = "{The Swift/BAT Hard X-Ray Transient Monitor}",
      journal = {\apjs},
         year = 2013,
        month = nov,
       volume = {209},
       number = {1},
          eid = {14},
        pages = {14},
          doi = {10.1088/0067-0049/209/1/14},
archivePrefix = {arXiv},
       eprint = {1309.0755},
 primaryClass = {astro-ph.HE},
       adsurl = {https://ui.adsabs.harvard.edu/abs/2013ApJS..209...14K}
}

@ARTICLE{santa_das2021,
       author = {{Das}, Santabrata and {Nandi}, Anuj and {Agrawal}, Vivek K. and {Dihingia}, Indu Kalpa and {Majumder}, Seshadri},
        title = "{Relativistic viscous accretion flow model for ULX sources: a case study for IC 342 X-1}",
      journal = {\mnras},
         year = 2021,
        month = oct,
       volume = {507},
       number = {2},
        pages = {2777-2781},
          doi = {10.1093/mnras/stab2307},
archivePrefix = {arXiv},
       eprint = {2108.02973},
 primaryClass = {astro-ph.HE},
       adsurl = {https://ui.adsabs.harvard.edu/abs/2021MNRAS.507.2777D}
}

@ARTICLE{chakrabarti_nandi2000,
       author = {{Chakrabarti}, Sandip K. and {Nandi}, Anuj},
        title = "{Fundamental States of Accretion/Jet Configuration and the Black Hole Candidate GRS1915+105}",
      journal = {arXiv e-prints},
         year = 2000,
        month = dec,
          eid = {astro-ph/0012526},
        pages = {astro-ph/0012526},
          doi = {10.48550/arXiv.astro-ph/0012526},
archivePrefix = {arXiv},
       eprint = {astro-ph/0012526},
 primaryClass = {astro-ph},
       adsurl = {https://ui.adsabs.harvard.edu/abs/2000astro.ph.12526C}
}

@ARTICLE{nandi_2001,
       author = {{Nandi}, A. and {Chakrabarti}, S.~K. and {Vadawale}, S.~V. and {Rao}, A.~R.},
        title = "{Ejection of the inner accretion disk in GRS 1915+105: The magnetic rubber-band effect}",
      journal = {\aap},
         year = 2001,
        month = dec,
       volume = {380},
        pages = {245-250},
          doi = {10.1051/0004-6361:20011444},
archivePrefix = {arXiv},
       eprint = {astro-ph/0402550},
 primaryClass = {astro-ph},
       adsurl = {https://ui.adsabs.harvard.edu/abs/2001A&A...380..245N}
}

@ARTICLE{chakrabarti2002,
       author = {{Chakrabarti}, Sandip K. and {Nandi}, Anuj and {Manickam}, S.~G. and {Mandal}, S. and {Rao}, A.~R.},
        title = "{Spectral Signature of Mass Loss from (and Mass Gain by) an Accretion Disk around a Black Hole}",
      journal = {\apjl},
         year = 2002,
        month = nov,
       volume = {579},
       number = {1},
        pages = {L21-L24},
          doi = {10.1086/344783},
archivePrefix = {arXiv},
       eprint = {astro-ph/0012515},
 primaryClass = {astro-ph},
       adsurl = {https://ui.adsabs.harvard.edu/abs/2002ApJ...579L..21C}
}

@ARTICLE{garcia2022,
       author = {{Garc{\'\i}a}, Federico and {Karpouzas}, Konstantinos and {M{\'e}ndez}, Mariano and {Zhang}, Liang and {Zhang}, Yuexin and {Belloni}, Tomaso and {Altamirano}, Diego},
        title = "{The evolving properties of the corona of GRS 1915+105: a spectral-timing perspective through variable-Comptonization modelling}",
      journal = {\mnras},
         year = 2022,
        month = jul,
       volume = {513},
       number = {3},
        pages = {4196-4207},
          doi = {10.1093/mnras/stac1202},
archivePrefix = {arXiv},
       eprint = {2204.13279},
 primaryClass = {astro-ph.HE},
       adsurl = {https://ui.adsabs.harvard.edu/abs/2022MNRAS.513.4196G}
}

@ARTICLE{mendez_nat2022,
       author = {{M{\'e}ndez}, Mariano and {Karpouzas}, Konstantinos and {Garc{\'\i}a}, Federico and {Zhang}, Liang and {Zhang}, Yuexin and {Belloni}, Tomaso M. and {Altamirano}, Diego},
        title = "{Coupling between the accreting corona and the relativistic jet in the microquasar GRS 1915+105}",
      journal = {Nature Astronomy},
         year = 2022,
        month = mar,
       volume = {6},
        pages = {577-583},
          doi = {10.1038/s41550-022-01617-y},
archivePrefix = {arXiv},
       eprint = {2203.02963},
 primaryClass = {astro-ph.HE},
       adsurl = {https://ui.adsabs.harvard.edu/abs/2022NatAs...6..577M}
}

@ARTICLE{chakrabarti2009,
       author = {{Chakrabarti}, Sandip K. and {Dutta}, Broja G. and {Pal}, P.~S.},
        title = "{Accretion flow behaviour during the evolution of the quasi-periodic oscillation frequency of XTE J1550-564 in 1998 outburst}",
      journal = {\mnras},
         year = 2009,
        month = apr,
       volume = {394},
       number = {3},
        pages = {1463-1468},
          doi = {10.1111/j.1365-2966.2008.14328.x},
archivePrefix = {arXiv},
       eprint = {0906.5068},
 primaryClass = {astro-ph.HE},
       adsurl = {https://ui.adsabs.harvard.edu/abs/2009MNRAS.394.1463C}
}

@ARTICLE{dutta_2010,
       author = {{Dutta}, Broja G. and {Chakrabarti}, Sandip K.},
        title = "{Evidence for two-component flows around the black hole candidate XTEJ1550-564 from spectral features during its 1998-1999 outburst}",
      journal = {\mnras},
         year = 2010,
        month = jun,
       volume = {404},
       number = {4},
        pages = {2136-2142},
          doi = {10.1111/j.1365-2966.2010.16428.x},
       adsurl = {https://ui.adsabs.harvard.edu/abs/2010MNRAS.404.2136D}
}

@ARTICLE{radhika_2016,
       author = {{Radhika}, D. and {Nandi}, A. and {Agrawal}, V.~K. and {Seetha}, S.},
        title = "{`Spectro-temporal' variabilities and possible physical mechanism for jet ejections}",
      journal = {\mnras},
         year = 2016,
        month = aug,
       volume = {460},
       number = {4},
        pages = {4403-4416},
          doi = {10.1093/mnras/stw1239},
archivePrefix = {arXiv},
       eprint = {1605.08351},
 primaryClass = {astro-ph.HE},
       adsurl = {https://ui.adsabs.harvard.edu/abs/2016MNRAS.460.4403R}
}

@ARTICLE{wijnands_1999_QPO-type,
       author = {{Wijnands}, Rudy and {Homan}, Jeroen and {van der Klis}, Michiel},
        title = "{The Complex Phase-Lag Behavior of the 3-12 HZ Quasi-Periodic Oscillations during the Very High State of XTE J1550-564}",
      journal = {\apjl},
         year = 1999,
        month = nov,
       volume = {526},
       number = {1},
        pages = {L33-L36},
          doi = {10.1086/312365},
archivePrefix = {arXiv},
       eprint = {astro-ph/9909515},
 primaryClass = {astro-ph},
       adsurl = {https://ui.adsabs.harvard.edu/abs/1999ApJ...526L..33W}
}

@ARTICLE{casella_2004_QPO-type,
       author = {{Casella}, P. and {Belloni}, T. and {Homan}, J. and {Stella}, L.},
        title = "{A study of the low-frequency quasi-periodic oscillations in the X-ray light curves of the black hole candidate <ASTROBJ>XTE J1859+226</ASTROBJ>}",
      journal = {\aap},
         year = 2004,
        month = nov,
       volume = {426},
        pages = {587-600},
          doi = {10.1051/0004-6361:20041231},
archivePrefix = {arXiv},
       eprint = {astro-ph/0407262},
 primaryClass = {astro-ph},
       adsurl = {https://ui.adsabs.harvard.edu/abs/2004A&A...426..587C}
}

@ARTICLE{sreehari_2025,
       author = {{Harikesh}, Sreehari and {Majumder}, Seshadri and {Das}, Santabrata and {Nandi}, Anuj},
        title = "{Evidence of oscillating 'compact' Comptonized corona in GRS 1915+105: insights into HFQPOs with AstroSat}",
      journal = {\mnras},
         year = 2025,
        month = jul,
       volume = {540},
       number = {4},
        pages = {2965-2974},
          doi = {10.1093/mnras/staf926},
archivePrefix = {arXiv},
       eprint = {2506.00935},
 primaryClass = {astro-ph.HE},
       adsurl = {https://ui.adsabs.harvard.edu/abs/2025MNRAS.540.2965H}
}

@ARTICLE{titarchuk_2004,
       author = {{Titarchuk}, Lev and {Fiorito}, Ralph},
        title = "{Spectral Index and Quasi-Periodic Oscillation Frequency Correlation in Black Hole Sources: Observational Evidence of Two Phases and Phase Transition in Black Holes}",
      journal = {\apj},
         year = 2004,
        month = sep,
       volume = {612},
       number = {2},
        pages = {988-999},
          doi = {10.1086/422573},
archivePrefix = {arXiv},
       eprint = {astro-ph/0405360},
 primaryClass = {astro-ph},
       adsurl = {https://ui.adsabs.harvard.edu/abs/2004ApJ...612..988T}
}

@ARTICLE{chakrabarti_nandi2004,
       author = {{Chakrabarti}, Sandip K. and {Nandi}, A. and {Choudhury}, Asit and {Chatterjee}, Utpal},
        title = "{Evidence of Class Transitions in GRS 1915+105 from Indian X-Ray Astronomy Experiment Data}",
      journal = {\apj},
         year = 2004,
        month = may,
       volume = {607},
       number = {1},
        pages = {406-409},
          doi = {10.1086/383235},
archivePrefix = {arXiv},
       eprint = {astro-ph/0501282},
 primaryClass = {astro-ph},
       adsurl = {https://ui.adsabs.harvard.edu/abs/2004ApJ...607..406C}
}

@ARTICLE{Garg_grs2022,
       author = {{Garg}, Akash and {Misra}, Ranjeev and {Sen}, Somasri},
        title = "{On the energy dependence of the QPO phenomenon in the black hole system MAXI J1535-571}",
      journal = {\mnras},
         year = 2022,
        month = aug,
       volume = {514},
       number = {3},
        pages = {3285-3293},
          doi = {10.1093/mnras/stac1490},
archivePrefix = {arXiv},
       eprint = {2205.11899},
 primaryClass = {astro-ph.HE},
       adsurl = {https://ui.adsabs.harvard.edu/abs/2022MNRAS.514.3285G}
}







\bsp	
\label{lastpage}
\end{document}